\documentclass[a4paper,10pt]{article}
 \usepackage{indentfirst, latexsym,bm}
 \usepackage{amsmath}

\usepackage[top=2.6cm,bottom=3cm, outer=2cm, inner=2cm]{geometry}

 \usepackage{pifont}
 \usepackage{amsfonts}
 \usepackage{mathrsfs}
 \usepackage{array}
 \usepackage{multirow}
 \usepackage{graphicx}
 \usepackage{subfigure}
 \usepackage{picinpar}

 \usepackage{booktabs}
 \usepackage{threeparttable}
 \usepackage{mathrsfs,amsfonts,amsmath}
 \usepackage[dvips]{color}

 \newcommand{\blue}{\color{blue}}

 \newtheorem{theorem}{Theorem}[section]
 \newtheorem{definition}[theorem]{Definition}
 \newtheorem{lemma}[theorem]{Lemma}
 \newtheorem{corollary}[theorem]{Corollary}
 \newtheorem{proposition}[theorem]{Proposition}
 \newtheorem{remark}[theorem]{Remark}
 \newtheorem{condition}[theorem]{Condition}
 \newtheorem{example}{Example}[section]

 \newtheorem{assumption}[theorem]{Assumption}

\def\blue{\color{blue}}
 
 \def\blemma{\begin{lemma}}\def\elemma{\end{lemma}}
 \def\bproposition{\begin{proposition}}\def\eproposition{\end{proposition}}
 \def\ttheorem{\begin{theorem}}\def\etheorem{\end{theorem}}
 \def\bcorollary{\begin{corollary}}\def\ecorollary{\end{corollary}}
 \def\bremark{\begin{remark}}\def\eremark{\end{remark}}
 \def\bcondition{\begin{condition}} \def\econdition{\end{condition}}

 \def\benumerate{\begin{enumerate}}\def\eenumerate{\end{enumerate}}
 \def\bitemize{\begin{itemize}}\def\eitemize{\end{itemize}}

 \def\beqlb{\begin{eqnarray}}\def\eeqlb{\end{eqnarray}}
 \def\beqnn{\begin{eqnarray*}}\def\eeqnn{\end{eqnarray*}}
 \def\ar{\!\!\!&}

 \def\proof{\noindent{\it Proof.~~}}\def\qed{\hfill$\Box$\medskip}

\begin{document}

\setlength\parindent{10pt}
\pagestyle{plain}

\title{A Scaling Limit for Limit Order Books Driven by Hawkes Processes}

\author{Ulrich Horst\footnote{Department of Mathematics, and School of Business and Economics, Humboldt-Universit\"at zu Berlin
         Unter den Linden 6, 10099 Berlin, Germany; email: horst@math.hu-berlin.de}~ and Wei Xu\footnote{Department of Mathematics,  Humboldt-Universit\"at zu Berlin
         Unter den Linden 6, 10099 Berlin, Germany; email: xuwei@math.hu-berlin.de}}
\maketitle

	\begin{abstract}
	In this paper we derive a scaling limit for an infinite dimensional limit order book model driven by Hawkes random measures. The dynamics of the incoming order flow is allowed to depend on the current market price as well as on a volume indicator. With our choice of scaling the dynamics converges to a coupled SDE-ODE system where limiting best bid and ask price processes follows a diffusion dynamics, the limiting volume density functions follows an ODE in a Hilbert space and the limiting order arrival and cancellation intensities follow a Volterra-Fredholm integral equation. 

	\end{abstract}

%
%
%

\renewcommand{\baselinestretch}{1.15}\normalsize

\section{Introduction}
A significant part of financial transactions is nowadays settled through electronic limit order books (LOBs). A LOB is a record of unexecuted orders awaiting execution. From a mathematical perspective, LOBs are infinite-dimensional complex interactive stochastic processes. Incoming limit orders can be placed at infinitely many different price levels, and incoming market orders are matched against standing limit orders according to a set of priority rules.

In this paper, we prove a novel scaling result for LOBs that are driven by Hawkes random measures. Hawkes processes have been extensively used in the financial mathematics literature in recent years to capture the empirically well documented clustering and cross-dependencies between different order arrivals and cancellations; see \cite{HLR2015, LuAbergel2018} and references therein. Hawkes random measures can be viewed as infinite-dimensional Hawkes processes. They appear tailor-made to describe the dynamics of limit order books when the order arrival dynamics depends on past order placements and cancellations. With our choice of scaling the limiting dynamics of the LOB can be described by a fully coupled SDE-ODE system. The dynamics of the best bid and ask prices follows an SDE, the dynamics of the volume density functions follows an ODE on a Hilbert space, and the dynamics of the order arrival and cancellation intensities follows a Volterra-Fredholm integral equation.

Scaling limits for limit order books have received considerable attention in the financial mathematics literature in recent years. When the analysis of the order book is limited to prices or prices and aggregate volumes (e.g. at the top of the book) as in \cite{AbergelJedidi2011, BayraktarHorstSircar, Cont1}, then the limiting dynamics can naturally be described by ordinary differential equations or real-valued diffusion processes, depending on the choice of scaling. The analysis of the full book including the distribution of standing volume across different price levels is much more complex. Horst and Paulsen \cite{HP} and Horst and Kreher \cite{HK1} were the first to obtain fluid limits for the full LOB dynamics. Starting from a microscopic event-by-event description of the LOB, they proved convergence of the price-volume process to coupled ODE-PDE systems. Their scaling limits required two time scales: a fast time scale for cancelations and limit order placements outside the spread, and a comparably slow time scale for market order arrivals and limit order placements in the spread. The different times scales had at least two drawbacks: first, they imply that the proportion of market orders and spread placements is negligible in the limit; second, as shown in the recent paper \cite{HK3}, they make it impossible to obtain a non-degenerate second-order approximation for the full LOB dynamics. Our scaling limit does not require different time scales.

A model similar to \cite{HK1, HP} has been studied by Gao and Deng \cite{Gao}. They derived a deterministic ODE limit using weak convergence in the space of positive measures on a compact interval. Lakner et al. \cite{Lakner1} derived a high frequency limit for a one-sided order book model under the assumption that on average investors place their limit orders above the current best ask price. The opposite case when orders are placed in the spread with higher probability is analyzed in \cite{Lakner2}, where the authors use a coupling between a simple one-sided limit order book model and a branching random walk to characterize the diffusion limit. Bayer et al. \cite{BHQ} extends the models in \cite{HK1, HP} by introducing additional noise terms in the pre-limit in which case the dynamics can then be approximated by an SPDE in the scaling limit. With a different choice of scaling an SPDE limit for LOB models has recently been established in \cite{HK2}. Macroscopic SPDE models of limit order markets were studied in \cite{Hubalek, Marvin}. These models describe the volume dynamics by exogenous SPDEs while \cite{HK2} endogenously derived a semi-martingale random measure driving volumes from a microscopic approach.

There is considerable empirical evidence that the state of the order book, especially order imbalance at the top of the book, has a noticeable impact on order dynamics; see \cite{Biais, Cebiroglu-Horst} and references therein. Many of the aforementioned LOB models therefore allow for a dependence of the order arrival dynamics on the current state of the book. There is also empirical evidence of clustering of and cross-dependencies between order arrivals; see \cite{Chakraborti2011, Hewlett2006, HLR2015, LuAbergel2018} and references therein. Hawkes processes provide a powerful tool to model clustering and cross-dependencies of events. They were first introduced in \cite{Hawkes1971a, Hawkes1971} and have since been applied in many areas, ranging from earthquake modelling \cite{Ogata1986} to financial analysis \cite{Embrechts2011}. Recently, they have been extensively used to model the dynamics of prices volumes in limit order markets \cite{Abergel2015, Bacry2015, Bacry2014, LuAbergel2018, Rosenbaum_Hawkes, Rambaldi2017,  Zheng2014}.

In this paper, we introduce a generalization of Hawkes processes, termed Hawkes random measures, and analyse a novel class of LOB models driven by such measures. Specifically, we analyze the limiting dynamics of the LOB models when the order and tick sizes tend to zero while order arrivals and cancellations tend to infinity. Under standard assumptions on the model parameters we prove that the sequence of prices, volumes and order arrival and cancellation intensities is tight as a sequence of processes taking values in a suitable Skorohood space, and that any weak accumulation point solves a certain dynamic stochastic system. Uniqueness of solutions to this  system can not be expected in general. Under additional conditions on the model parameters we prove that the limiting LOB model always has a strictly positive spread. From this, we deduce that the limiting stochastic system is non-degenerate and that it hence has a unique solution. In order to characterize weak accumulation points as solutions to the stochastic systems we prove that any accumulation point solves the martingale problem associated with the path-dependent generator of the limit stochastic system. Several special cases can be solved in closed form.

Our framework allows for a dependence of the probability of order placements and/or cancellations at different price levels on past price changes. A dependence of order arrivals on price {\sl changes} allows us to model the arrival of large market orders that exceed the liquidity at the top of the book. A market order that exceeds the liquidity at the top of the book is typically split by the exchange into a series of smaller orders that are consecutively executed against standing volumes at less competitive prices. This may be viewed as a series of ``child market orders'' triggered by the arrival of some ``parent market order''.  Our probabilistic framework is flexible enough to capture such dynamics. Our framework also allows us to model so-called peg orders. Peg orders follow the best bid, when buying a stock, and the best offer, when selling a stock at a fixed distance. As such they are typically cancelled and immediately resubmitted after a price change occurred. In our framework this corresponds to an increase of the cancellation rate at a particular price level triggered by a price change, which then triggers an increase of the limit order arrival intensity at a different price level. Such a spatial dependence of arrival intensities on price changes can not be captured by the Markovian LOB models in \cite{HK1, HP}.

Several testable hypotheses can be inferred from our limit result. In particular, our model predicts that increasing cross-dependencies between order arrivals as well as increasing limit order arrivals and cancellations (e.g.~at the top of the book) increase price and volume volatility. Moreover, positive correlations between squared price increments and hence volatility clustering may result from cross-dependencies between different order types. An empirical verification of these hypotheses and/or an empirical analysis of the Hawkes kernels is beyond the scope of this paper, though.

The remainder of this paper is organized as follows. Hawkes random measures and a sequence of LOB models driven by Hawkes random measures is introduced in Section \ref{Sec2}. Section \ref{Sec3} states the main result of this paper, namely the characterisation of weak accumulation points as solutions to a certain stochastic system. In Section \ref{Sec4} we state additional conditions under which the limiting spread is strictly positive from which we then deduce uniqueness of solution to the limiting system. Section \ref{Sec5} establishes tightness of the state sequences and hence the existence of a weak accumulation point. In Section \ref{Sec6} we prove our result on the characterization of accumulation points.


 \section{LOB models driven by Hawkes random measures}\label{Sec2}
 \setcounter{equation}{0}

The goal of this paper is to establish a scaling limit for a sequence of limit order book models driven by Hawkes random measures. Hawkes random measures can be viewed as an extension of the Hawkes processes introduced in \cite{Hawkes1971a,Hawkes1971}; they are introduced in the following section. Subsequently, we introduce a class of LOB models driven by Hawkes random measures and state our main convergence results.

  \subsection{Hawkes random measures}\label{Sec1}
 \setcounter{equation}{0}


 Let $(\Omega,\mathscr{F},\mathbf{P})$ be a complete probability space endowed with filtration $\{\mathscr{F}_t\}_{t\geq 0}$ that satisfies the usual hypotheses. Let $(U,\mathscr{U})$ be a measurable space endowed with a base measure $\mathbf{m}(du)$.
 A real-valued two-parameter process $\{h(t,x): t \geq  0, x \in U\}$ is said to be \textit{$(\mathscr{F}_t)$-progressive} if for every $t\geq 0$ the mapping $(\omega, s, x)\mapsto h(\omega,s, x)$ restricted to $\Omega\times[0,t] \times U$ is measurable relative to $\mathscr{F}_t\times\mathscr{B}([0, t])\times \mathscr{U}$.
 Let $\mathbf{p}_t$ be a $(\mathscr{F}_t)$-point process on $U$ and $N(dt,du)$ be a random point measure on $[0,\infty)\times U$ defined as follows:
 \beqnn
 N(I,A)=\#\{s\in I: \mathbf{p}_s\in A\},\quad I \in \mathscr{B}(\mathbb{R}_+), A\in\mathscr{U}.
 \eeqnn

 \begin{definition}
 A nonnegative, $(\mathscr{F}_t)$-progressive process $\lambda(t,u)$ is called the {\rm intensity process} of $N(dt,du)$ with respect to the base measure $\mathbf{m}(du)$ if for any nonnegative $(\mathscr{F}_t)$-predictable process $H(t,u)$ on $U$,
     \beqnn
     \mathbf{E}\Big[\int_0^t\int_U H(s,u)N(ds,du)\Big]=\mathbf{E}\Big[\int_0^t ds\int_U H(s,u)\lambda(s,u)\mathbf{m}(du)\Big].
     \eeqnn
 \end{definition}

 For any nonnegative, $(\mathscr{F}_t)$-progressive process $\lambda(t,u)$ defined on $U$, we can construct a random point measure $N(dt,du)$ on $[0,\infty)\times U$ with intensity process $\lambda(t,u)$ as follows:
 \beqnn
 N([0,t],A)=\int_0^t\int_{A}\int_0^\infty \mathbf{1}_{\{z\leq \lambda(s,u)\}}N_0(ds,du,dz), \quad t\geq 0, A\in\mathscr{U},
 \eeqnn
 where $N_0(ds,du,dz)$ is a Poisson random measure on $[0,\infty)\times U\times [0,\infty)$ with intensity $ds\mathbf{m}(du)dz$.

 \begin{definition}
 We say that $N(dt,du)$ is a {\rm Hawkes random measure} on $[0,\infty)\times U$ if its intensity process $\lambda(t,u)$ can be written as
 \beqlb\label{eqn1.01}
 \lambda(t,u)=\mu(t,u)+\int_0^t\int_U \phi(s,u,v,t-s)N(ds,dv),
 \eeqlb
 where $\mu(t,u):[0,\infty)\times U \mapsto [0,\infty)$ and $\phi(t,u,v,r):[0,\infty)\times U^2\times [0,\infty) \mapsto [0,\infty) $ are $(\mathscr{F}_t)$-progressive.
 \end{definition}

 The processes $\mu(t,u)$ and $\phi(t,u,v,r)$ are called the \textit{exogenous intensity} and \textit{kernel} of the Hawkes random measure $N(dt,du)$, respectively. Let $\tilde{N}(ds,du)$ be the compensated random measure of $N(ds,du)$ defined by,
 \beqnn
 \int_0^t\int_U f(s,u)\tilde{N}(ds,du)\ar:=\ar \int_0^t\int_U f(s,u)N(ds,du)- \int_0^t\int_U f(s,u)\lambda(s,u)ds\mathbf{m}(du),
 \eeqnn
 for any bounded function $f(t,u)$. The compensated random measure is a purely discontinuous martingale with bracket processes
 \beqnn
 \Big[\int_0^\cdot\int_U f(t,u)\tilde{N}(ds,du)\Big]_t= \int_0^t|f(s,u)|^2\int_UN(ds,du).
 \eeqnn
 We always assume that there exists some $C_0>0$ such that, for any $t\in[0,T]$,
 \beqlb\label{eqn2.2}
 \int_U\mu(t,u)\mathbf{m}(du)+ \sup_{v\in U} \int_U\phi(t,u,v,s) \mathbf{m}(du)\leq C_0.
 \eeqlb
 Then,
 \beqnn
 \mathbf{E}\Big[\int_{U}\lambda(t,u)\mathbf{m}(du)\Big]\ar=\ar \mathbf{E}\Big[\int_{U}\mu(t,u)\mathbf{m}(du)\Big] + \mathbf{E}\Big[\int_0^t\int_{U} \int_{U} \phi(s,u,v,t-s)\mathbf{m}(du)N(ds,dv)\Big]\cr
 \ar\leq\ar C_0+C_0 T\mathbf{E}\Big[\int_0^t\int_U\lambda(s,v)\mathbf{m}(dv)ds\Big].
 \eeqnn
 and by Gr\"onwall's inequality,  $\mathbf{E}[\int_{U}\lambda(t,u)\mathbf{m}(du)]<\infty$ and $\mathbf{E}[N([0,T],U)]<\infty$.

 The following lemma proves the existence of a Hawkes random measure for any $(\mathscr{F}_t)$-progressive processes $\mu(t,u)$ and $\phi(t,u,v,r)$.
 \begin{lemma}
 For any nonnegative, $(\mathscr{F}_t)$-progressive processes $\mu(t,u)$ and $\phi(t,u,v,r)$ satisfying (\ref{eqn2.2}), there exists a Hawkes random measure with intensity process defined by (\ref{eqn1.01}).
 \end{lemma}
 \proof Let $N_0(ds,du,dz)$ be the Poisson random measure introduced above. For any $t\geq 0$ and $u\in U$, define $\lambda_{-1}(t,u)=0$, $\lambda_{0}(t,u)=\mu(t,u)$ and for any $n\geq 1$,
 \beqnn
 \lambda_n(t,u)=\mu(t,u)+\sum_{m=1}^n\int_0^t\int_U \phi(s,u,v,t-s)N_m(ds,dv),
 \eeqnn
 where
 \beqlb\label{eqn1.02}
 N_m(I,A)=\int_I\int_A\int_0^\infty \mathbf{1}_{\{\lambda_{m-2}(s,u)\leq z< \lambda_{m-1}(s,u)\}}N_0(ds,du,dz),\quad  I \in \mathscr{B}(\mathbb{R}_+),A\in\mathscr{U}.
 \eeqlb
 The integral intervals in (\ref{eqn1.02}) are disjoint for $m=1,2,\cdots$ and the random measures $\{N_m(dt,du):m=1,2,\cdots\}$ are independent. It is easy to see the following limit is well defined:
 \beqnn
 N(I,A):=\sum_{m=1}^\infty N_m(I,A),\quad  I \in \mathscr{B}(\mathbb{R}_+),A\in\mathscr{U}.
 \eeqnn
 For any fixed $t$ and $u$, the sequence $\lambda_n(t,u)$ $(n=1,2, \cdots)$ is nondecreasing and the following limit exists:
 \beqnn
 \lambda(t,u)=\lim_{n\to \infty }\lambda_n(t,u)\ar=\ar\mu(t,u)+\sum_{m=1}^\infty\int_0^t\int_U \phi(s,u,v,t-s)N_m(ds,dv)\cr
 \ar=\ar \mu(t,u)+\int_0^t\int_U \phi(s,u,v,t-s)N(ds,dv).
 \eeqnn
 Now, for any nonnegative $(\mathscr{F}_t)$-predictable process $H(t,u)$ on $U$,
 \beqnn
 \mathbf{E}\Big[\int_0^t\int_U H(s,u)N(ds,du)\Big]\ar=\ar\mathbf{E}\Big[\sum_{m=1}^\infty\int_0^t\int_U H(s,u)N_m(ds,du)\Big]\cr
 \ar=\ar\mathbf{E}\Big[\sum_{m=1}^\infty\int_0^t\int_U\int_0^\infty H(s,u)\mathbf{1}_{\{\lambda_{m-2}(s,u)< z\leq \lambda_{m-1}(s,u)\}}N_0(ds,du,dz)\Big]\cr
 \ar=\ar \mathbf{E}\Big[\sum_{m=1}^\infty\int_0^tds\int_U H(s,u)[\lambda_{m-1}(s,u)-\lambda_{m-2}(s,u)]\mathbf{m}(du)\Big]\cr
 \ar=\ar \mathbf{E}\Big[\int_0^tds\int_U H(s,u)\lambda(s,u)\mathbf{m}(du)\Big],
 \eeqnn
 Thus $N(ds,dv)$ is the desired Hawkes random measure with intensity process $\lambda(s,u)$.
 \qed

 \begin{example}{ \rm(Multi-variate Hawkes processes)}
 Assume that $U=\{1,\cdots,d\}$ and that $\mathbf{m}(\{i\})=1$ for $i\in U$. Let $N_i(t)=N([0,t],\{i\})$, $\lambda_i(t)=\lambda(t,i)$, $\mu_i(t)=\mu(t,i)$ and $\phi_{ij}(t)=\phi(i,j,t)$.  Then (\ref{eqn1.01}) can be written as
 \beqnn
 \lambda_i(t)=\mu_i(t)+\sum_{j=1}^d\int_0^t\phi_{ij}(t-s)dN_j(s)
 \eeqnn
 and $\{N_i(t):t\geq 0, i=1,\cdots,d\}$ is a multi-variate Hawkes process; see \cite{Hawkes1971}.
 \end{example}

 \begin{example}{\rm (Marked Hawkes processes)}
 Consider a Hawkes random measuare $N(dt,du)$ on $[0,\infty)\times U$ with intensity process $\lambda(t)f(u)$, where $f(u)$ is a nonnegative function on $U$ and
 \beqnn
 \lambda(t)= \mu(t)+\int_0^t\int_U \phi(u,t-s)N(ds,du).
 \eeqnn
 The counting processes $N(t):=N([0,t],U)$ is called {\sl Marked Hawkes process}. It was first introduced in \cite{Ogata1986} to describe the occurrences of earthquakes of different magnitudes.
 \end{example}

 \begin{example}{\rm (Exponential kernel)} Consider a Hawkes random measure $N(dt,du)$ on $[0,\infty)\times U$ with non-random exogenous intensity $\mu(t,u)$ and exponential kernel
 $\phi(u,v)\beta e^{-\beta t}$,
 where $\beta>0$ and $\phi(u,v)$ is a nonnegative function on $U^2$.
 From (\ref{eqn1.01}), the density process is easily identified as:
 \beqnn
 \lambda(t,u)=\mu(t,u)+\int_0^t \beta (\mu(s,u)-\lambda(s,u))ds+\int_0^t \int_{U}\beta \phi(u,v)N(ds,dv).
 \eeqnn
 In this case, $\{(\lambda(t,\cdot),N([0,t],\cdot)):t\geq 0\}$ is a Markov process.
 \end{example}

\subsection{The LOB model}

In this subsection, we introduce a class of LOB models driven by Hawkes random measures and state the main assumptions on the modelling parameters and the main convergence results.
The event-by-event dynamics of the order book follows \cite{HP}, to which we refer for any unspecified modelling details. Throughout, all random processes are defined on a filtered probability space $(\Omega, \mathscr{F}, \{\mathscr{F}_t\}_{t \in [0,T]}, \mathbf{P})$.

\subsubsection{The book}

For a given time horizon $T>0$, the dynamics of the $n$-th order book model is described by a  continuous-time stochastic process $\left(\mathbf{S}^{(n)}(t)\right)_{0 \leq t \leq T}$ taking values in the Hilbert space
\[
	\mathcal{S}:=\mathbb{R}^2\times (L^2(\mathbb{R};\mathbb{R}_+))^2, \quad \|S\|^2_{\mathcal{S}^2}:=|p_a|^2+|p_b|^2+\|v_a\|^2_{L^2}+\|v_b\|^2_{L^2}.
\]
The state of the book changes due to arriving market and limit orders and cancellations. The state at time $t \in [0,T]$ is denoted
\[
	S^{(n)}(t) := \left( P^{(n)}_a(t),P^{(n)}_b(t),V^{(n)}_a(t),V^{(n)}_b(t) \right).
\]
The $\mathbb{R}$-valued process $P^{(n)}_{a/b}$ denotes the {\it best ask/bid price process}, that is, lowest/highest price at which a single unit of a stock can be bought/sold; the $L^2$-valued function $V^{(n)}_{a/b}$ denotes the {\it volume density function} at the ask/bid side of the order book. The {\it tick size}, i.e.~the minimum price movement is denoted $\delta^{(n)}_x$. The price grid is $\left\{x_j^{(n)},\ j\in\mathbb{Z}\right\}$, where $x^{(n)}_j:=j\cdot\delta_x^{(n)}$ for $j \in \mathbb{Z}$ and $n \in \mathbb{N}$.  For all $n\in\mathbb{N}$ and $x\in\mathbb{R}$ the price interval that contains $x$ is denoted
\[
	\Delta^{(n)}(x):=[x^{(n)}_j,x^{(n)}_{j+1}) \quad \mbox{for} \quad x^{(n)}_j \leq x < x^{(n)}_{j+1}.
\]	
For every $t \in [0,T]$, the volume density functions $V^{(n)}_{a/b}(t,\cdot)$ are c\`adl\`ag step function on the price grid. The volume available for trading at the price $x^{(n)}_j$ at time $t \in [0,T]$ is given by the integral of $V^{(n)}_{a/b}(t,\cdot)$ over the interval $[x^{(n)}_j,x^{(n)}_{j+1})$.
 The state $S^{(n)}(0)$ of the book at time $t=0$  is deterministic for all $n\in\mathbb{N}$.

\begin{remark} Following \cite{BHQ, HK1, HP} prices and volume density functions are defined on the whole real line. The restrictions
\[
	\left\{ V^{(n)}_a(t,y) : y \geq P^{(n)}_a(t) \right\} \quad \mbox{and} \quad \left\{ V^{(n)}_b(t,y) : y \leq P^{(n)}_b(t) \right\}
\]	
of the functions $V^{(n)}_{a/b}(t,\cdot)$ to the respective intervals $[P_a^{(n)}, \infty)$ and $(-\infty, P_b^{(n)}]$ correspond to the actual ask, respectively bid side of the order book at time $t \in [0,T]$. The respective complements specify the size of spread placements; see Remark \ref{rem-shadow} below.
\end{remark}

\subsubsection{Event types and dynamics}

We assume that there are eight events -- labeled ${\bf(A1) - (A4)}$ and ${\bf(P1) - (P4)}$ -- that change the state of the book:
\begin{align*}
	\textbf{A1}&\ldots \text{buy market orders}& \textbf{A2}& \ldots
        \text{sell limit orders placed in the spread}\\
	\textbf{A3}&\ldots \text{sell market ordres}& \textbf{A4}&
        \ldots \text{buy limit orders placed in spread} \\
	\textbf{P1}&\ldots \text{sell limit orders}& \textbf{P2}& \ldots \text{cancellations of sell volume}\\
	\textbf{P3}&\ldots \text{buy limit orders}& \textbf{P4}&
        \ldots \text{cancellations of buy volume}.
\end{align*}
Following \cite{BHQ, HK1, HP} we assume that market orders match precisely against the volume at the top of the book and that limit orders placed into the spread are placed at the first best price increment. In particular, market orders and spread placements change prices by exactly one tick. We refer to market orders and limit order placements in the spread as {\sl active orders}.

\begin{remark} \label{rem-shadow}
	Defining the volume density functions on the whole real line allows for a convenient modelling of spread placements. The restrictions
\[
	\left\{ V^{(n)}_a(t,y) : y < P^{(n)}_a(t) \right\} \quad \mbox{and} \quad \left\{ V^{(n)}_b(t,y) : y > P^{(n)}_b(t) \right\}
\]	
of the ask and bid side volume density functions to the intervals $(-\infty, P_a^{(n)})$ and $(P_b^{(n)}, \infty)$ specify the volumes placed into the spread should such events occur next. For example, if an ask side spread placement occurs at time $0 < t < T$, then the ask-side volume density function at that time is
\[
	\left\{ V^{(n)}_a(t,y) : y \geq P^{(n)}_a(t) \right\} = \left\{ V^{(n)}_a(t-,y) : y \geq P^{(n)}_a(t-) -  \delta^{(n)}_x \right\}.
\]	
We refer to \cite{HP} for further details on the modelling of spread placements.
\end{remark}

The assumption that spread placements occur at the first best price increment is not restrictive. The assumption that market orders match the liquidity at the top of the book is made for mathematical convenience.
{Mathematically}, a market order that does not lead to a price change may be viewed as a cancellation at the top of the book while a cancellation that leads to a price change may be treated as market order\footnote{Alternatively, we could add two additional event types that describe market order arrivals that do not lead to price changes and two additional event types that describe the arrivals of cancellations that lead to price changes. This would increase the number of events from eight to twelve but would not change our mathematical arguments.}. Moreover, a market order that exceeds the liquidity at the top of the book is typically split by the exchange into a series of smaller orders that are consecutively executed against standing volumes at less competitive prices. Mathematically, this may be viewed as a series of ``child market orders'' triggered by the arrival of some ``parent market order''.  Our probabilistic framework is flexible enough to capture such dynamics.

In what follows, we put $\mathcal{I}=\{a,b\}$ (``ask side'', ``bid side''), $\mathcal{J}=\{M,L\}$ (``market order'', ``limit order placed in spread''), and $\mathcal{K}=\{L,C\}$ (``limit order placement outside the spread'', ``cancellation''). When $I,i$, $J,j$ and $K,k$ appear as subscripts, it is always assumed that $I,i\in\mathcal{I}$, $J,j\in\mathcal{J}$ and $K,k\in\mathcal{K}$.

\begin{assumption}
Market buy/sell orders arrive according to an $(\mathscr{F}_t)$-random point measure $N^{(n)}_{a/bM}(dt)$ on $\mathbb{R}_+$ with intensity $\rho^{(n)}_{a/bM}(\mathbf{S}^{(n)}(t)) \mu^{(n)}_{a/bM}(t)dt$ and sell/buy limit orders placed in the spread arrive according to an $(\mathscr{F}_t)$-random point measure $N^{(n)}_{a/bL}(dt)$ on $\mathbb{R_+}$ with intensity $\rho^{(n)}_{a/bL}(\mathbf{S}^{(n)}(t)) \mu^{(n)}_{a/bL}(t)dt$.  Here $t \in [0,T]$ represents the event arrival time, $\{\rho^{(n)}_{IJ}(S)\}_{I\in\mathcal{I},J\in\mathcal{J}}$ are deterministic nonnegative mappings defined on $\mathcal{S}$ and $\{\mu^{(n)}_{IJ}(t)\}_{I\in\mathcal{I},J\in\mathcal{J}}$ are nonnegative and $(\mathscr{F}_t)$-progressive processes.
\end{assumption}

The deterministic functions $\rho^{(n)}_{IJ}$ can be chosen so as to guarantee that bid and ask prices never cross; cf.~equation (\ref{eqn.CNon}) below. The progressively measurable random processes $\mu^{(n)}_{IJ}$ capture the (non-Markovian) dependence of the price dynamics on past price changes. Their precise dynamics will be introduced at the end of this section.

As in \cite{BHQ, HK1, HP}  we assume that limit order placements outside the spread and cancellations of standing volume do not change prices. We refer to these order types as {\sl passive orders}. Cancellations occur at random distances from the same side best price for random {\sl proportions} of the standing volume, and limit order placements outside the spread occur at random distances from the same side best price for random {\sl volumes}. This guarantees that volumes are always non-negative.

 \begin{assumption}
 Sell/buy limit orders of size $z$ at the distance $x$ from the best ask/bid price arrive according to an $(\mathscr{F}_t)$-random point measure $M^{(n)}_{a/bL}(dt,dx,dz)$ on $\mathbb{R_+}\times\mathbb{R}\times\mathbb{R}_+$ with intensity $\lambda^{(n)}_{a/bL}(t,x)dtdx\nu_{a/bL}(dz)$ and cancellation of sell/buy volume at the distance $x$ from the ask/best bid price arrive according to an $(\mathscr{F}_t)$-random point measure $M^{(n)}_{a/bC}(dt,dx,dz)$ on $\mathbb{R_+}\times\mathbb{R}\times\mathbb{R}_+$ with intensity $\lambda^{(n)}_{a/bC}(t,x)dtdx\nu_{a/bC}(dz)$.
 Here $(t,x,z)$ represents the event arrival time, the distance from the top of the book where a placement or cancellation takes place, and the size of a cancellation or placement, respectively. The processes $\{\lambda^{(n)}_{IK}(t,\cdot)\}_{I\in\mathcal{I},K\in\mathcal{K}}$ are $(\mathscr{F}_t)$-progressive, nonnegative function-valued and \{$\nu_{IK}(dz)\}_{I\in\mathcal{I},K\in\mathcal{K}}$ are probability measures on $\mathbb{R}_+$ satisfying $\nu_{IK}(|e^z-1|^4)<\infty$ for each $n \in \mathbb{N}$.
 \end{assumption}

The deterministic measures $\nu_{IK}(dz)$ specify the sizes of limit order placements and cancellations. If $\nu_{IK}(dz)$ is a Dirac measure, then $M^{(n)}_{a/bL}(dt,dx,dz)$ is a Hawkes random measure in the sense of the previous section. The random processes $\lambda^{(n)}_{IK}(t,\cdot)$ describe the intensities of limit order arrivals and cancellations at different price levels as functions of past events. Their precise dynamics will be specified below.

\subsubsection{The LOB dynamics}

Since prices move by exactly one tick when market orders are spread placement arrive the dynamics of the ask and bid price processes can be described as follows:
 \begin{equation}\label{DLOB01}
 \begin{split}
 P^{(n)}_a(t) &=  P^{(n)}_a(0)
      +\int_0^t\delta^{(n)}_xN^{(n)}_{aM}(ds)-\int_0^t\delta^{(n)}_xN^{(n)}_{aL}(ds),\\
 P^{(n)}_b(t) & =  P^{(n)}_b(0)
      -\int_0^t\delta^{(n)}_xN^{(n)}_{bM}(ds)+\int_0^t\delta^{(n)}_xN^{(n)}_{bL}(ds).
\end{split}
\end{equation}

Since the active arrival intensities are of the multiplicative form $\rho^{(n)}_{IJ} \mu^{(n)}_{IJ}$, the following assumption guarantees that the best ask price is never smaller than the best bid price.

 \begin{condition}\label{Noncrossing}
 For any $S=(p_a,p_b,v_a,v_b)\in\mathcal{S}$ with $p_a-p_b< \delta^{(n)}_x$ it holds that $\rho^{(n)}_{aL}(S)=\rho^{(n)}_{bL}(S)=0.$
 \end{condition}

We denote by $\delta^{(n)}_v$ a scaling parameter that determines the size of an individual order/cancellation in the $n$-th model. We will later analyze the high-frequency limit where order and tick sizes tend to zero but order arrival intensities tend to infinity as $n \to \infty$. Since limit order placements and cancellations occur at random distances from the same side best prices, and because limit order placements are additive while cancellations are proportional in standing volumes,
%
%
the dynamics of the volume density functions (in absolute coordinates) is given by:\footnote{The factor $|\delta^{(n)}_x|^ {-1}$ captures the fact that volumes at a given price level are given by integral of the volume density function over an interval of length $\delta^{(n)}_x$. Integrating over the interval $\Delta^{(n)}(x-P^{(n)}_a(s-))$ captures the fact the the measures $M^{(n)}$ describe volume placements and cancellations at random distances from the same side best price. Finally, expressing the added/cancelled volume in exponential terms allows us to view the measures $M^{(n)}$ as measures on $\mathbb{R}_+$ in the third variable.}
\begin{equation}\label{DLOB03}
 \begin{split}
 V^{(n)}_{a}(t,x) & = V^{(n)}_a (0,x)
      +\int_0^t\int_{\Delta^{(n)}(x-P^{(n)}_a(s-))}\int_{\mathbb{R}_+}\frac{\delta^{(n)}_v}{\delta^{(n)}_x} (e^{z}-1)M^{(n)}_{aL}(ds,dy,dz)\\
      & ~~~ +\int_0^t\int_{\Delta^{(n)}(x-P^{(n)}_a(s-))}\int_{\mathbb{R}_+}\frac{\delta^{(n)}_v}{\delta^{(n)}_x} V^{(n)}_{a}(s-,y+P^{(n)}_a(s-)) (e^{-z}-1)M^{(n)}_{aC}(ds,dy,dz), \\
 V^{(n)}_{b}(t,x) & = V^{(n)}_{b}(0,x)
      +\int_0^t\int_{\Delta^{(n)}(P^{(n)}_b(s-)-x)}\int_{\mathbb{R}_+}\frac{\delta^{(n)}_v}{\delta^{(n)}_x} (e^{z}-1)M^{(n)}_{bL}(ds,dy,dz) \\
      & ~~~ +\int_0^t\int_{\Delta^{(n)}(P^{(n)}_b(s-)-x)}\int_{\mathbb{R}_+}\frac{\delta^{(n)}_v}{\delta^{(n)}_x} V^{(n)}_{b}(s-,P^{(n)}_b(s-)-y)(e^{-z}-1)M^{(n)}_{bC}(ds,dy,dz).
 \end{split}
 \end{equation}

In order to obtain a diffusive limiting dynamics for the price processes and a deterministic limiting dynamics for the volume density functions we assume that active orders arrive at a rate $|\delta_x^{(n)} |^{-2}$ while passive orders arrive at a rate $|\delta_v^{(n)}|^{-1}$. To capture clustering and cross-dependencies between order arrivals we assume that the event arrival intensities depend on the past price movements as well as past limit order placements and cancellations. Specifically, we assume that the arrival intensities take the form:
 \begin{equation}\label{Dintensity01}
 \begin{split}
%
 \mu^{(n)}_{IJ}(t) &= \frac{1}{|\delta_x^{(n)}|^2} \hat{\mu}^{(n)}_{IJ}(t,\mathbf{S}^{(n)}(t-))
     +\sum_{i\in\mathcal{I},j\in\mathcal{J}}\int_0^t \phi^{(n)}_{IJ,ij}(t-s)N^{(n)}_{ij}(ds) \\
     & ~~~ +\sum_{i\in\mathcal{I},k\in\mathcal{K}}\int_0^t\int_{\mathbb{R}}\int_{\mathbb{R}} \frac{\delta_v^{(n)}}{|\delta_x^{(n)}|^2} \Phi^{(n)}_{IJ,ik}(y,t-s)M^{(n)}_{ik}(ds,dy,dz),
\end{split}
\end{equation}
and
 \begin{equation}\label{Dintensity02}
 \begin{split}
 \lambda^{(n)}_{IK}(t,x) &= \frac{1}{\delta_v^{(n)}} \hat{\lambda}_{IK}(t,\mathbf{S}^{(n)}(t-),x)
     +\sum_{i\in\mathcal{I},j\in\mathcal{J}}\int_0^t \frac{|\delta_x^{(n)}|^2}{\delta_v^{(n)}}  \psi_{IK,ij}(x,t-s)N^{(n)}_{ij}(ds) \\
     & ~~~ +\sum_{i\in\mathcal{I},k\in\mathcal{K}}\int_0^t\int_{\mathbb{R}}\int_{\mathbb{R}} \Psi_{IK,ik}(x,y,t-s)M^{(n)}_{ik}(ds,dy,dz).
 \end{split}
 \end{equation}

 Here, $\hat{\mu}^{(n)}_{IJ}$ and $\hat{\lambda}_{IK}$ are exogenous densities that depend on the current state of the book only. The kernels  $\phi^{(n)}_{IJ,ij},\Phi^{(n)}_{IJ,ik}$  measure the impact of past active/passive events on the price dynamics while the kernels $\psi_{IK,ij},\Psi_{IK,ik}$ measure the impact of past passive events on placements/cancellations. For instance, $\phi^{(n)}_{aM,aL}(t-s)$ measures the impact of a market order arrival at time $s$ on the intensity of a market order arrival at time $t$. Depending on the choice of that kernel, this allows us to model the arrival of ``child market orders triggered by the arrival of a parent market order''.

 The quantities $\psi_{aL,aL}(x,t-s)$ and $\psi_{aC,aL}(x,t-s)$ measure the impact of an ask-side limit order placement at a distance $x$ from the best ask price at the time $s$  on the arrival intensity of an ask-side limit order placement/cancellation at the same distance from the then prevailing price at time $t$. For $x \in (-\delta_x^{(n)},0 )$ this amounts to an idealized modelling of peg orders. Finally, for any $x$ and $y$, the quantity $\Psi_{aC,aL}(x,y,t-s)$ measures the impact of an ask-side limit order placement at  price level $\Delta^{(n)}(y)$ (the price interval that contains $y$)at the time $s$ on the arrival of an ask-side limit order cancellation at the level $\Delta^{(n)}(y)$ at time $t$. The choice of the scaling constants in (\ref{Dintensity01}) and (\ref{Dintensity02}) reflects the arrival intensities of active and passive orders. Tables 1 and 2 summarize the notation.

%

  \begin{table}
  \centering \small
  \begin{tabular}{|m{5em}|c|c |c|c | }
   \hline
  {\bf Type}&{\bf A1}&{\bf A2}&{\bf A3}&{\bf A4 } \\ \hline
 {\bf Notation} &$N^{(n)}_{aM}(dt)$ &$N^{(n)}_{aL}(dt)$&$N^{(n)}_{bM}(dt)$& $N^{(n)}_{bL}(dt)$\\ \hline
 {\bf Space} &$\mathbb{R}_+$&$\mathbb{R}_+$&$\mathbb{R}_+$&$\mathbb{R}_+$\\ \hline
 {\bf Intensity} & $\rho^{(n)}_{aM}(S)\mu^{(n)}_{aM}(t)$ &{$\rho^{(n)}_{aL}(S)\mu^{(n)}_{aL}(t)$}&
 {$\rho^{(n)}_{bM}(S)\mu^{(n)}_{bM}(t)$}& {$\rho^{(n)}_{bL}(S)\mu^{(n)}_{bL}(t)$}\\ \hline
 {\bf Exogenous density}&$\hat{\mu}^{(n)}_{aM}(t,S)$ & $\hat{\mu}^{(n)}_{aL}(t,S)$ & $\hat{\mu}^{(n)}_{bM}(t,S)$ & $\hat{\mu}^{(n)}_{bL}(t,S)$\\ \hline
 \multirow{2}*{\bf Kernel}&$\phi^{(n)}_{aM,ij}(t)$ &$\phi^{(n)}_{aL,ij}(t)$&$\phi^{(n)}_{bM,ij}(t)$&$\phi^{(n)}_{bL,ij}(t)$\\
 &$\Phi^{(n)}_{aM,ik}(y,t)$ &$\Phi^{(n)}_{aL,ik}(y,t)$&$\Phi^{(n)}_{bM,ik}(y,t)$&$\Phi^{(n)}_{bL,ik}(y,t)$\\ \hline
 \multirow{5}*{\bf Differences}
   & \multicolumn{2}{c|}{$\beta_a^{(n)}(t):= \delta_x^{(n)} \left( \mu_{aM}^{(n)}(t)- \mu_{aL}^{(n)}(t) \right)$} &\multicolumn{2}{c|}{$ \beta_b^{(n)}(t):= |\delta_x^{(n)}|^{-1} \left(  \mu_{bM}^{(n)}(t,S)- \mu_{bL}^{(n)}(t,S) \right)$}\\
   \cline{2-5}
   & \multicolumn{2}{c|}{$\hat \beta_a^{(n)}(t):= |\delta_x^{(n)}|^{-1} \left( \hat \mu_{aM}^{(n)}(t,S)- \hat \mu_{aL}^{(n)}(t,S) \right)$} &\multicolumn{2}{c|}{$\hat \beta_b^{(n)}(t):= |\delta_x^{(n)}|^{-1} \left( \hat \mu_{bM}^{(n)}(t,S)- \hat \mu_{bL}^{(n)}(t,S) \right)$}\\
   \cline{2-5}
   &\multicolumn{2}{c|}{$\varrho^{(n)}_a(S):=|\delta_x^{(n)}|^{-1}\left(\rho^{(n)}_{aM}(S)-\rho^{(n)}_{aL}(S)\right) $} &\multicolumn{2}{c|}{$\varrho^{(n)}_b(S):=|\delta_x^{(n)}|^{-1}\left(\rho^{(n)}_{bM}(S)-\rho^{(n)}_{bL}(S)\right) $} \\
   \cline{2-5}
   & \multicolumn{2}{c|}{$\theta^{(n)}_{a,ij}(t)
   :=|\delta_x^{(n)}|^{-1}\Big(\phi^{(n)}_{aM,ij}(t)-\phi^{(n)}_{aL,ij}(t)\Big)$} &\multicolumn{2}{c|}{$\Theta^{(n)}_{b,ik}(y,t) := |\delta_x^{(n)}|^{-1}\Big(\Phi^{(n)}_{bM,ik}(y,t)-\Phi^{(n)}_{bL,ik}(y,t)\Big)$}\\
 \hline
  \end{tabular}
  \caption{Active events}\label{Table01}
 \end{table}

 \begin{table}
  \centering
  \small\begin{tabular}{|m{5em}|c|c|c|c|}
   \hline
 {\bf Type}&{\bf P1}&{\bf P2}&{\bf P3}& {\bf P4}  \\
 \hline
 {\bf Notation} &$M_{aL}^{(n)}(dt,dx,dz)$ &$M_{aC}^{(n)}(dt,dx,dz)$&$M_{bL}^{(n)}(dt,dx,dz)$&   $M_{bC}^{(n)}(dt,dx,dz)$\\ \hline
 {\bf Space} &$\mathbb{R_+}\times\mathbb{R}\times\mathbb{R}_+$&$\mathbb{R_+}\times\mathbb{R}\times\mathbb{R}_-$&$\mathbb{R_+}\times\mathbb{R}\times\mathbb{R}_+$&$\mathbb{R_+}\times\mathbb{R}\times\mathbb{R}_-$\\ \hline
 {\bf Intensity} & { $\lambda_{aL}^{(n)}(t,x)dtdx\nu_{aL}(dz)$} & {$\lambda_{aC}^{(n)}(t,x)dtdx\nu_{aC}(dz)$} &
 {$\lambda_{bL}^{(n)}(t,x)dtdx\nu_{aL}(dz)$} & {$\lambda_{bC}^{(n)}(t,x)dtdx\nu_{bC}(dz)$}\\
  \hline
  {\bf Exogenous density}&$\hat{\lambda}_{aL}(t,S,x)$ & $\hat{\lambda}_{aC}(t,S,x)$ & $\hat{\lambda}_{bL}(t,S,x)$ & $\hat{\lambda}_{bC}(t,S,x)$\\ \hline
 \multirow{2}*{\bf Kernel} &$\psi_{aL,ij}(x,t)$ &$\psi_{aC,ij}(x,t)$&$\psi^{(n)}_{bL,ij}(x,t)$&$\psi^{(n)}_{bC,ij}(x,t)$\\
 &$\Psi_{aL,ik}(x,y,t)$ &$\Psi_{aC,ik}(x,y,t)$&$\Psi^{(n)}_{bL,ik}(x,y,t)$&$\Psi^{(n)}_{bC,ik}(x,y,t)$\\ \hline
  \end{tabular}
  \caption{Passive events}\label{Table02}
 \end{table}


 \subsection{Scaling conditions and the limiting dynamics}\label{Sec3}


 In this section we state assumptions on the arrival intensities and the Hawkes kernels that guarantee that the sequence of LOB models described by (2.1)-(2.4) converges in law to the unique solution of a certain coupled SDE-ODE system. The SDE will describe the limiting price dynamics while the ODE will describe the limiting volume dynamics. We start with a moment condition and convergence assumption on the initial states.

\begin{condition}\label{C3.1}
There exists a constant $C_0>0$ such that for any $n>1$ and $I\in\mathcal{I} $,
 \beqlb\label{BInitial01}
 \mathbf{E}\Big[\|\mathbf{S}^{(n)}(0)\|^2_{\mathcal{S}^2}\Big]+\mathbf{E}\Big[\|V_I^{(n)}(0,\cdot)\|^4_{L^4}\Big]\leq C_0.
 \eeqlb
Moreover, there exists {\blue an} $\mathcal{S}$-valued random variable $\mathbf{S}(0)$ such that as $n\to \infty$
 \beqlb\label{CInitial01}
 \mathbf{E}\Big[\|\mathbf{S}^{(n)}(0)-\mathbf{S}(0)\|^2_{\mathcal{S}^2}\Big] \to 0.
 \eeqlb
\end{condition}

\subsubsection{Scaling assumptions}


Let us first consider the benchmark case of a purely Markovian LOB dynamics where all the Hawkes kernels vanish.
Multiplying both sides of the equation (\ref{Dintensity01}) by $|\delta^{(n)}_x|^{2}$ and both sides of the equation (\ref{Dintensity02}) by $\delta^{(n)}_v $ shows that some form of convergence of the sequence $\{ \hat \mu_{IJ}^{(n)}(t,S)\}_{n\geq 0}$ is required  ($\hat\lambda(t,S,x)$ is independent of $n$).

 Since the active order arrival intensities are of the product form $\rho^{(n)}_{IJ}\hat \mu^{(n)}_{IJ}$ we need to impose additional conditions to guarantee the convergence of the drift and the volatility of the price processes. The expected ask and bid price increments are given by the differences between market order and spread placement arrival intensities as
 \beqnn
 |\delta_x^{(n)}|^{-1} \left( \rho^{(n)}_{IM}(S) \hat \mu^{(n)}_{IM}(t,S) -\rho^{(n)}_{IL}(S) \hat \mu^{(n)}_{IL}(t,S) \right).
 \eeqnn
 This can be rewritten into
 \begin{equation*}
  \varrho^{(n)}_I(S) \hat \mu^{(n)}_{IM}(t,S) +\rho^{(n)}_{IL}(S) \hat  \beta_I^{(n)}(t),
 \end{equation*}
 where
 \begin{equation} \label{hatbeta}
 	\varrho^{(n)}_I(S):=|\delta_x^{(n)}|^{-1}\left(\rho^{(n)}_{IM}(S)-\rho^{(n)}_{IL}(S)\right)
 	\quad \mbox{and} \quad
 	\hat \beta_I^{(n)}(t):= |\delta_x^{(n)}|^{-1} \left( \hat \mu_{IM}^{(n)}(t,S)- \hat \mu_{IL}^{(n)}(t,S) \right).
 \end{equation}
 This representation motivates the following two conditions. The first condition guarantees the convergence of the first factors of the two summands above to  continuous limit.

\begin{condition}\label{C3.2}
\begin{itemize}
\item[i)]  The functions $(\rho^{(n)}_{IJ},\varrho_I^{(n)})$ 
are uniformly bounded.
\item[ii)]  The functions $\{(\rho^{(n)}_{IJ},\varrho_I^{(n)})\}_{n\geq 0}$ converge uniformly to Lipschitz continuous functions $(\rho_{IJ},\varrho_I)$.
   %
 \end{itemize}
\end{condition}

As an immediate consequence from the preceding condition we obtain that
 \begin{equation}
 	\rho_I:=\rho_{IM}=\rho_{IL}.
\end{equation}
The second condition guarantees the convergence of the rescaled (net) arrival rates to a continuous limit. In view of (\ref{hatbeta}) it implies that spread placements and market orders are equally likely on average:
\begin{equation}
	\hat\mu_I:=\hat\mu_{IM}=\hat\mu_{IL}.
\end{equation}

\begin{condition}\label{C3.3}
\begin{itemize}
 	\item[i)] There exists a constant $C_0>0$ such that for any $p\in\{1,2,4\}$,
 \beqlb\label{BIntensity01}
 \sup_{t\in[0,T],S\in\mathcal{S}}\Big\{|\hat{\mu}^{(n)}(t,S)|+|\hat\beta_I^{(n)}(t,S)|
 + \|\hat{\lambda}_{IK}(t,S,\cdot)\|_{L^p} \Big\}\leq C_0.
 \eeqlb
 and for any $\epsilon>0$, $t,t'\in[0,T]$, $S,S'\in\mathcal{S}$
 \beqlb\label{eqn2.8}
\|\hat{\lambda}_{IK}(t',S',\cdot+\epsilon)-\hat{\lambda}_{IK}(t,S,\cdot)\|_{L^p}\leq C_0(\epsilon+|t-t'|+\|S-S'\|_{\mathcal{S}^2}).
 \eeqlb
	\item[ii)] There exist Lipschitz continuous functions $\hat\mu_{IJ}(t,S)$ and $ \hat\beta_I(t,S)$ 
 such that  
 \beqlb\label{CIntensity01}
 \sup_{t\in[0,T],S\in\mathcal{S}}\Big\{|\hat\mu^{(n)}_{IJ}(t,S)-\hat\mu_{IJ}(t,S)|+\big|\hat\beta_I^{(n)}(t,S)-\hat\beta_I(t,S)\big|\Big\}\to 0.
 \eeqlb
 \end{itemize}
\end{condition}



It remains to state scaling conditions on the Hawkes kernels. Compared to the Markovian case, the expected price increments comprise the following additional terms (up to the multiplicative bounded processes $\rho^{(n)}$) resulting from the impact of past events on the active order arrival dynamics:
  \begin{align*}
   \theta^{(n)}_{I,ij}(t):=|\delta_x^{(n)}|^{-1}\Big(\phi^{(n)}_{IM,ij}(t)-\phi^{(n)}_{IL,ij}(t)\Big)\quad \mbox{and}\quad
 \Theta^{(n)}_{I,ik}(y,t) := |\delta_x^{(n)}|^{-1}\Big(\Phi^{(n)}_{IM,ik}(y,t)-\Phi^{(n)}_{IL,ik}(y,t)\Big).
 \end{align*}
The next condition states regularity conditions on the Hawkes kernels that specify the impact of past events on limit order placement and cancellation arrivals and guarantees the convergence of the (rescaled) Hawkes kernels that specify the impact of past events on prices to sufficiently regular functions.


 \begin{condition}\label{C3.4}
 \begin{itemize}
\item[i)] There exists a constant $C_0>0$ such that for any $\epsilon>0$, $p\in\{1,2,4\}$
\beqnn
 \sup_{t\in[0,T],y\in\mathbb{R}} \Big\{\|\psi_{IK,ij}(\cdot,t)\|_{L^p}+\|\Psi_{IK,ik}(\cdot,y,t)\|_{L^p}\Big\}<C_0
\eeqnn
and
\beqnn
\sup_{t\in[0,T],y\in\mathbb{R}} \Big\{\|\psi_{IK,ij}(\cdot+\epsilon,t)-\psi_{IK,ij}(\cdot,t)\|_{L^p}+\|\Psi_{IK,ik}(\cdot+\epsilon,y,t)-\Psi_{IK,ik}(\cdot,y,t)\|_{L^p}\Big\}\leq C_0 \epsilon.
\eeqnn
\item[ii)] The functions
\[
	\kappa^{(n)}(y,t):= \left( \phi^{(n)}_{IJ,ij}(t),\Phi^{(n)}_{IJ,ij}(y,t),\theta^{(n)}_{I,ik}(t),\Theta^{(n)}_{I,ik}(y,t) \right)_{I,i\in\mathcal{I},J\in\mathcal{J},k\in\mathcal{K}}
\]
are uniformly bounded and converge uniformly to functions
\[
	\kappa(y,t)= \left( \phi_{IJ,ij}(t),\Phi_{IJ,ij}(y,t),\theta_{I,ik}(t),\Theta_{I,ik}(y,t) \right)_{I,i\in\mathcal{I},J\in\mathcal{J},k\in\mathcal{K}}
\]	
that are uniformly Lipschitz continuous in the time variable:
%
 \beqlb\label{CIntensity02}
  \sup_{t\in[0,T],y\in\mathbb{R}}|\kappa^{(n)}(y,t)-\kappa(y,t)|\to 0.
 \eeqlb
 \end{itemize}
 \end{condition}
 From the definitions of $\theta_{I,ij}^{(n)}$ and $\Theta_{I,ik}^{(n)}$, the preceding condition implies that the limiting impact of same side market orders and spread placements is the same:
 \beqnn
  \phi_{I,ij}:=\phi_{IM,ij}= \phi_{IL,ij}\quad\mbox{and}\quad \Phi_{I,ik}:=\Phi_{IM,ik}= \Phi_{IL,ik}.
 \eeqnn

 \subsubsection{Existence and uniqueness of solutions of the limiting system}

 Let $\alpha_{IL}=\nu_{IL}(e^z-1)$, $\alpha_{IC}=\nu_{IC}(e^{-z}-1)$ and
   $$
 \tilde{\phi}_{Ii}=\phi_{I,iM}+\phi_{I,iL},\quad\tilde\psi_{IK,i}=\psi_{IK,iM}+\psi_{IK,iL},\quad\tilde\theta_{Ii}= \theta_{I,iM}+\theta_{I,iL}
 .$$
 Here $\tilde{\phi}_{Ii}$ measures the total impact of active events on themselves. Moreover, $\tilde\psi_{IK,i}$ and $\tilde\theta_{Ii}$ measure the total impact of active events on passive events and price dynamics respectively.
In order to state the main result in this paper we further introduce the functions
\begin{equation}
	 \beta_I^{(n)}(t):= \delta_x^{(n)}\Big( \mu_{IM}^{(n)}(t)- \mu_{IL}^{(n)}(t)\Big)
\end{equation}
and
\begin{equation}
	{\mathbf{D}}^{(n)}(t,S) := \left( |\delta_x^{(n)}|^2\mu_{ij}^{(n)}(t,S), \delta_v^{(n)}	\lambda_{ik}^{(n)}(t,S,\cdot) \right)_{i\in\mathcal{I},j\in\mathcal{J},k\in\mathcal{K}}.
\end{equation}
 The preceding vector $\mathbf{D}^{(n)}(t,S)$ belongs to the space $\mathcal{D}:=\mathbb{R}^4\times \left( L^1(\mathbb{R};\mathbb{R}_+)\cap L^2(\mathbb{R};\mathbb{R}_+) \right)^4$ for every $n \in \mathbb{N}$.
The space is a Banach space when endowed with the norm  $\|\cdot\|_{\mathcal{D}_{1,2}^2}:=\|\cdot\|_{\mathcal{D}_{1}^2}+\|\cdot\|_{\mathcal{D}_{2}^2}$, where $\|\cdot\|_{\mathcal{D}_{q}^p}$ ($p,q\in\mathbb{Z}_+$) is defined for any $D:=(D_1,\cdots,D_8)\in\mathcal{D}$ by
 \beqnn
 \|D\|^p_{\mathcal{D}_q^p}=\sum_{k=1}^4|D_k|^p+\sum_{k=5}^8 \|D_k\|_{L^q}^p.
 \eeqnn

 We are now ready to state the main result of this paper. Its proof is given in Section {\blue 4-5} below.

 \begin{theorem}\label{T1} Suppose Conditions~\ref{C3.1}-\ref{C3.4} hold.
 Then,
 $$ \left( \mathbf{{S}}^{(n)},\mathbf{D}^{(n)},\beta^{(n)}_a,\beta_b^{(n)} \right)\Rightarrow \left( \mathbf{S},\mathbf{D},\beta_a,\beta_b \right)$$
 weakly in $\mathbb{D}(\mathbb{R}_+,\mathcal{S}\times \mathcal{D}\times\mathbb{R}^2)$, where $\mathbf{S}=(P_a,P_b,V_a,V_b)$ and $\mathbf{D}=(\mu_{ij},\lambda_{ik})_{i\in\mathcal{I},j\in\mathcal{J},k\in\mathcal{K}}$ with $\mu_i:=\mu_{iM}=\mu_{iL}$ for $i\in\mathcal{I}$.
 Moreover, the limit is a solution to the following stochastic dynamic system:
\begin{equation}\label{LLOB01}
\begin{split}
 P_a(t) & =  P_a(0)
     +\int_0^t\Big[\rho_a(\mathbf{S}(s))\beta_a(s)+\varrho_a(\mathbf{S}(s))\mu_a(s)\Big]ds +\int_0^t\sqrt{2\rho_a(\mathbf{S}(s))\mu_a(s)}dB_a(s), \\
 P_b(t) & =  P_b(0)
     -\int_0^t\Big[\rho_b(\mathbf{S}(s))\beta_b(s)+\varrho_b(\mathbf{S}(s))\mu_b(s)\Big]ds +\int_0^t\sqrt{2\rho_b(\mathbf{S}(s))\mu_b(s)}dB_b(s), \\ 
 V_a(t,x) & = V_a(0,x)
     +\int_0^t\Big[\alpha_{aL}\lambda_{aL}(s,x-P_a(s)) +\alpha_{aC}\lambda_{aC}(s,x-P_a(s))V_a(s,x)\Big]ds,\\
 V_b(t,x) & = V_b(0,x)
     +\int_0^t\Big[\alpha_{bL}\lambda_{bL}(s,P_b(s)-x) +\alpha_{bC}\lambda_{bC}(s,P_b(s)-x)V_b(s,x)\Big]ds, 
\end{split}
\end{equation}
where $(B_a,B_b)$ is a standard two-dimensional Brownian motion, and
 \beqlb\label{Lintensity01}
 \mu_I(t)\ar=\ar \hat\mu_I(t,\mathbf{S}(t))
     +\sum_{i\in\mathcal{I}}\int_0^t\tilde{\phi}_{Ii}(t-s)\rho_i(\mathbf{S}(s))\mu_i(s)ds\cr
     \ar\ar
     +\sum_{i\in\mathcal{I},k\in\mathcal{K}}\int_0^t\int_\mathbb{R}\Phi_{I,ik}(y,t-s)\lambda_{ik}(s,y) dsdy, \\
 \lambda_{IK}(t,x)\ar=\ar \hat\lambda_{IK}(t,\mathbf{S}(t),x)
     +\sum_{i\in\mathcal{I}}\int_0^t\tilde\psi_{IK,i}(x,t-s)\rho_i(\mathbf{S}(s))\mu_i(s)ds\cr
     \ar\ar
     +\sum_{i\in\mathcal{I},k\in\mathcal{K}}\int_0^t\int_{\mathbb{R}}\Psi_{IK,ik}(x,y,t-s)\lambda_{ik}(s,y) ds dy, \label{Lintensity02}\\
 \beta_{I}(t)\ar=\ar \hat{\beta}(t,\mathbf{S}(t))
     +\sum_{i\in\mathcal{I}}\int_0^t\tilde\theta_{Ii}(t-s)\rho_i(\mathbf{S}(s))\mu_i(s)ds\cr
     \ar\ar +\sum_{i\in\mathcal{I},k\in\mathcal{K}}\int_0^t\int_{\mathbb{R}}\Theta_{I,ik}(y,t-s)\lambda_{ik}(s,y) dsdy. \label{Lintensity03}
 \eeqlb
 \end{theorem}

\begin{remark}
Since the system (\ref{Lintensity01})-(\ref{Lintensity02}) can be viewed as the solution to the linear Volterra-Fredholm integral equation  (see \cite{Corduneanu1991}) the limiting intensities can be approximated in terms of recursively defined linear operators.  In order to see this, let us denote by $l(dz)=\mathbf{1}_{\mathbb{R}}(z)dz+\mathbf{1}_{\infty}(dz)$ a measure on $\bar{\mathbb{R}} := \mathbb{R} \cup \{+ \infty\} $, put
  \beqnn
  L^p_l(\bar{\mathbb{R}},\mathbb{R}):=\Big\{f:\bar{\mathbb{R}}\mapsto \mathbb{R}: \|f\|_{L^p_l}^p:=\int_{\mathbb{R}}|f(z)|^pdz+f(\infty)<\infty\Big\}
  \eeqnn
and define linear operators $\{\mathrm{T} = \mathrm{T}(x,y,S,t,s):S\in\mathcal{S}, 0\leq s\leq t<\infty\}$ from  $L_l^2(\bar{\mathbb{R}},\mathbb{R})$ to $L_l^2(\bar{\mathbb{R}},\mathbb{R})$ by 
 \beqnn
\mathrm{T}:= \left(
 \begin{array}{cc}
 \Big(
 \tilde{\phi}_{Ii}(t-s)\rho_i(S)\mathbf{1}_{(x,y)=(\infty,\infty)}
 \Big)_{I,i\in\mathcal{I}}
 &  \Big(
 \Phi_{I,ik}(y,t-s)\mathbf{1}_{(x,y)\in(\infty,\mathbb{R})}
 \Big)_{I,i\in\mathcal{I};k\in\mathcal{K}}
 \\
 \Big(
 \tilde\psi_{IK,i}(x,t-s)\rho_i(S)\mathbf{1}_{(x,y)\in(\mathbb{R},\infty)}
 \Big)_{I,i\in\mathcal{I};k\in\mathcal{K}}
 &  \Big(
 \Psi_{IK,ik}(x,y,t-s)\mathbf{1}_{(x,y)\in(\mathbb{R},\mathbb{R})}
 \Big)_{I,i\in\mathcal{I};K,k\in\mathcal{K}}
 \end{array}
 \right)_{6\times 6}.
 \eeqnn
 Let $\hat{\mathbf{D}}(t,S,x)=(\hat\mu_I(t,S)\mathbf{1}_{x=\infty},\hat\lambda_{IK}(t,S,x)\mathbf{1}_{x\in\mathbb{R}})_{I\in\mathcal{I},K\in\mathcal{K}}$. Then,
 \beqlb \label{eqn3.21}
 \mathbf{D}(t,x)=  \hat{\mathbf{D}}(t,\mathbf{S}(t),x)+\int_0^t ds \int_{\bar{\mathbb{R}}} \mathrm{T}(x,y,\mathbf{S}(s),t,s)\mathbf{D}(s,y)l(dy).
 \eeqlb
The solution to this linear Volterra-Fredholm integral equation is given by
       \beqnn
        \mathbf{D}(t,x)=  \hat{\mathbf{D}}(t,\mathbf{S}(t),x)+\int_0^t ds \int_{\bar{\mathbb{R}}} \mathbf{T}(x,y,\mathbf{S},t,s)\hat{\mathbf{D}}(s,\mathbf{S}(s),y)l(dy),
       \eeqnn
       where $\mathbf{T}(x,y,\mathbf{S},s,t)=\sum_{k=1}^\infty \mathrm{T}_n(x,y,\mathbf{S},t,s)$ and $\mathrm{T}_1(x,y,\mathbf{S},t,s)=\mathrm{T}(x,y,\mathbf{S}(s),t,s)$
       \beqnn
       \mathrm{T}_n(x,y,\mathbf{S},t,s)\ar=\ar\int_s^t dr \int_{\bar{\mathbb{R}}} \mathrm{T}_{n-1}(x,z,\mathbf{S},t,r)\mathrm{T}(z,y,\mathbf{S}(s),r,s)l(dz)\cr
       \ar=\ar\int_s^t dr \int_{\bar{\mathbb{R}}} \mathrm{T}(z,y,\mathbf{S}(r),t,r)\mathrm{T}_{n-1}(x,z,\mathbf{S},r,s)l(dz).
       \eeqnn
Hence the limiting intensities can be approximated in terms of the recursively defined operators $T_n$.
\end{remark}

The uniqueness of solutions to stochastic systems of the form (\ref{LLOB01})-(\ref{Lintensity03}) is an open problem in general.
 The following theorem establishes a uniqueness result under a mild additional condition on the price dynamics. The condition is satisfied if, for instance, $\rho_I(S)=(p_a-p_b)^+$ and $\|\varrho_I\|_\infty\geq 1$. It implies strict positiveness of the spread from which we shall then deduce strong uniqueness of the solution to (\ref{LLOB01})-(\ref{Lintensity03}) and hence convergence in law of our LOB models to a unique limit.


 \begin{theorem}\label{T2}
 Suppose conditions in Theorem \ref{T1} hold and there exists $\epsilon>0$ such that for any $S\in\mathcal{S}$ with $p_a-p_b\in(0,\epsilon)$ have
 \beqlb\label{eqn.CNon}
 0<\rho_I(S)\leq\varrho_I(S)(p_a-p_b),\quad I\in\mathcal{I}.
  \eeqlb
 Then there exists a unique strong solution to (\ref{LLOB01})-(\ref{Lintensity03}).
 \end{theorem}


\subsubsection{Examples and discussion}

Our model predicts that cross-dependencies between order arrivals as well as increasing limit order arrivals and cancellations increase price volatility. Moreover, cross-dependencies in order arrivals may generate positive correlations in the price increments over small time periods and hence volatility clustering as illustrated by the following example.

\begin{example}\label{ex1}
Let us consider the one-sided order book model with\footnote{We ignore the boundedness assumption on the $\rho$ processes for simplicity. In fact, this assumption can be weakened to locally bounded processes.}
 \beqlb\label{eqn.ex01}
 P(t)\ar=\ar P(0)+\int_0^t \sqrt{|P(s)|^2\mu(s)}dB(s),\\
 \mu(t)\ar=\ar \sigma^2+\int_0^t \phi(t-s)|P(s)|^2\mu(s)ds.\label{eqn.ex02}
 \eeqlb
 The benchmark case of a geometric Brownian motion model under a risk-neutral probability measure corresponds to the kernel $\phi(t)\equiv 0$. In that case, the square increments of the log price process are uncorrelated. Let us now fix $\epsilon > 0$, put
 \[
 	\Delta_\epsilon \log P(\cdot) := \log P(\cdot + \epsilon) - \log P(\cdot)
\]	
and assume that
 \beqnn
 \inf_{t\geq 0}\frac{\phi(t+r)}{\phi(t)}>\underline{C}(r) > 0.
 \eeqnn
For the special case of an exponential kernel $\underline{C}(r) = e^{-r}$; for $\phi(t) = \sqrt{1+t}$ we have $\underline{C}(r) = \mathcal{O}(r)$.
%
%
%
%
Then, for $\epsilon$ and $r_0$ small enough, and for all $0 \leq r \leq r_0$,
 \beqnn
\ar \ar  {\rm Cov}((\Delta_\epsilon \log P(t))^2,(\Delta_\epsilon \log P(t+r))^2) \cr
 \ar \approx \ar \epsilon^2{\rm Cov}\left(\int_0^t \phi(t-s)|P(s)|^2\mu(s)ds,\int_0^{t+r} \phi(t+r-s)|P(s)|^2\mu(s)ds\right)\cr
 \ar\geq\ar \underline{C}(r)\epsilon^2{\rm Var}\left(\int_0^t \phi(t-s)|P(s)|^2\mu(s)ds\right)\cr
 \ar\ar -\epsilon^2\int_0^t \phi(t-s)\mathbf{E}[|P(s)|^2\mu(s)]ds\int_0^{r} \phi(r-s)\mathbf{E}[|P(t+s)|^2\mu(t+s)]ds >  0.
 \eeqnn
 \end{example}

 For specific choices of the Hawkes kernels, the price dynamics can be given in closed form.

 \begin{example}
 Consider again the price dynamics of Example \ref{ex1}. For the exponential kernel $\phi(t)=e^{-\kappa t}$ and $\kappa>0$, it is easy to see that (\ref{eqn.ex02}) can be rewritten into
 \beqnn
 e^{\kappa t}\mu(t)\ar=\ar \sigma^2+\int_0^t e^{\kappa s}|P(s)|^2\mu(s)ds.
 \eeqnn
 Applying It\^o's formula to $(|P(t)|^2,\mu(t))$, we have
 \beqnn
  |P(t)|^2\ar=\ar |P(0)|^2+\int_0^t |P(s)|^2\mu(s)ds +\int_0^t 2|P(s)|^2\sqrt{\mu(s)}dB(s),\cr
 \mu(t)\ar=\ar \sigma^2+\int_0^t[\kappa \sigma^2+(|P(s)|^2-\kappa)\mu(s)]ds.
 \eeqnn
 Solving the second equation,
 \beqnn
 \mu(t)\ar=\ar \sigma^2\exp\left\{\int_0^t (|P(r)|^2-\kappa)dr\right\}+\kappa\sigma^2 \int_0^t \exp\left\{\int_s^t(|P(r)|^2- \kappa)dr\right\}ds.
 \eeqnn

\end{example}

We close this section with a simple example where the dynamics of a one-sided book can be given in closed form.

 \begin{example}
 Let us consider a one-sided order books defined by:
 \beqnn
 P(t)\ar=\ar P(0)+\int_0^t \sqrt{\mu(s)}dB(s),\cr
 V(t,x)\ar=\ar V(0,x)+\int_0^t [\lambda(s,x)-\lambda(s,x)V(s,x)]ds
 \eeqnn
 and
 \beqnn
 \mu(t)\ar=\ar |P(t)|^2 +\int_0^t \phi(t-s)\mu(s)ds+\int_0^t\int_{\mathbb{R}}\sqrt{\frac{\pi}{2}}\phi(t-s)e^{-y^2}\lambda(s,y)dsdy,\cr
 \lambda(t,x)\ar=\ar |P(t)|^2e^{-x^2} +\int_0^t \phi(t-s)e^{-x^2}\mu(s)ds+\int_0^t\int_{\mathbb{R}}\sqrt{\frac{\pi}{2}}\phi(t-s)e^{-x^2-y^2}\lambda(s,y)dsdy.
 \eeqnn
 Solving these equations, we have
 \beqnn
 P(t)\ar=\ar P(0)+\int_0^t \sqrt{|P(s)|^2+\mathbf{K}*|P|^2(s)}dB(s),\cr
 V(t,x)\ar=\ar 1+[V(0,x)-1]\exp\left\{-e^{-x^2}\int_0^t \left[ | P(s)|^2+\mathbf{K}*|P|^2(s) \right]ds\right\},
 \eeqnn
 where $*$ denotes the convolution operator and $\mathbf{K}(t)$ is the unique solution to
 $$\mathbf{K}(t)=\phi(t)+\mathbf{K}*\phi(t).$$
 When $\phi(t)$ is constant, exponential or Gamma kernel, then
 \beqnn
 \mathbf{K}(t)=\left\{ \begin{array}{ll}
 2ce^{2ct},&  \mbox{if } \phi(t)=c;\cr
 2ce^{-(\kappa-2c)t},& \mbox{if }\phi(t)=ce^{-\kappa t};\cr
 \sqrt{2c}e^{-\kappa t}\sin(\sqrt{2c}t), & \mbox{if }\phi(t)=ce^{-\kappa t}t.
 \end{array}
 \right.
 \eeqnn
 \end{example}

  \section{Uniqueness of accumulation points}\label{Sec4}
 \setcounter{equation}{0}

In this section, we prove the pathwise uniqueness of solutions to the stochastic dynamic system (\ref{LLOB01})-(\ref{Lintensity03}) under the assumptions of Theorem \ref{T2}. We first prove the positivity of the spread
\[
	\bar{P}(t) := P_a(t) - P_b(t).
\]
This result is then used to prove the uniqueness of solutions. In what follows we assume that $\bar{P}(0)>0$ and that $\hat{\mu}_I(0,S)>0$ for any $S\in\mathcal{S}$.

\subsection{Positivity of the spread}\label{uniqueness1}

We start with the following simple result on the non-negativity of degenerate diffusion processes. The proof follows immediately from the continuity of the sample paths.

\begin{lemma}\label{lemma-non-neg}
	 Let $x(0)>0$ be a $\mathscr{F}_0$-measurable random variable and $b(t,x)\in\mathbb{R}$ and $\sigma(t,x) \geq 0$ be $(\mathscr{F}_t)$-progressive processes such that the diffusion process
 \beqlb\label{CBI2}
 x(t)=x(0)+\int_0^t b(s,x_s)ds+\int_0^t \sigma(s,x_s) dB(s),
 \eeqlb
is well defined and continuous.  If $b(t,x) \geq 0$ and $\sigma(t,x) = 0$ for any $x\leq 0$, then $\mathbf{P}\{x(t)\geq0,t\geq 0\}=1$.
\end{lemma}

 \begin{corollary}[Non price-crossing]\label{NonPC} Suppose that Conditions~\ref{Noncrossing} holds. Then, for any solution $(\mathbf{S},\mathbf{D},\beta_a,\beta_b)$ to the stochastic dynamic system  (\ref{LLOB01})-(\ref{Lintensity03}), we have
 $$\mathbf{P}\{\bar{P}(t)\geq 0:t\geq 0\}=1.$$
 Moreover, if $\varrho_a(S)+\varrho_b(S)>0$ for any $S\in\mathcal{S}$ with $p_a=p_b$,  then the process $\{\bar{P}(t):t\geq 0\}$ reflects at zero.
 \end{corollary}
 \proof  When  $p_a\leq p_b$, Condition~\ref{Noncrossing} implies that
  \beqnn
  \rho_a(S)=\rho_b(S)=\rho^{(n)}_{aL}(S)=\rho^{(n)}_{bL}(S)=0
  \eeqnn
 Moreover, (\ref{hatbeta}) implies that
 \beqnn
 \varrho^{(n)}_I(S)=\frac{\rho^{(n)}_{IM}(S)-\rho^{(n)}_{IL}(S)}{\delta_x^{(n)}}=\frac{\rho^{(n)}_{IM}(S)}{\delta_x^{(n)}}\geq 0\quad \mbox{and}\quad \varrho_I(S)=\lim_{n\to\infty}\varrho^{(n)}_I(S) \geq 0.
 \eeqnn
 The first statement follows from Lemma \ref{lemma-non-neg}.
 For the second statement, define
 \[
 	\tau_-=\inf\{t\geq 0: \bar{P}(t)= 0\} \quad \mbox{and} \quad \tau_+=\inf\{t> \tau_-: \bar{P}(t)> 0\}.
 \]	
 It suffices to prove that $\tau_+=\tau_-$ almost surely. 
 From the continuity of $\bar{P}$, we have that $\mathbb{P}\{\bar{P}(t)=0:t\in[\tau_-,\tau_+]\}=1$.  Assume that $\tau_+(\omega)>\tau_-(\omega)$ for some $\omega\in\Omega$. Then, using that $\mu_a,\mu_b>0$ and that $\varrho_a(S)+\varrho_b(S)>0$ for any $S\in\mathcal{S}$ with $P_a=P_b$, we have for any $t>0$ that
 \beqnn
 \bar{P}((\tau_-+t)\wedge\tau_+)\ar=\ar \int_{\tau_-}^{(\tau_-+t)\wedge\tau_+}\Big[\varrho_a(\mathbf{S}(s))\mu_a(s)
     +\varrho_b(\mathbf{S}(s))\mu_b(s)\Big]ds>0.
 \eeqnn
 This contradicts the assumption that $\tau_+(\omega)>\tau_-(\omega)$ and hence proves the desired result.
 \qed

We proceed with the following lemma from which we shall deduce the strict positivity of the spread.

 \begin{lemma}\label{ThmCBI}
 Let $x(0)>0$ be a $\mathscr{F}_0$-measurable random variable and $a(t)\geq c(t) \geq 0$, $b(t)\in\mathbb{R}$ are  $(\mathscr{F}_t)$-progressive processes. If $\{(x(t),B(t)):t\geq 0\}$ is a weak solution to the following stochastic equation:
 \beqlb\label{CBI}
 x(t)=x(0)+\int_0^t(a(s)-b(s)x(s))ds+\int_0^t \sqrt{2c(s)x(s)}dB(s),
 \eeqlb
 then  $\mathbf{P}\{x(t)>0,t\geq 0\}=1$.
 \end{lemma}
 \proof  Applying It\^o's formula to $\hat{z}(t):=e^{\int_0^tb(s)ds}x(t)$, we have
 \beqnn
 \hat{z}(t)\ar=\ar x(0)+\int_0^ta(s)e^{\int_0^sb(r)dr}ds+\int_0^t \sqrt{2c(s)x(s)}e^{\int_0^sb(r)dr}dB(s)\cr
 \ar=\ar x(0)+\int_0^ta(s)e^{\int_0^sb(r)dr}ds+\int_0^t \sqrt{2c(s)\hat{z}(s)}e^{\frac{1}{2}\int_0^sb(r)dr}dB(s).
 \eeqnn
 Let $\tau_t$ be a strictly increasing process defined as follows:
 $$\tau_t:=\int_0^t[c(s)+\mathbf{1}_{\{c(s)=0\}}]e^{\int_0^sb(r)dr}ds.$$
 Let $\sigma_t:=\tau_t^{-1}$.
 Then
 \beqnn
  d\sigma_t=\frac{e^{-\int_0^{\sigma_t}b(r)dr}}{c(\sigma_t)+\mathbf{1}_{\{c(\sigma_t)=0\}}}dt
 \eeqnn
  and $z(t):=\hat{z}(\sigma_t)$ satisfies the following equation:
 \beqnn
  z(t)\ar=\ar x(0)+\int_0^{\sigma_t}a(s)e^{\int_0^sb(r)dr}ds+\int_0^{\sigma_t} \sqrt{2c(s)\hat{z}(s)}e^{\frac{1}{2}\int_0^sb(r)dr}dB(s)\cr
  \ar=\ar x(0)+\int_0^t a(\sigma_s)e^{\int_0^{\sigma_s}b(r)dr}d{\sigma_s}+\int_0^{t} \sqrt{2c(\sigma_s)\hat{z}(\sigma_s)}e^{\frac{1}{2}\int_0^{\sigma_s}b(r)dr}dB(\sigma_s)\cr
  \ar=\ar x(0)+\int_0^t \Big[\frac{a(\sigma_s)}{c(\sigma_s)}\mathbf{1}_{\{c(\sigma_s)>0\}} +a(\sigma_s)\mathbf{1}_{\{c(\sigma_s)=0\}}\Big]ds\cr
  \ar\ar +\int_0^{t} \sqrt{2\cdot\mathbf{1}_{\{c(\sigma_s)>0\}}z(s)}
  \sqrt{e^{\int_0^{\sigma_s}b(r)dr}[c(\sigma_s)+\mathbf{1}_{\{c(\sigma_s)=0\}}]}dB(\sigma_s).
 \eeqnn
 Obviously, $W(t):=\int_0^t\sqrt{e^{\int_0^{\sigma_s}b(r)dr}[c(\sigma_s)+\mathbf{1}_{\{c(\sigma_s)=0\}}]}dB(\sigma_s)$ is a standard Brownian motion and $(z(t),W(t))$ is a weak solution to
 \beqlb\label{CBI01}
 z(t)=z(0)+\int_0^t \Big[\frac{a(\sigma_s)}{c(\sigma_s)}\mathbf{1}_{\{c(\sigma_s)>0\}} +a(\sigma_s)\mathbf{1}_{\{c(\sigma_s)=0\}}\Big]ds+ \int_0^t \sqrt{2\cdot\mathbf{1}_{\{c(\sigma_s)>0\}}z(s)}dW(s).
 \eeqlb
  Since $a_s\geq c_s$, the comparison theorem \cite[Theorem 1.1]{Yamada1973} yields $\mathbb{P}\{z(t)\geq \tilde{z}(t),t\geq 0\}=1$, where $\tilde{z}(t)$ is the unique solution to
 \beqnn
 \tilde{z}(t)=z(0)+\int_0^t \mathbf{1}_{\{c(\sigma_s)>0\}}ds+ \int_0^t \sqrt{2\cdot\mathbf{1}_{\{c(\sigma_s)>0\}}\tilde{z}(s)}dW(s).
 \eeqnn
 From \cite[p.442]{Revuz1999}, $\mathbf{P}\{ \tilde{z}(t)>0,t\geq 0\}=1$. Hence the desired result follows from the definition of $\tilde z$.
 \qed

 \begin{proposition}\label{StrPos02}
 Suppose conditions in Theorem~\ref{T2} hold. Then any solution $(\mathbf{S},\mathbf{D},\beta_a,\beta_b)$ to the stochastic dynamic system  (\ref{LLOB01})-(\ref{Lintensity03}) satisfies
 \beqnn
 \mathbf{P}\{P_a(t)>P_b(t), t\geq 0\}=1.
 \eeqnn
 \end{proposition}
 \proof
 For any $S\in\mathcal{S}$ with $ p_a-p_b>0$, (\ref{eqn.CNon}) yields,
\begin{equation} \label{est1}
	\hat\rho_a(S):= \frac{\rho_a(S)}{p_a-p_b}\leq \varrho_a(S)\quad \mbox{and}\quad \hat\rho_b(S):= 		\frac{\rho_b(S)}{p_a-p_b}\leq \varrho_b(S).
\end{equation}
 Let $B'(s)$ be another Brownian motion independent to $B_{a/b}$ and put
 \beqnn
 W(t)\ar:=\ar\int_0^t\mathbf{1}_{\{\rho_a(\mathbf{S}(s))\mu_a(s)+\rho_b(\mathbf{S}(s))\mu_b(s)>0\}}\frac{\sqrt{\rho_a(\mathbf{S}(s))\mu_a(s)}dB_a(s)+\sqrt{\rho_b(\mathbf{S}(s))\mu_b(s)}dB_b(s)}{\sqrt{\rho_a(\mathbf{S}(s))\mu_a(s)+\rho_b(\mathbf{S}(s))\mu_b(s)}}\cr
 \ar\ar+\int_0^t\mathbf{1}_{\{\rho_a(\mathbf{S}(s))\mu_a(s)+\rho_b(\mathbf{S}(s))\mu_b(s)=0\}}dB'(s).
 \eeqnn
 Then, $W(t)$ is a standard Brownian motion and
$\bar{P}(t)$ satisfies,
 \beqnn
 \bar{P}(t)\ar=\ar \bar{P}_a(0)
     +\int_0^t\Big[\varrho_a(\mathbf{S}(s))\mu_a(s)+\varrho_b(\mathbf{S}(s))\mu_b(s)+\rho_a(\mathbf{S}(s))\beta_a(s)+\rho_b(\mathbf{S}(s))\beta_b(s)\Big]ds \cr
 \ar\ar+\int_0^t\sqrt{2\rho_a(\mathbf{S}(s))\mu_a(s)}dB_a(s) +\int_0^t\sqrt{2\rho_b(\mathbf{S}(s))\mu_b(s)}dB_b(s)\cr
 \ar=\ar \bar{P}_a(0)
     +\int_0^t\Big[\varrho_a(\mathbf{S}(s))\mu_a(s)+\varrho_b(\mathbf{S}(s))\mu_b(s)+\rho_a(\mathbf{S}(s))\beta_a(s)+\rho_b(\mathbf{S}(s))\beta_b(s)\Big]ds \cr
 \ar\ar+ \int_0^t\sqrt{\rho_a(\mathbf{S}(s))\mu_a(s)+\rho_b(\mathbf{S}(s))\mu_b(s)}\mathbf{1}_{\{\rho_a(\mathbf{S}(s))\mu_a(s)+\rho_b(\mathbf{S}(s))\mu_b(s)>0\}}\cr
 \ar\ar \quad\times\frac{\sqrt{\rho_a(\mathbf{S}(s))\mu_a(s)}dB_a(s)+\sqrt{\rho_b(\mathbf{S}(s))\mu_b(s)}dB_b(s)}{\sqrt{\rho_a(\mathbf{S}(s))\mu_a(s)+\rho_b(\mathbf{S}(s))\mu_b(s)}}\cr
 \ar=\ar \bar{P}(0)+\int_0^t\Big[\varrho_a(\mathbf{S}(s))\mu_a(s)+\varrho_b(\mathbf{S}(s))\mu_b(s)+(\hat\rho_a(\mathbf{S}(s))\beta_a(s)+\hat\rho_b(\mathbf{S}(s))\beta_b(s))\bar{P}(s)\Big]ds\cr
 \ar\ar +\int_0^t\sqrt{2[\hat\rho_a(\mathbf{S}(s))\mu_a(s)+\hat\rho_b(\mathbf{S}(s))\mu_b(s)]\bar{P}(s)}dW(s).
 \eeqnn 
 Hence,  the desired result follows from \eqref{est1} and Lemma~\ref{ThmCBI}.
 \qed

\subsection{Pathwise uniqueness}\label{uniqueness2}

 We are now going to prove the uniqueness of solutions to the stochastic dynamic system (\ref{LLOB01})-(\ref{Lintensity03}).
 From Condition~\ref{C3.2}-\ref{C3.4}, we can see that there exists a constant $C_0>0$ such that for any $t\in[0,T]$, $S\in\mathcal{S}$, $y\in\bar{\mathbb{R}}$
 \beqlb\label{eqn4.5}
 |\rho_I(S)|+|\varrho_I(S)|+|\hat\beta_I(t,S)|+|\hat\mu_I(t,S)|+|\kappa(y,t)|\leq C_0.
 \eeqlb
 From this, (\ref{Lintensity01})-(\ref{Lintensity02}) and Gr\"onwall's inequality, we have
 \beqlb\label{eqn4.6}
 \|\mathbf{D}(t)\|_{\mathcal{D}_{1}^1}\ar\leq\ar C_0+ C_0 \int_0^t  \|\mathbf{D}(s)\|_{\mathcal{D}_{1}^1} ds\quad \mbox{and}\quad
 \sup_{t\in[0,T]}\|\mathbf{D}(t)\|_{\mathcal{D}_{1}^1}\leq C_0.
 \eeqlb

 \textit{Proof for Theorem~\ref{T2}:}
 By \cite[Theorem 1.1, p.163-166]{IW}, distributional existence and pathwise uniqueness imply strong existence. It suffices to prove that pathwise uniqueness holds.
 Define  $\tilde{\mathbf{D}}:=(\tilde{\mu}_I,\tilde{\lambda}_{IK})_{I\in\mathcal{I},J\in\mathcal{J},K\in\mathcal{K}}$ and $\tilde{\beta}_I:=\beta_I-\hat\beta_I$, where $\tilde{\mu}_I=\mu_I-\hat{\mu}_I$ and $\tilde{\lambda}_{IK}=\lambda_{IK}-\hat{\lambda}_{IK}$. Then (\ref{Lintensity01})-(\ref{Lintensity03}) can be written as
 \beqlb\label{MLintensity01}
 \tilde{\mu}_I(t)\ar=\ar
     \sum_{i\in\mathcal{I}}\int_0^t\tilde{\phi}_{Ii}(t-s)\rho_i(\mathbf{S}(s))(\tilde\mu_i(s)+\hat\mu_i(s,\mathbf{S}(s)))ds\cr
     \ar\ar
     +\sum_{i\in\mathcal{I},k\in\mathcal{K}}\int_0^t\int_\mathbb{R}\Phi_{I,ik}(y,t-s)(\tilde{\lambda}_{ik}(s,y)+\hat\lambda_{ik}(s,\mathbf{S}(s),y)) dsdy, \nonumber \\
 \tilde{\lambda}_{IK}(t,x)\ar=\ar
     \sum_{i\in\mathcal{I}}\int_0^t\tilde\psi_{IK,i}(x,t-s)\rho_i(\mathbf{S}(s))(\tilde\mu_i(s)+\hat\mu_i(s,\mathbf{S}(s)))ds\cr
     \ar\ar
     +\sum_{i\in\mathcal{I},k\in\mathcal{K}}\int_0^t\int_{\mathbb{R}}\Psi_{IK,ik}(x,y,t-s)(\tilde{\lambda}_{ik}(s,y)+\hat\lambda_{ik}(s,\mathbf{S}(s),y)) ds dy, \nonumber \\\label{MLintensity02}
 \tilde{\beta}_{I}(t)\ar=\ar \sum_{i\in\mathcal{I}}\int_0^t\tilde\theta_{Ii}(t-s)\rho_i(\mathbf{S}(s))(\tilde\mu_i(s)+\hat\mu_i(s,\mathbf{S}(s)))ds\cr
     \ar\ar +\sum_{i\in\mathcal{I},k\in\mathcal{K}}\int_0^t\int_{\mathbb{R}}\Theta_{I,ik}(y,t-s)(\tilde{\lambda}_{ik}(s,y)+\hat\lambda_{ik}(s,\mathbf{S}(s),y)) dsdy. \nonumber\label{MLintensity03}
 \eeqlb
  Suppose $(\mathbf{S}^{(1)},\tilde{\mathbf{D}}^{(1)},\beta^{(1)}_a,\beta^{(1)}_b)$ and $(\mathbf{S}^{(2)},\tilde{\mathbf{D}}^{(2)},\beta^{(2)}_a,\beta^{(2)}_b)$ are two solutions. Let
 $$(\bar{\mathbf{S}},\bar{\mathbf{D}},\bar{\beta}_a,\bar{\beta}_b):=(\mathbf{S}^{(1)},\tilde{\mathbf{D}}^{(1)},\tilde\beta^{(1)}_a,\tilde\beta^{(1)}_b)-(\mathbf{S}^{(2)},\tilde{\mathbf{D}}^{(2)},\tilde\beta^{(2)}_a,\tilde\beta^{(2)}_b).$$ From (\ref{eqn4.6}) and the Lipschitz continuity of $\rho_I$, we deduce that
 \beqlb\label{eqn3.6}
 \|\bar{\mathbf{D}}(t)\|_{\mathcal{D}_{1}^2}^2 +|\bar{\beta}_I(t)|^2
  \ar\leq\ar C_0\int_0^t[\|\bar{\mathbf{S}}(s)\|_{\mathcal{S}^2}^2+\|\bar{\mathbf{D}}(s)\|_{\mathcal{D}^2_1}^2]ds.
 \eeqlb
 By H\"older's inequality,
 \beqnn
 |\bar{\lambda}_{IK}(t,x)|^2\ar\leq\ar
     C_0\sum_{i\in\mathcal{I}}\int_0^t|\tilde\psi_{IK,i}(x,t-s)|^2|\rho_i(\mathbf{S}^{(1)}(s))-\rho_i(\mathbf{S}^{(2)}(s))|^2ds\cr
     \ar\ar +C_0\sum_{i\in\mathcal{I}}\int_0^t|\tilde\psi_{IK,i}(x,t-s)|^2[|\bar\mu_i(s)|^2+|\hat\mu_i(s,\mathbf{S}^{(1)}(s))-\hat\mu_i(s,\mathbf{S}^{(2)}(s))|^2]ds\cr
     \ar\ar
     +C_0\sum_{i\in\mathcal{I},k\in\mathcal{K}}\int_0^t ds\int_{\mathbb{R}}|\bar{\lambda}_{ik}(s,y)|dy\int_{\mathbb{R}}|\Psi_{IK,ik}(x,y,t-s)|^2|\bar{\lambda}_{ik}(s,y)|dy\cr
     \ar\ar\cr
     \ar\ar+ C_0\sum_{i\in\mathcal{I},k\in\mathcal{K}}\int_0^t ds\int_{\mathbb{R}}|\Psi_{IK,ik}(x,y,t-s)|^2 |\hat\lambda_{ik}(s,\mathbf{S}^{(1)}(s),y)-\hat\lambda_{ik}(s,\mathbf{S}^{(2)}(s),y)|dy\cr
     \ar\ar \qquad\qquad\times \int_{\mathbb{R}}|\hat\lambda_{ik}(s,\mathbf{S}^{(1)}(s),y)-\hat\lambda_{ik}(s,\mathbf{S}^{(2)}(s),y)| dy
 \eeqnn
 and
 \beqlb\label{eqn3.9}
 \|\bar{\lambda}_{IK}(t,\cdot)\|_{L^2}^2
     \ar\leq\ar C_0\int_0^t[|\bar{\mathbf{S}}(s)|_{\mathcal{S}^2}^2+\|\bar{\mathbf{D}}(s)\|_{\mathcal{D}^2_1}^2]ds.  \nonumber  \eeqlb
 In order to estimate the square of the norm of the price difference, we denote, for any $\varepsilon>0$,
 $$\tau_{\varepsilon}:=\inf\left\{t\geq 0: \sqrt{2\rho_I(\mathbf{S}^{(l)}(s))\mu^{(l)}_I(s)}\leq \varepsilon, I\in\mathcal{I},l=1,2\right\}.$$
 From Proposition~\ref{StrPos02} and the continuity of $\rho_I$ and $\mu^{(l)}_I$ (see (\ref{eqn2.8}) and (\ref{Lintensity01})), we see that $\tau_\varepsilon\to\infty$ a.s. as $\varepsilon\to 0$. Hence, it is enough to consider $t\in [0,\tau_\varepsilon]$. In particular, for all such $t$
 \beqnn
 	& & \int_0^t\Big|\sqrt{2\rho_a(\mathbf{S}^{(1)}(s))\mu^{(1)}_a(s)}-\sqrt{2\rho_a(\mathbf{S}^{(2)}(s))\mu^{(2)}_a(s)}\Big|^2ds \\
	& \leq & \frac{1}{2 \epsilon^2}\int_0^t\Big[|\rho_a(\mathbf{S}^{(1)}(s))-\rho_a(\mathbf{S}^{(2)}(s))|^2+|\bar\mu_a(s)|^2+|\hat\mu_a(s,\mathbf{S}^{(1)}(s))-\hat\mu_a(s,\mathbf{S}^{(2)}(s))|^2\Big]ds.
 \eeqnn
Thus, an application of It\^o's formula to $|\bar{P}_a(t)|^2$ yields,
 \beqnn
  |\bar{P}_a(t)|^2
     \ar\leq\ar   C_0\int_0^t|\bar{P}_a(s)|^2ds+ C_0\int_0^t\Big|\rho_a(\mathbf{S}^{(1)}(s))-\rho_a(\mathbf{S}^{(2)}(s))\Big|^2ds\cr
     \ar\ar+C_0\int_0^t\Big[|\bar\beta_a(s)|^2+|\hat\beta_a(s,\mathbf{S}^{(1)}(s))-\hat\beta_a(s,\mathbf{S}^{(2)}(s))|^2\Big]ds\cr
     \ar\ar
     +C_0\int_0^t\Big[|\varrho_a(\mathbf{S}^{(1)}(s))-\varrho_a(\mathbf{S}^{(2)}(s))|^2+|\bar\mu_a(s)|^2+|\hat\mu_a(s,\mathbf{S}^{(1)}(s))-\hat\mu_a(s,\mathbf{S}^{(2)}(s))|^2\Big]ds \cr
     \ar\ar
     +\frac{2}{\varepsilon^2}\int_0^t\Big[|\rho_a(\mathbf{S}^{(1)}(s))-\rho_a(\mathbf{S}^{(2)}(s))|^2+|\bar\mu_a(s)|^2+|\hat\mu_a(s,\mathbf{S}^{(1)}(s))-\hat\mu_a(s,\mathbf{S}^{(2)}(s))|^2\Big]ds\cr
     \cr
     \ar\ar +\int_0^t2\bar{P}_I(s)\Big[\sqrt{2\rho_a(\mathbf{S}^{(1)}(s))\mu^{(1)}_a(s)}-\sqrt{2\rho_a(\mathbf{S}^{(2)}(s))\mu^{(2)}_a(s)}\Big]dB_a(s),
 \eeqnn
 and hence
 \beqnn
  \mathbf{E}[|\bar{P}_a(t)|^2]\ar\leq\ar
    C_0(1+2/\varepsilon^2)\int_0^t\Big[\mathbf{E}[\|\bar{\mathbf{S}}(s)\|_{\mathcal{S}^2}^2]+\mathbf{E}[|\bar\beta_a(s)|^2]+\mathbf{E}[\|\bar{\mathbf{D}}(s)\|_{\mathcal{D}_2^2}^2]\Big]ds.
  \eeqnn
   The following estimate allows us to estimate the norms of the volume density functions. For any $\epsilon>0$,
 \begin{equation}\label{eqn5.15}
 \begin{split}
 & \lefteqn{\|\tilde{\lambda}^{(l)}_{IK}(s,\cdot+\epsilon)-\tilde{\lambda}^{(l)}_{IK}(s,\cdot)\|_{L^2}^2} \\
 \leq\ &
     C_0\sum_{i\in\mathcal{I},k\in\mathcal{K}}\int_0^t ds\int_{\mathbb{R}}|\lambda_{ik}(s,y)| dy \int_{\mathbb{R}}\|\Psi_{IK,ik}(\cdot+\epsilon,y,t-s)-\Psi_{IK,ik}(\cdot,y,t-s)\|_{L^2}^2|\lambda_{ik}(s,y)| dy\\
     & +C_0\sum_{i\in\mathcal{I}}\int_0^t\|\tilde\psi_{IK,i}(\cdot+\epsilon,t-s)-\tilde\psi_{IK,i}(\cdot,t-s)\|_{L^2}^2ds\\
     \leq\ & C\delta^2
     +C_0\delta^2\sum_{i\in\mathcal{I},k\in\mathcal{K}}\int_0^t \|{\lambda}_{ik}(s,\cdot)\|^2_{L^1} ds\leq  C_0\epsilon^2.
 \end{split}
\end{equation}
Thus, by direct computation we verify that
 \beqnn
  \|\bar{V}_a(t,\cdot)\|_{L^2}^2
   \ar\leq\ar C_0\int_0^t[\|\bar{\mathbf{S}}(s) \|_{\mathcal{S}^2}^2+\|\bar{\mathbf{D}}(s) \|_{\mathcal{D}_2^2}^2]ds.
      \eeqnn
 As a result,
 \beqnn
 \mathbf{E}[\|\bar{\mathbf{S}}(t) \|_{\mathcal{S}^2}^2+\|\bar{\mathbf{D}}(t) \|_{\mathcal{D}_{1,2}^2}^2] \leq C_0\int_0^t\mathbf{E}[\|\bar{\mathbf{S}}(s) \|_{\mathcal{S}^2}^2+\|\bar{\mathbf{D}}(s) \|_{\mathcal{D}_{1,2}^2}^2]ds.
 \eeqnn
 By Gr\"onwall's inequality, this yields
 \beqnn
 \mathbf{E}[\|\bar{\mathbf{S}}(t) \|_{\mathcal{S}^2}^2+\|\bar{\mathbf{D}}(t) \|_{\mathcal{D}_{1,2}^2}^2]=0.
 \eeqnn
 Along with the continuity of the solutions this yields the desired pathwise uniqueness.
 \qed

  \section{Tightness of the LOB models}\label{Sec5}
 \setcounter{equation}{0}

 In this section, we prove the tightness of the processes $(\mathbf{S}^{(n)},\mathbf{D}^{(n)},\beta_a^{(n)},\beta_b^{(n)})$ by showing that the pointwise moment conditions on the state sequence and the moment conditions on the increments of the state sequence in Kurtz's tightness criterion hold; see \cite[Theorem 6.8]{Walsh1986}.
 In what follows we assume without loss of generality that all constants in Condition~\ref{C3.1}-\ref{C3.4} are equal to $1$.

 \subsection{Pointwise norm estimates}

 Our norm estimates use the following quantity: for any $t\geq 0$,
 \beqnn
 J^{(n)}(t)= 1+\sum_{i\in\mathcal{I},j\in\mathcal{J}}\int_0^t |\delta_x^{(n)}|^2N_{ij}^{(n)}(ds) +\sum_{i\in\mathcal{I},k\in\mathcal{K}}\int_0^t\int_{\mathbb{R}}\int_{\mathbb{R}} \delta_v^{(n)}M_{ik}^{(n)}(ds,dy,dz).
 \eeqnn
 \begin{lemma}\label{BL01}
 There exists a constant $C_0>0$ such that for any $0\leq r\leq t\leq T$ and $p\in\{1,2,4\}$
 \beqnn
 \mathbf{E}_{\mathscr{F}_r}\Big[\|\mathbf{D}^{(n)}(t)\|^p_{\mathcal{D}_1^p}\Big]\leq C_0|J^{(n)}(r)|^p\quad \mbox{and}\quad \mathbf{E}\Big[\sup_{t\in[0,T]}\|\mathbf{D}^{(n)}(t)\|^p_{\mathcal{D}_1^p}\Big]\leq C_0.
 \eeqnn
 \end{lemma}
 \proof  From Condition~\ref{C3.3} i) and \ref{C3.4} ii),
 \beqlb\label{Bound.01}
 |\delta_x^{(n)}|^2\mu_{IJ}^{(n)}(t)\ar=\ar \hat{\mu}_{IJ}^{(n)}(t,\mathbf{S}^{(n)}(t-))
     +\sum_{i\in\mathcal{I},j\in\mathcal{J}}\int_0^t \phi^{(n)}_{IJ,ij}(t-s)|\delta_x^{(n)}|^2N_{ij}^{(n)}(ds) \cr
     \ar\ar +\sum_{i\in\mathcal{I},k\in\mathcal{K}}\int_0^t\int_{\mathbb{R}}\int_{\mathbb{R}} \Phi^{(n)}_{IJ,ik}(y,t-s)\delta_v^{(n)}M_{ik}^{(n)}(ds,dy,dz)\leq  J^{(n)}(t)
 \eeqlb
 and
 \beqnn
 \mathbf{E}_{\mathscr{F}_r}\left[ |\delta_x^{(n)}|^2\mu_{IJ}^{(n)}(t)\right]
 \ar\leq \ar  1
     +\sum_{i\in\mathcal{I},j\in\mathcal{J}}\int_0^r |\delta_x^{(n)}|^2N_{ij}^{(n)}(ds) +\sum_{i\in\mathcal{I},k\in\mathcal{K}}\int_0^r\int_{\mathbb{R}}\int_{\mathbb{R}} \delta_v^{(n)}M_{ik}^{(n)}(ds,dy,dz)\cr
     \ar\ar
     +\sum_{i\in\mathcal{I},j\in\mathcal{J}}\int_r^t \mathbf{E}_{\mathscr{F}_r}\left[|\delta_x^{(n)}|^2\mu_{ij}^{(n)}(s) +\sum_{i\in\mathcal{I},k\in\mathcal{K}}\int_{\mathbb{R}} \delta_v^{(n)}\lambda_{ik}^{(n)}(s,y)dy\right]ds\cr
     \ar\leq \ar J^{(n)}(r)
     +\int_r^t \mathbf{E}_{\mathscr{F}_r}\left[\|\mathbf{D}^{(n)}(s)\|_{\mathcal{D}_1^1}\right]ds.
 \eeqnn
 Similarly, we also have
 \beqlb\label{Bound.01.1}
 \|\delta_v^{(n)}\lambda_{IK}^{(n)}(t,\cdot)\|_{L^1} \ar\leq\ar J^{(n)}(t)
 \eeqlb
 and
 \beqnn
 \mathbf{E}_{\mathscr{F}_r}\left[ \|\delta_v^{(n)}\lambda_{IK}^{(n)}(t,\cdot)\|_{L^1}\right]\ar\leq\ar J^{(n)}(r)+\int_r^t \mathbf{E}_{\mathscr{F}_r}\left[\|\mathbf{D}^{(n)}(s)\|_{\mathcal{D}_1^1}\right]ds.
 \eeqnn
 Hence,
 \beqnn
 \mathbf{E}_{\mathscr{F}_r}\left[\|\mathbf{D}^{(n)}(t)\|_{\mathcal{D}_1^1}\right]\leq 8J^{(n)}(r)+8\int_r^t\mathbf{E}_{\mathscr{F}_r}\left[\|\mathbf{D}^{(n)}(s)\|_{\mathcal{D}_1^1}\right]ds.
 \eeqnn
 From Gr\"onwall's inequality, we have
 \beqlb\label{Bound.02}
 \mathbf{E}_{\mathscr{F}_r}\left[\|\mathbf{D}^{(n)}(t)\|_{\mathcal{D}_1^1}\right]\leq 8e^{8(t-r)}J^{(n)}(r).
 \eeqlb
For $p=4$ (the case $p=2$ is similar),
 \beqnn
 \Big||\delta_x^{(n)}|^2\mu_{IJ}^{(n)}(t)\Big|^4
 \ar\leq \ar  C_0|J^{(n)}(r)|^4
 +C_0(t-r)^3\int_r^t\|\mathbf{D}^{(n)}(s)\|^4_{\mathcal{D}_1^4}ds
 +C_0\sum_{i\in\mathcal{I},j\in\mathcal{J}}\Big|\int_r^t|\delta_x^{(n)}|^2\tilde{N}_{ij}^{(n)}(ds)\Big|^4 \cr
 \ar\ar +C_0\sum_{i\in\mathcal{I},k\in\mathcal{K}}\Big|\int_r^t\int_{\mathbb{R}} \int_{\mathbb{R}} \delta_v^{(n)}\tilde{M}_{ik}^{(n)}(ds,dy,dz)\Big|^4
 \eeqnn
 and
 \beqnn
  \|\delta_v^{(n)}\lambda_{IK}^{(n)}(t,\cdot)\|_{L^1}^4\ar\leq\ar C_0|J^{(n)}(r)|^4 +C_0(t-r)^3\int_r^t\|\mathbf{D}^{(n)}(s)\|^4_{\mathcal{D}_1^4}ds
  +C_0\sum_{i\in\mathcal{I},j\in\mathcal{J}}\Big|\int_r^t|\delta_x^{(n)}|^2\tilde{N}_{ij}^{(n)}(ds)\Big|^4 \cr
  \ar\ar +C_0\sum_{i\in\mathcal{I},k\in\mathcal{K}}\Big|\int_r^t\int_{\mathbb{R}} \int_{\mathbb{R}} \delta_v^{(n)}\tilde{M}_{ik}^{(n)}(ds,dy,dz)\Big|^4,
 \eeqnn
 where $\tilde{N}(ds)$ and $\tilde{M}(ds,dy,dz)$ are the compensated random measures of $N(ds)$ and $M(ds,dy,dz)$, respectively.
 By the Burkholder-Davis-Gundy inequality,  the Cauchy-Schwarz inequality and H\"older's inequality,
 \beqnn
 \mathbf{E}_{\mathscr{F}_r}\left[\Big|\int_r^t|\delta_x^{(n)}|^2\tilde{N}_{ij}^{(n)}(ds)\Big|^4\right]\ar\leq\ar C_0\mathbf{E}_{\mathscr{F}_r}\left[\Big|\int_r^t|\delta_x^{(n)}|^4N_{ij}^{(n)}(ds)\Big|^2\right]\cr
\ar=\ar C_0\mathbf{E}_{\mathscr{F}_r}\left[\Big|\int_r^t|\delta_x^{(n)}|^4\mu_{ij}^{(n)}(s)ds+\int_r^t|\delta_x^{(n)}|^4\tilde{N}_{ij}^{(n)}(ds)\Big|^2\right]\cr
 \ar\leq\ar C_0\mathbf{E}_{\mathscr{F}_r}\left[\Big|\int_r^t|\delta_x^{(n)}|^4\mu_{ij}^{(n)}(s)ds\Big|^2\right]+C_0\mathbf{E}_{\mathscr{F}_r}\left[\Big|\int_r^t|\delta_x^{(n)}|^4\tilde{N}_{ij}^{(n)}(ds)\Big|^2\right]\cr
 \ar\leq\ar C_0\mathbf{E}_{\mathscr{F}_r}\left[\Big|\int_r^t|\delta_x^{(n)}|^4\mu_{ij}^{(n)}(s)ds\Big|^2\right]+C_0\mathbf{E}_{\mathscr{F}_r}\left[\int_r^t|\delta_x^{(n)}|^8\mu_{ij}^{(n)}(s)ds\right]\cr
 \ar\leq\ar  C_0\int_r^t\mathbf{E}_{\mathscr{F}_r}\left[|\delta_x^{(n)}|^8|\mu_{ij}^{(n)}(s)|^2+|\delta_x^{(n)}|^8\mu_{ij}^{(n)}(s)\right]ds
 \eeqnn 
 and
 \beqnn
 \lefteqn{\mathbf{E}_{\mathscr{F}_r}\left[\Big|\int_r^t\int_{\mathbb{R}} \int_{\mathbb{R}} \delta_v^{(n)}\tilde{M}_{ik}^{(n)}(ds,dy,dz)\Big|^4\right]}\qquad\qquad\qquad\qquad\quad\ar\ar\cr
 \ar\leq\ar C_0\mathbf{E}_{\mathscr{F}_r}\left[\Big|\int_r^t\int_{\mathbb{R}} \int_{\mathbb{R}} |\delta_v^{(n)}|^2M_{ik}^{(n)}(ds,dy,dz)\Big|^2\right]\cr
 \ar\leq\ar  C_0\int_r^t\mathbf{E}_{\mathscr{F}_r}\left[\||\delta_v^{(n)}|^2\lambda_{ik}^{(n)}(s,\cdot)\|_{L^1}^2+\||\delta_v^{(n)}|^4\lambda_{ik}^{(n)}(s,\cdot)\|_{L^1}\right]ds.
 \eeqnn
 Hence,
 \beqnn
 \mathbf{E}_{\mathscr{F}_r} \left[\|\mathbf{D}^{(n)}(t)\|^4_{\mathcal{D}_1^4}\right]
  \ar\leq\ar C_0|J^{(n)}(r)|^4 +C_0(t-r)^3\int_r^t\mathbf{E}_{\mathscr{F}_r} \left[\|\mathbf{D}^{(n)}(s)\|^4_{\mathcal{D}_1^4}\right]ds\cr
 \ar\ar
  +C_0\int_r^t\mathbf{E}_{\mathscr{F}_r}\left[|\delta_x^{(n)}|^8|\mu_{ij}^{(n)}(s)|^2+|\delta_x^{(n)}|^8\mu_{ij}^{(n)}(s)\right]ds\cr
  \ar\ar +C_0\int_r^t\mathbf{E}_{\mathscr{F}_r}\left[\||\delta_v^{(n)}|^2\lambda_{ik}^{(n)}(s,\cdot)\|_{L^1}^2+\||\delta_v^{(n)}|^4\lambda_{ik}^{(n)}(s,\cdot)\|_{L^1}\right]ds\cr
  \ar\leq\ar  C_0|J^{(n)}(r)|^4 +C_0\int_r^t\mathbf{E}_{\mathscr{F}_r} \left[\|\mathbf{D}^{(n)}(s)\|^4_{\mathcal{D}_1^4}\right]ds.
 \eeqnn
 The first result now follows from Gr\"onwall's inequality.
 The second result follows from (\ref{Bound.01})-(\ref{Bound.01.1}) together with the first result with $r=0$.
 \qed

The following estimate of the conditional moments of the increment of $J^{(n)}(t)$ follows directly from the above proof and is hence omitted.

 \begin{proposition}\label{BP02}
 There exists a constant $C_0>0$ such that for any $0\leq r\leq t\leq T$ and $p\in\{1,2,4\}$
 \beqnn
 \mathbf{E}_{\mathscr{F}_r}\left[|J^{(n)}(t)-J^{(n)}(r)|^p\right]\leq C_0 |J^{(n)}(r)|^p|t-r| \quad\mbox{and}\quad \mathbf{E}\left[|J^{(n)}(T)|^p\right]\leq C_0.
 \eeqnn

 \end{proposition}

Using the same arguments as in the proof of Lemma~\ref{BL01}, the following moment estimates of the drift follow from Condition~\ref{C3.3} i).

 \begin{proposition}\label{BP03}
 There exists a constant $C_0>0$ such that for any $0\leq r\leq t\leq T$ and $I\in\mathcal{I}$ have
 \beqnn
 \mathbf{E}_{\mathscr{F}_r}\Big[|\beta_I^{(n)}(t)|^2\Big]\leq C_0|J^{(n)}(r)|^2 \quad \mbox{and}\quad \mathbf{E}\Big[\sup_{t\in[0,T]}|\beta_I^{(n)}(t)|^2\Big]\leq C_0.
 \eeqnn
 \end{proposition}

Next, we establish moment estimates for the intensities of passive order arrivals.

  \begin{lemma}\label{BL04}
  There exists a constant $C_0>0$ such that for any $0\leq r\leq t\leq T$ and $p\in\{2,4\}$
  \beqnn
  \mathbf{E}_{\mathscr{F}_r}\Big[\|\delta_v^{(n)}\lambda_{IK}^{(n)}(t,\cdot)\|^p_{L^p}\Big]\leq C_0|J^{(n)}(r)|^p\quad{and}\quad  \mathbf{E}\Big[\sup_{t\in[0,T]}\|\delta_v^{(n)}\lambda_{IK}^{(n)}(t,\cdot)\|^p_{L^p}\Big] \leq  C_0.
  \eeqnn
 \end{lemma}
 \proof Here we just prove this result with $p=4$. From Condition~\ref{C3.4} i) and H\"older's inequality,  we have
 \beqnn
 \lefteqn{\int_{\mathbb{R}}\Big|\int_0^t \psi_{IK,ij}(x,t-s)|\delta_x^{(n)}|^2N_{ij}^{(n)}(ds)\Big|^4dx }\qquad\ar\ar\cr
 \ar\leq\ar\int_0^t \int_{\mathbb{R}}|\psi_{IK,ij}(x,t-s)|^4dx|\delta_x^{(n)}|^2N_{ij}^{(n)}(ds) \times\Big| \int_0^t |\delta_x^{(n)}|^2N_{ij}^{(n)}(ds)\Big|^3\cr
 \ar\leq\ar \Big| \int_0^t |\delta_x^{(n)}|^2N_{ij}^{(n)}(ds)\Big|^4 \cr
 \eeqnn
 and
 \beqnn
 \lefteqn{\int_{\mathbb{R}}\Big|\int_0^t\int_{\mathbb{R}}\int_{\mathbb{R}} \Psi_{IK,ik}(x,y,t-s) \delta_v^{(n)}M_{ik}^{(n)}(ds,dy,dz)\Big|^4 dx }\qquad\ar\ar\cr
 \ar\leq\ar
 \int_0^t\int_{\mathbb{R}}\int_{\mathbb{R}} \int_{\mathbb{R}}|\Psi_{IK,ik}(x,y,t-s)|^4dx \delta_v^{(n)}M_{ik}^{(n)}(ds,dy,dz)\times\Big|\int_0^t\int_{\mathbb{R}}\int_{\mathbb{R}} \delta_v^{(n)}M_{ik}^{(n)}(ds,dy,dz)\Big|^3\cr
 \ar\leq\ar  \Big|\int_0^t\int_{\mathbb{R}}\int_{\mathbb{R}} \delta_v^{(n)}M_{ik}^{(n)}(ds,dy,dz)\Big|^4.
 \eeqnn
 By (\ref{BIntensity01}) and Cauchy-Schwarz inequality,
  \beqnn
 \|\delta_v^{(n)}\lambda_{IK}^{(n)}(t,x)\|_{L^4}^4\ar\leq\ar C_0
     +C_0\sum_{i\in\mathcal{I},j\in\mathcal{J}}\Big| \int_0^t |\delta_x^{(n)}|^2N_{ij}^{(n)}(ds)\Big|^4 \cr
     \ar\ar +C_0\sum_{i\in\mathcal{I},k\in\mathcal{K}}\Big|\int_0^t\int_{\mathbb{R}}\int_{\mathbb{R}} \delta_v^{(n)}M_{ik}^{(n)}(ds,dy,dz)\Big|^4 \leq C_0 |J^{(n)}(t)|^4.
 \eeqnn
 The desired result now follows as in the proof of Lemma~\ref{BL01}. 
 \qed

The conditional moment estimates on the volume density functions use the following observation. Since $$\mathbf{1}_{\Delta^{(n)}(x-P_a^{(n)}(s))}(y)=\mathbf{1}_{\Delta^{(n)}(y+P_a^{(n)}(s))}(x), \quad x,y\in\mathbb{R},$$
it follows from Fubini's theorem, for any integrable function $g(y)$,
 \beqlb\label{eqn3.4}
 \int_{\mathbb{R}}\int_{\Delta^{(n)}(x-P_a^{(n)}(s))} g(y)dydx \ar=\ar \int_{\mathbb{R}}\int_{\mathbb{R}} g(y)\mathbf{1}_{\Delta^{(n)}(x-P_a^{(n)}(s))}(y)dydx\cr
 \ar=\ar\int_{\mathbb{R}}\int_{\mathbb{R}} g(y)\mathbf{1}_{\Delta^{(n)}(y+P_a^{(n)}(s))}(x)dydx\cr
 \ar=\ar\int_{\mathbb{R}} g(y)dy\int_{\mathbb{R}}\mathbf{1}_{\Delta^{(n)}(y+P_a^{(n)}(s))}(x)dx= \delta_x^{(n)}\int_{\mathbb{R}} g(y)dy.
 \eeqlb

   \begin{lemma}\label{BL05}
 There exists a constant $C_0>0$ such that for any $0\leq r\leq t\leq T$ and $I\in\mathcal{I}$,
 \beqnn
 \mathbf{E}_{\mathscr{F}_r}\left[\|V_I(t,\cdot)\|^4_{L^4} \right]\leq C_0[\|V^{(n)}_I (r,x)\|_{L^4}^4 + |J^{(n)} (r)|^4]\quad\mbox{and}\quad  \mathbf{E}\left[\sup_{t\in[0,T]}\|V_I(t,\cdot)\|^4_{L^4} \right]\leq C_0.
 \eeqnn
 \end{lemma}
 \proof Since the third term on the right side of the third equations in (\ref{LLOB01}) is non-positive and $V_{a}^{(n)}(t,x)$ is always nonnegative, it follows from H\"older's inequality that
 \beqnn
 |V_{a}^{(n)}(t,x)|^4\ar\leq\ar C_0|V^{(n)}_a (r,x)|^4 +C_0\Big|\int_r^t\int_{\Delta^{(n)}(x-P_a^{(n)}(s))} \alpha^{(n)}_{aL}\frac{\delta_v^{(n)}}{\delta_x^{(n)}} \lambda^{n}_{aL}(s,y) dsdy\Big|^4\cr
 \ar\ar+C_0\Big|\int_r^t\int_{\Delta^{(n)}(x-P_a^{(n)}(s))}\int_{\mathbb{R}_+}\frac{\delta_v^{(n)}}{\delta_x^{(n)}} (e^{z}-1)\tilde{M}_{aL}^{(n)}(ds,dy,dz)\Big|^4\cr
 \ar\leq \ar C_0|V^{(n)}_a (r,x)|^4 +C_0\frac{|\alpha_{aL}|^4}{\delta_x^{(n)}}|t-r|^3\int_r^t\int_{\Delta^{(n)}(x-P_a^{(n)}(s))} \Big|\delta_v^{(n)} \lambda^{n}_{aL}(s,y)\Big|^4dy ds\cr
 \ar\ar+C_0\Big|\int_r^t\int_{\Delta^{(n)}(x-P_a^{(n)}(s))}\int_{\mathbb{R}_+}\frac{\delta_v^{(n)}}{\delta_x^{(n)}} (e^{z}-1)\tilde{M}_{aL}^{(n)}(ds,dy,dz)\Big|^4.
 \eeqnn
 From the Burkholder-Davis-Gundy inequality, there exists a constant $C_0>0$ such that
 \beqnn
 \lefteqn{\mathbf{E}_{\mathscr{F}_r}\left[\Big|\int_r^t\int_{\Delta^{(n)}(x-P_a^{(n)}(s))}\int_{\mathbb{R}_+}\frac{\delta_v^{(n)}}{\delta_x^{(n)}} (e^{z}-1)\tilde{M}_{aL}^{(n)}(ds,dy,dz)\Big|^4\right]}\ar\ar\cr
 \ar\leq \ar  C\mathbf{E}_{\mathscr{F}_r}\left[\Big|\int_r^t\int_{\Delta^{(n)}(x-P_a^{(n)}(s))}\int_{\mathbb{R}_+}\Big|\frac{\delta_v^{(n)}}{\delta_x^{(n)}} (e^{z}-1)\Big|^2M_{aL}^{(n)}(ds,dy,dz)\Big|^2\right]\cr
 \ar\leq\ar
 C_0\mathbf{E}_{\mathscr{F}_r}\left[\Big|\int_r^t\int_{\Delta^{(n)}(x-P_a^{(n)}(s))}\Big|\frac{\delta_v^{(n)}}{\delta_x^{(n)}} \Big|^2\lambda_{aL}^{(n)}(s,y)dyds\Big|^2\right]\cr
 \ar\leq\ar
 C_0\frac{|\delta_v^{(n)}|^2}{|\delta_x^{(n)}|^3}|t-r|\mathbf{E}_{\mathscr{F}_r}\left[\int_r^t\int_{\Delta^{(n)}(x-P_a^{(n)}(s))}|\delta_v^{(n)} \lambda_{aL}^{(n)}(s,y)|^2 dyds\right]\cr
 \ar\ar +C_0\mathbf{E}_{\mathscr{F}_r}\left[\int_r^t\int_{\Delta^{(n)}(x-P_a^{(n)}(s))}\Big|\frac{\delta_v^{(n)}}{\delta_x^{(n)}} \Big|^4\lambda_{aL}^{(n)}(s,y)dyds\right].
 \eeqnn
 From (\ref{eqn3.4}) and Fubini's theorem,
 \beqnn
 \mathbf{E}_{\mathscr{F}_r}\left[\|V_a(t,\cdot)\|^4_{L^4} \right]\ar\leq\ar C_0\|V^{(n)}_a (r,x)\|_{L^4}^4 +C_0|\alpha_{aL}|^4|t-r|^3\int_r^t \mathbf{E}_{\mathscr{F}_r}\left[\|\delta_v^{(n)} \lambda^{n}_{aL}(s,\cdot)\|_{L^4}^4\right] ds\cr
 \ar\ar +C_0\Big|\frac{\delta_v^{(n)}}{\delta_x^{(n)}} \Big|^2|t-r|\int_r^t\mathbf{E}_{\mathscr{F}_r}\left[\|\delta_v^{(n)} \lambda_{aL}^{(n)}(s,\cdot)\|_{L^2}^2\right] ds\cr
 \ar\ar +C_0\Big|\frac{\delta_v^{(n)}}{\delta_x^{(n)}} \Big|^3\int_r^t\mathbf{E}_{\mathscr{F}_r}\left[\|\delta_v^{(n)} \lambda_{aL}^{(n)}(s,\cdot)\|_{L^1}\right]ds.
 \eeqnn
 The first result now follows from Lemma~\ref{BL01} and Lemma~\ref{BL04}. The second result follows from:
 \beqnn
  \lefteqn{\mathbf{E} \left[\sup_{t\in[0,T]}\|V_a(t,\cdot)\|^4_{L^4} \right]}\ar\ar\cr
  \ar\leq\ar C_0\|V^{(n)}_a (0,x)\|_{L^4}^4 + C_0\mathbf{E} \left[\int_\mathbb{R}\Big|\int_0^T\int_{\Delta^{(n)}(x-P_a^{(n)}(s))}\int_{\mathbb{R}_+}\frac{\delta_v^{(n)}}{\delta_x^{(n)}} (e^{z}-1)M_{aL}^{(n)}(ds,dy,dz)\Big|^4dx\right]\cr
   \ar\leq\ar C_0\|V^{(n)}_a (0,x)\|_{L^4}^4+C_0\sum_{p\in\{1,2,4\}}\int_0^T \mathbf{E} \left[\|\delta_v^{(n)} \lambda^{n}_{aL}(s,\cdot)\|_{L^p}^p\right] ds\leq C_0.
 \eeqnn
 \qed

\subsection{Moment estimates for the increments}

 We start to prove moment estimates for the increments of the state processes. We rewrite (\ref{DLOB01})-(\ref{DLOB03}) into
 \beqnn
 P_a^{(n)}(t)\ar=\ar P_a^{(n)}(0)+ \int_0^t\Big[\rho_{bM}^{(n)}(\mathbf{S}^{(n)}(s))\beta_a^{(n)}(s)+\varrho_a^{(n)}(\mathbf{S}^{(n)}(s))|\delta_x^{(n)}|^2\mu_{aL}^{(n)}(s)\Big]ds\cr
 \ar\ar+\int_0^t\delta_x^{(n)}\tilde{N}_{bM}^{(n)}(ds)-\int_0^t\delta_x^{(n)}\tilde{N}_{aL}^{(n)}(ds),\cr
 V_{a}^{(n)}(t,x)\ar=\ar V^{(n)}_a (0,x) +\int_0^t\int_{\Delta^{(n)}(x-P_a^{(n)}(s-))}\int_{\mathbb{R}_+}\frac{\delta_v^{(n)}}{\delta_x^{(n)}} (e^{z}-1)\tilde{M}_{aL}^{(n)}(ds,dy,dz)\cr
 \ar\ar +\int_0^t\int_{\Delta^{(n)}(x-P_a^{(n)}(s-))}\int_{\mathbb{R}_+}\frac{\delta_v^{(n)}}{\delta_x^{(n)}} V_{a}^{(n)}(s-,y+P_a^{(n)}(t)) (e^{-z}-1)\tilde{M}_{aC}^{(n)}(ds,dy,dz)\cr
 \ar\ar +\int_0^t\int_{\Delta^{(n)}(x-P_a^{(n)}(s))} \frac{\delta_v^{(n)}}{\delta_x^{(n)}}\Big[\alpha_{aL} \lambda_{aL}(s,y) +\alpha_{aC} V_{a}^{(n)}(s,y+P_a^{(n)}(t)) \lambda_{aC}(s,y)\Big]dsdy,\cr
 \eeqnn
with $P_b^{(n)}(t)$ and $V_{b}^{(n)}(t,x)$ being rewritten analogously.

   \begin{lemma}\label{BL06}
  There exists a constant $C_0>0$ such that for any $0\leq r\leq t\leq T$
 	\beqnn
    \mathbf{E}_{\mathscr{F}_r}\left[\|\mathbf{S}^{(n)}(t)-\mathbf{S}^{(n)}(r)\|^2_{\mathcal{S}^2}\right]\leq C_0\big(\sum_{I\in\mathcal{I}}\|V^{(n)}_I (r,x)\|_{L^4}^4 + |J^{(n)} (r)|^4\big)(|t-r|^2+|t-r|)
 	\eeqnn
 and
 \beqnn
 \mathbf{E}\Big[\sup_{t\in[0,T]}\|\mathbf{S}^{(n)}(t)\|^2_{\mathcal{S}^2}\Big]\leq C_0.
 \eeqnn
 \end{lemma}
 \proof  It suffices to prove the first inequality, since the second one follows directly from the first one with $r=0$ and (\ref{BInitial01}).
 From Lemma~\ref{BL01} and Proposition~\ref{BP03}, we have
 \beqnn
 \mathbf{E}_{\mathscr{F}_r}\left[|P_a^{(n)}(t)-P_a^{(n)}(r)|^2\right]\ar\leq\ar C_0(t-r)\int_r^t\mathbf{E}_{\mathscr{F}_r}\left[|\beta^{(n)}_a(s)|^2+\Big||\delta_x^{(n)}|^2\mu_{aL}^{(n)}(s)\Big|^2\right]ds\cr
 \ar\ar +C_0\mathbf{E}_{\mathscr{F}_r}\left[\Big|\int_r^t\delta_x^{(n)}\tilde{N}_{bM}^{(n)}(ds)\Big|^2\right]+C_0\mathbf{E}_{\mathscr{F}_r}\left[\Big|\int_r^t\delta_x^{(n)}\tilde{N}_{aL}^{(n)}(ds)\Big|^2\right]\cr
 \ar\leq\ar C_0(t-r)\int_r^t\mathbf{E}_{\mathscr{F}_r}\left[|\beta^{(n)}_a(s)|^2+||\delta_x^{(n)}|^2\mu_{aL}^{(n)}(s)|^2\right]ds\cr
 \ar\ar +C_0\int_r^t\mathbf{E}_{\mathscr{F}_r}\left[|\delta_x^{(n)}|^2\mu_{bM}^{(n)}(s)+|\delta_x^{(n)}|^2\mu_{aL}^{(n)}(s)\right]ds\cr
 \ar\ar\cr
 \ar\leq\ar C_0|J^{(n)}(r)|^2(|t-r|^2+|t-r|).
 \eeqnn
 Moreover, by Markov's inequality,
 \beqnn
 \lefteqn{|V_{a}^{(n)}(t,x)-V_{a}^{(n)}(r,x)|^2\leq C_0 \frac{t-r}{\delta_x^{(n)}}\int_r^t\int_{\Delta^{(n)}(x-P_a^{(n)}(s))} |\delta_v^{(n)}\lambda^{(n)}_{aL}(s,y)|^2dy ds}\qquad\qquad\ar\ar\cr
 \ar\ar +C_0\frac{t-r}{\delta_x^{(n)}}\int_r^t \int_{\Delta^{(n)}(x-P_a^{(n)}(s))} |V_{a}^{(n)}(s-,y+P_a^{(n)}(t))|^2 |\delta_v^{(n)}\lambda^{(n)}_{aC}(s,y)|^2dsdy\cr
 \ar\ar +C_0\Big|\int_r^t\int_{\Delta^{(n)}(x-P_a^{(n)}(s-))}\int_{\mathbb{R}_+}\frac{\delta_v^{(n)}}{\delta_x^{(n)}} (e^{z}-1)\tilde{M}_{aL}^{(n)}(ds,dy,dz)\Big|^2\cr
 \ar\ar +C_0\Big|\int_r^t\int_{\Delta^{(n)}(x-P_a^{(n)}(s-))}\int_{\mathbb{R}_+}\frac{\delta_v^{(n)}}{\delta_x^{(n)}} V_{a}^{(n)}(s-,y+P_a^{(n)}(t)) (e^{-z}-1)\tilde{M}_{aC}^{(n)}(ds,dy,dz)\Big|^2.
 \eeqnn
Hence
 \beqnn
 \lefteqn{\mathbf{E}_{\mathscr{F}_r}\left[\|V_{a}^{(n)}(t,\cdot)-V_{a}^{(n)}(r,\cdot)\|_{L^2}^2\right]}\qquad\ar\ar\cr
 \ar\leq\ar C_0 |t-r|\int_r^t\mathbf{E}_{\mathscr{F}_r}\left[ \|\delta_v^{(n)}\lambda^{(n)}_{aL}(s,\cdot)\|_{L^2}^2+ \|V_{a}^{(n)}(s,\cdot+P_a^{(n)}(t)) \delta_v^{(n)}\lambda^{(n)}_{aC}(s,\cdot)\|_{L^2}^2\right]ds\cr
 \ar\ar +C_0\frac{\delta_v^{(n)}}{\delta_x^{(n)}}\int_r^t\mathbf{E}_{\mathscr{F}_r}\left[\|\delta_v^{(n)} \lambda_{aL}^{(n)}(s,y)\|_{L^1}+\||V_{a}^{(n)}(s,\cdot+P_a^{(n)}(t))|^2\delta_v^{(n)}\lambda_{aC}^{(n)}(s,\cdot)\|_{L^1}\right]ds\cr
 \ar\leq\ar C_0 |t-r|\int_r^t\mathbf{E}_{\mathscr{F}_r}\left[ \|\delta_v^{(n)}\lambda^{(n)}_{aL}(s,\cdot)\|_{L^2}^2+ \|\delta_v^{(n)}\lambda^{(n)}_{aC}(s,\cdot)\|_{L^4}^4+ \|V_{a}^{(n)}(s,\cdot) \|_{L^4}^4\right]ds\cr
 \ar\ar +C_0\frac{\delta_v^{(n)}}{\delta_x^{(n)}}\int_r^t\mathbf{E}_{\mathscr{F}_r}\left[\|\delta_v^{(n)} \lambda_{aL}^{(n)}(s,y)\|_{L^1}+\|\delta_v^{(n)}\lambda_{aC}^{(n)}(s,\cdot)\|^2_{L^2}+\|V_{a}^{(n)}(s,\cdot)\|^4_{L^4}\right]ds\cr
 \ar\leq\ar   C_0[\|V^{(n)}_a (r,x)\|_{L^4}^4 + |J^{(n)} (r)|^4](|t-r|^2+|t-r|).
 \eeqnn
Here the last inequality follows from Lemma~\ref{BL01}, \ref{BL04} and \ref{BL05}. We can get the similar results for the other terms.
 In conclusion,
 $$
 \mathbf{E}_{\mathscr{F}_r}\left[\|\mathbf{S}^{(n)}(t)-\mathbf{S}^{(n)}(r)\|^2_{\mathcal{S}^2}\right]\leq C_0\Big[\sum_{I\in\mathcal{I}}\|V^{(n)}_I (r,x)\|_{L^4}^4+ |J^{(n)} (r)|^4\Big](|t-r|^2+|t-r|).
 $$
 The second result can be proved using similar arguments as in the proof of Lemma~\ref{BL05}.
 \qed

 From (\ref{CIntensity02}), there exists a sequence $\{\gamma_n\}_n\geq 1$ vanishing as $n\to \infty$, such that
 \beqnn
 \sup_{t\in[0,T],y\in\mathbb{R}}|\kappa^{(n)}(y,t)-\kappa(y,t)|\leq\gamma_n.
 \eeqnn

  \begin{lemma}\label{BL07}
 There exists a constant $C_0>0$ such that for any $0\leq r\leq t\leq T$
 \beqnn
 \mathbf{E}_{\mathscr{F}_r}\left[\|\mathbf{D}^{(n)}(t)-\mathbf{D}^{(n)}(r)\|^2_{\mathcal{D}_{1,2}^2}\right]\leq C_0\Big[\sum_{I\in\mathcal{I}}\|V^{(n)}_I (r,\cdot)\|_{L^4}^4+ |J^{(n)} (r)|^4\Big][\gamma_n+|t-r| +|t-r|^2].
 \eeqnn
 \end{lemma}
 \proof Here we just deal with $\mathbf{E}_{\mathscr{F}_r}[\|\mathbf{D}^{(n)}(t)-\mathbf{D}^{(n)}(r)\|^2_{\mathcal{D}_{2}^2}]$.  From Condition~\ref{C3.4} ii) and H\"older's inequality,
 \beqnn
 \lefteqn{\Big|\int_0^r |\phi^{(n)}_{IJ,ij}(t-s)-\phi^{(n)}_{IJ,ij}(r-s)||\delta_x^{(n)}|^2N_{ij}^{(n)}(ds)\Big|^2}\ar\ar\cr
 \ar\leq\ar \int_0^r |\phi^{(n)}_{IJ,ij}(t-s)-\phi^{(n)}_{IJ,ij}(r-s)|^2|\delta_x^{(n)}|^2N_{ij}^{(n)}(ds)
 \times\int_0^r |\delta_x^{(n)}|^2N_{ij}^{(n)}(ds)\cr
 \ar\leq\ar C_0[\gamma_n+|t-r|]
 \Big|\int_0^r |\delta_x^{(n)}|^2N_{ij}^{(n)}(ds)\Big|^2
 \eeqnn
 and
 \beqnn
 \lefteqn{\Big|\int_0^r\int_{\mathbb{R}}\int_{\mathbb{R}} \Big|\Phi^{(n)}_{IJ,ik}(y,t-s)-\Phi^{(n)}_{IJ,ik}(y,r-s)\Big|\delta_v^{(n)}M_{ik}^{(n)}(ds,dy,dz)\Big|^2}\ar\ar\cr
 \ar\leq\ar \int_0^r\int_{\mathbb{R}}\int_{\mathbb{R}} \Big| \Phi^{(n)}_{IJ,ik}(y,t-s)-\Phi^{(n)}_{IJ,ik}(y,r-s)\Big|^2\delta_v^{(n)}M_{ik}^{(n)}(ds,dy,dz) \times \int_0^r\int_{\mathbb{R}}\int_{\mathbb{R}} \delta_v^{(n)}M_{ik}^{(n)}(ds,dy,dz)\cr
 \ar\leq\ar  C_0[\gamma_n+\gamma(t-r)]\Big|\int_0^r\int_{\mathbb{R}}\int_{\mathbb{R}} \delta_v^{(n)}M_{ik}^{(n)}(ds,dy,dz)\Big|^2.
 \eeqnn
  From the above inequalities, Condition~\ref{C3.3} ii) and Cauchy-Schwarz inequality,
  \beqnn
 \Big||\delta_x^{(n)}|^2\mu_{IJ}^{(n)}(t)-|\delta_x^{(n)}|^2\mu_{IJ}^{(n)}(r)\Big|^2\ar\leq\ar C_0\|\mathbf{S}^{(n)}(t)-\mathbf{S}^{(n)}(r)\|^2_{\mathcal{S}^2}
  +C_0[\gamma_n+|t-r|] |J^{(n)}(r)|^2\cr
  \ar\ar + C_0[J^{(n)}(t)-J^{(n)}(r)]^2.
 \eeqnn
 Similarly, from Condition~\ref{C3.4} i)
 \beqnn
 \lefteqn{\int_{\mathbb{R}}  \Big|\int_0^r |\psi_{IK,ij}(x,t-s)-\psi_{IK,ij}(x,r-s)||\delta_x^{(n)}|^2N_{ij}^{(n)}(ds) \Big|^2 dx}\ar\ar\cr
 \ar\leq\ar\int_0^r \int_{\mathbb{R}}  \Big||\psi_{IK,ij}(x,t-s)-\psi_{IK,ij}(x,r-s)| \Big|^2 dx |\delta_x^{(n)}|^2N_{ij}^{(n)}(ds)\times \int_0^r |\delta_x^{(n)}|^2N_{ij}^{(n)}(ds)\cr
 \ar\leq\ar |t-r| \Big|\int_0^r |\delta_x^{(n)}|^2N_{ij}^{(n)}(ds)\Big|^2
 \eeqnn
 and
 \beqnn
 \lefteqn{\int_{\mathbb{R}}\Big|\int_0^r\int_{\mathbb{R}}\int_{\mathbb{R}} |\Psi_{IK,ik}(x,y,t-s)-\Psi_{IK,ik}(x,y,r-s)|\delta_v^{(n)}M_{ik}^{(n)}(ds,dy,dz)\Big|^2dx}\ar\ar\cr
 \ar\leq\ar \int_0^r\int_{\mathbb{R}}\int_{\mathbb{R}} \delta_v^{(n)}M_{ik}^{(n)}(ds,dy,dz) \int_{\mathbb{R}}|\Psi_{IK,ik}(x,y,t-s)-\Psi_{IK,ik}(x,y,r-s)|^2 dx\cr
 \ar\ar\qquad\times \int_0^r\int_{\mathbb{R}}\int_{\mathbb{R}}\delta_v^{(n)}M_{ik}^{(n)}(ds,dy,dz)\cr
 \ar\leq\ar |t-r|\Big|\int_0^r\int_{\mathbb{R}}\int_{\mathbb{R}}\delta_v^{(n)}M_{ik}^{(n)}(ds,dy,dz)\Big|^2.
 \eeqnn
 From the above inequalities and Cauchy-Schwarz inequality,
  \beqnn
\|\delta_v^{(n)}\lambda_{IK}^{(n)}(t,\cdot)-\delta_v^{(n)}\lambda_{IK}^{(n)}(r,\cdot)\|_{L^2}^2 \leq
C_0\|\mathbf{S}^{(n)}(t)-\mathbf{S}^{(n)}(r)\|^2_{\mathcal{S}^2}
  +C_0|t-r||J^{(n)}(r)|^2
+ C_0[J^{(n)}(t)-J^{(n)}(r)]^2.
 \eeqnn
 In conclusion, from Proposition~\ref{BP02} and Lemma~\ref{BL06},
 \beqnn
 \mathbf{E}_{\mathscr{F}_r}\left[\|\mathbf{D}^{(n)}(t)-\mathbf{D}^{(n)}(r)\|^2_{\mathcal{D}_2^2}\right]\ar\leq\ar C_0[\gamma_n+|t-r|] |J^{(n)}(r)|^2 + C_0\mathbf{E}_{\mathscr{F}_r}\left[\|\mathbf{S}^{(n)}(t)-\mathbf{S}^{(n)}(r)\|^2_{\mathcal{S}^2}\right]
  \cr
  \ar\ar + C_0\mathbf{E}_{\mathscr{F}_r}\left[|J^{(n)}(t)-J^{(n)}(r)|^2\right]\cr
  \ar\leq\ar C_0[\gamma_n +|t-r|] |J^{(n)}(r)|^2 \cr
  \ar\ar + C_0\Big[\sum_{I\in\mathcal{I}}\|V^{(n)}_I (r,x)\|_{L^4}^4+ |J^{(n)} (r)|^4\Big](|t-r|^2+|t-r|).
  \eeqnn
  \qed

Using Condition~\ref{C3.2} and \ref{C3.4} ii) the following result can be proved similarly to the previous one.

 \begin{lemma}\label{BL08}
 There exists a constant $C_0>0$ such that for any $0\leq r\leq t\leq T$ and $I\in\mathcal{I}$ have
 \beqnn
 \mathbf{E}_{\mathscr{F}_r}\left[|\beta_I^{(n)}(t)-\beta_I^{(n)}(r)|^2\right]\leq C_0\Big[\sum_{I\in\mathcal{I}}\|V^{(n)}_I (r,x)\|_{L^4}^4+ |J^{(n)} (r)|^4\Big][\gamma_n+|t-r|+|t-r|^2].
 \eeqnn
 \end{lemma}

We are ready to prove the tightness of $(\mathbf{S}^{(n)},\mathbf{D}^{(n)},\beta_a^{(n)},\beta_b^{(n)})$ and the continuity
 of the cluster points.

 \begin{proposition}\label{PL01}
 Suppose Condition~\ref{C3.1}-\ref{C3.4} hold. Then the sequence $(\mathbf{S}^{(n)},\mathbf{D}^{(n)},\beta_a^{(n)},\beta_b^{(n)})$ is tight as a sequence in $\mathbb{D}([0,\infty),\mathcal{S}\times\mathcal{D}\times\mathbb{R}^2)$. Moreover, any cluster point $(\mathbf{S}^*,\mathbf{D}^*,\beta_a^*,\beta_b^*)$ is continuous, i.e.
 \beqnn
 \mathbf{P}\left\{(\mathbf{S}^*,\mathbf{D}^*,\beta_a^*,\beta_b^*)\in C([0,\infty),\mathcal{S}\times\mathcal{D}\times\mathbb{R}^2)\right\}=1.
 \eeqnn
 \end{proposition}
 \proof The tightness of $(\mathbf{S}^{(n)},\mathbf{D}^{(n)},\beta_a^{(n)},\beta_b^{(n)})$ follows from Lemmas~\ref{BL01}, \ref{BP03}, \ref{BL06}, \ref{BL07} and \ref{BL08} using Kurtz's criterion.
 It remains to prove the continuity of the cluster points. From the Skorokhod representation theorem \cite[Theorem 2.7]{IW}, we can construct a new sequence of processes defined on a common space and  with the same law as the initial sequence such that they converge almost surely. Continuity of the price processes follows from standard arguments. In order to prove the continuity of $V_a^*$, let $f\in C_b^2(\mathbb{R})$. Then
  \beqnn
  \ar\ar\mathbf{E}[\sum_{s\leq t}\langle V_{a}^{(n)}(s,\cdot)-V_{a}^{(n)}(s-,\cdot),f\rangle^2]\cr
  \ar\leq\ar
  C_0\delta_v^{(n)}\mathbf{E}\Big[\int_0^t\int_{\mathbb{R}} \Big(\frac{\int_{\Delta^{(n)}(y-P_a^{(n)}(s))}f(x)dx}{\delta_x^{(n)}}\Big)^2\delta_v^{(n)} \lambda_{aL}^{(n)}(s,y)dyds\Big]\cr
  \ar\ar +\delta_v^{(n)}\mathbf{E}\Big[\int_0^t\Big(\frac{\int_{\Delta^{(n)}(y-P_a^{(n)}(s))}f(x)dx}{\delta_x^{(n)}}\Big)\int_{\mathbb{R}} V_{a}^{(n)}(s,y+P_a^{(n)}(t))^2\delta_v^{(n)}\lambda_{aC}^{(n)}(s,y)dyds\Big]\cr
  \ar\leq \ar C_0\delta_v^{(n)}\int_0^t\mathbf{E}[\|\delta_v^{(n)} \lambda_{aL}^{(n)}(s,\cdot)\|_{L^1}+\| V_{a}^{(n)}(s,\cdot)\|^4_{L^4}+\|\delta_v^{(n)}\lambda_{aC}^{(n)}(s,\cdot)\|^2_{L^2}]ds\leq C_0\delta_v^{(n)}.
  \eeqnn
   For any $m\geq 1$, let $g_m(x)=x^2\mathbf{1}_{|x|<m}$.
  By the monotone convergence theorem and Fatou's lemma,
  \beqnn
  \mathbf{E}\Big[\sum_{s\leq t}\langle V^*_{a}(s)-V^*_{a}(s-),f\rangle^2\Big]\ar=\ar \mathbf{E}\Big[\sum_{s\leq t}\lim_{m\to\infty}g_m(\langle V^*_{a}(s)-V^*_{a}(s-),f\rangle)\Big]\cr
  \ar=\ar \lim_{m\to\infty}\mathbf{E}\Big[\sum_{s\leq t}g_m(\langle V^*_{a}(s)-V^*_{a}(s-),f\rangle)\Big]\cr
  \ar=\ar \lim_{m\to\infty}\mathbf{E}\Big[\liminf_{n\to\infty}\sum_{s\leq t}g_m(\langle V^{(n)}_{a}(s)-V^{(n)}_{a}(s-),f\rangle)\Big]\cr
  \ar\leq\ar \lim_{m\to\infty}\liminf_{n\to\infty}\mathbf{E}\Big[\sum_{s\leq t}g_m(\langle V^{(n)}_{a}(s)-V^{(n)}_{a}(s-),f\rangle)\Big]\cr
  \ar\leq\ar \lim_{m\to\infty}\liminf_{n\to\infty}\mathbf{E}\Big[\sum_{s\leq t}\langle V^{(n)}_{a}(s)-V^{(n)}_{a}(s-),f\rangle^2\Big]\cr
  \ar\leq\ar \liminf_{n\to\infty} C_0\delta_v^{(n)}= 0.
  \eeqnn 
    Hence the continuity follows from standard arguments.
 Likewise,
  \beqnn
   \lefteqn{\left||\delta_x^{(n)}|^2[\mu_{IJ}^{(n)}(t)-\hat\mu_{IJ}^{(n)}(t,\mathbf{S}^{(n)}(t))]-|\delta_x^{(n)}|^2[\mu_{IJ}^{(n)}(t-)-\hat\mu_{IJ}^{(n)}(t-,\mathbf{S}^{(n)}(t-))]\right|^2}\ar\ar\cr
  \ar=\ar  \sum_{i\in\mathcal{I},j\in\mathcal{J}} \Big||\delta_x^{(n)}|^2\phi^{(n)}_{IJ,ij}(0)\Big|^2N_{ij}^{(n)}(dt) +\sum_{i\in\mathcal{I},k\in\mathcal{K}}\int_{\mathbb{R}}\int_{\mathbb{R}} ||\delta_x^{(n)}|^2\Phi^{(n)}_{IJ,ik}(y,0)|^2M_{ik}^{(n)}(dt,dy,dz)\cr
  \ar\leq\ar \sum_{i\in\mathcal{I},j\in\mathcal{J}} |\delta_x^{(n)}|^4N_{ij}^{(n)}(dt) +\sum_{i\in\mathcal{I},k\in\mathcal{K}}\int_{\mathbb{R}}\int_{\mathbb{R}} |\delta_v^{(n)}|^2M_{ik}^{(n)}(dt,dy,dz).
  \eeqnn
  and as $n\to \infty$
  \beqnn
  \lefteqn{\mathbf{E}\left[\sum_{t\leq T}\left||\delta_x^{(n)}|^2[\mu_{IJ}^{(n)}(t)-\hat\mu_{IJ}^{(n)}(t,\mathbf{S}^{(n)}(t))]-|\delta_x^{(n)}|^2[\mu_{IJ}^{(n)}(t-)-\hat\mu_{IJ}^{(n)}(r,\mathbf{S}^{(n)}(r))]\right|^2\right]}
  \ar\ar\cr
     \ar\leq \ar  C_0\mathbf{E}\left[\sum_{i\in\mathcal{I},j\in\mathcal{J}}\int_0^T |\delta_x^{(n)}|^4N_{ij}^{(n)}(dt) +\sum_{i\in\mathcal{I},k\in\mathcal{K}}\int_0^T \int_{\mathbb{R}}\int_{\mathbb{R}} |\delta_v^{(n)}|^2M_{ik}^{(n)}(dt,dy,dz)\right]\cr
     \ar\leq\ar  C_0 (|\delta_x^{(n)}|^2\vee \delta_v^{(n)})\mathbf{E}\left[J^{(n)}(T) \right]\to 0 .
  \eeqnn
  Hence we educe the continuity of $\mu^*_{IJ}(t)-\hat\mu_{IJ}(t,\mathbf{S}(t))$. The continuity of $\mu^*_{IJ}(t)$ follows  from the continuity of $\mathbf{S}(t)$ and $\hat\mu_I(t,S)$; see Condition~\ref{C3.3} ii).  The continuity of other terms can be proved similarly.
  \qed


  \section{Proof of the characterization result}\label{Sec6}
 \setcounter{equation}{0}
 \medskip

 In this section, we characterize the weak accumulation points of the sequence of LOB models.

 \begin{definition}
 Let $\mathscr{G}_t=\sigma\{X(s):s\in[0,t],X\in C([0,\infty),\mathcal{S}\times\mathcal{D}\times\mathbb{R}^2)\}$ and $\mathscr{G}=\bigcup_{t\geq 0}\mathscr{G}_t$. We say that a mapping $Z: [0,\infty)\times C([0,\infty),\mathcal{S}\times\mathcal{D}\times\mathbb{R}^2)\mapsto \mathcal{D}\times\mathbb{R}^2$ is a {\rm progressively measurable functional}, if for any $t\geq 0 $ fixed, $Z$ restricted to $[0,t]\times C([0,\infty),\mathcal{S}\times\mathcal{D}\times\mathbb{R}^2)$ is $\mathscr{B}([0,t])\times \mathscr{G}_t/\mathscr{B}(\mathcal{D}\times\mathbb{R}^2)$ measurable.
 \end{definition}

 The process $\mathbf{D}(t)$ depends on the whole trajectory $\{\mathbf{S}(s):s\in[0,t]\}$. We rewrite $\mathbf{D}(t)$ as $\mathbf{D}(t,\mathbf{S})$. For any $\mathrm{S}=(p_a,p_b,v_a,v_b)\in C([0,\infty),\mathcal{S})$, define
 $$
 h_I(t,\mathrm{S}):=\rho_I(\mathrm{S}(t))\beta_I(t,\mathrm{S})+\varrho_I(\mathrm{S}(t))\mu_I(t,\mathrm{S}),\quad \sigma_I(t,\mathrm{S}):=  \rho_I(\mathrm{S}(t))\mu_I(t,\mathrm{S}),
 $$
 $$\eta_a(t,\mathrm{S},x):=\alpha_{aL}\lambda_{aL}(t,\mathrm{S},x-p_a(t)) +\alpha_{aC}\lambda_{aC}(t,\mathrm{S},x-p_a(t))v_a(s,x),$$
 $$\eta_b(t,\mathrm{S},x):=\alpha_{bL}\lambda_{bL}(t,\mathrm{S},p_b(t)-x) +\alpha_{bC}\lambda_{bC}(t,\mathrm{S},p_b(t)-x)v_b(t,x),$$
 which are progressively measurable functionals. Then (\ref{LLOB01}) turns to be
 \begin{equation} \label{NLLOB01}
 \begin{split}
 P_a(t) &=   P_a(0)
     +\int_0^th_a(s,\mathbf{S})ds +\int_0^t\sqrt{2\sigma_a(s,\mathbf{S})}dB_a(s),\\
 V_a(t,x) &= V_a(0,x)
     +\int_0^t\eta_a(s,\mathbf{S},x)ds 
 \end{split}
 \end{equation}
with $P_b$ and $V_b$ being represented analogously. For any $f_a(x),f_b(x)\in L^2(\mathbb{R})$, define $S^f:=(p_a,p_b,v_a^f,v_b^f)\in\mathbb{R}^4$, where
 \beqnn
  v^f_I(t):=\int_{\mathbb{R}} v_I(t,x)f_I(x)dx.
 \eeqnn
 For any $t\geq 0$, we introduce a second-order differential operator: for any $G\in C^2(\mathbb{R}^4)$ and $\mathrm{S}\in C([0,\infty),\mathcal{S})$
 \beqnn
 \mathscr{A}^f_tG(S)=\sum_{I\in\mathcal{I}}h_I(t,S)\frac{\partial G(S^f(t))}{\partial p_I}+\sum_{I\in\mathcal{I}}\sigma_I(t,S)\frac{\partial^2 G(S^f(t))}{\partial p_I^2}+\sum_{I\in\mathcal{I}}\eta^f_I(t,S)\frac{\partial G(S^f(t))}{\partial v^f_I},
 \eeqnn
 where $$\eta^f_I(t,S)=\int_\mathbb{R}\eta_I(t,S,x)f_I(x)dx.$$
 \begin{definition} A probability measure $\mathbf{P}$ on $C([0,\infty),\mathcal{S})$ is called a {\rm solution} to the martingale problem associated to $\mathscr{A}^f_t$, if for any $f_a(x),f_b(x)\in L^2(\mathbb{R})$ and $G\in C^2(\mathbb{R}^4)$, under $\mathbf{P}$
 \beqnn
 \mathcal{M}^f_G(t):=G(S^f(t))-\int_0^t\mathscr{A}^f_sG(S)ds; \quad t\geq 0,
 \eeqnn
 is a continuous, local martingale. If $\mathbf{P}$ is induced by a $C([0,\infty),\mathcal{S})$-valued random variable $\mathbf{S}$, we also say  $\mathbf{S}$ is a {\rm solution} to the martingale problem associated to $\mathscr{A}^f_t$.
 \end{definition}

 \begin{theorem}\label{ThmMart}
 For some two-dimensional Brownian motion $(B_a,B_b)$, if $(\mathbf{S},B_a,B_b)$ is a weak solution to the equation (\ref{NLLOB01}), then $\mathbf{S}$ is a solution to the martingale problem associated to $\mathscr{A}^f_t$. Conversely, if $\mathbf{S}$ is a solution to the martingale problem associated to $\mathscr{A}^f_t$, then there exists a two-dimensional Brownian motion $(B_a,B_b)$ defined on an enlarged probability space, such that $(\mathbf{S},B_a,B_b)$ is a weak solution to (\ref{NLLOB01}).
 \end{theorem}
 \proof The result can be considered as an extension of Problem~4.3 and Proposition~4.6 in \cite[p.313-315]{Karatzas2012}. We just give a brief outline.
 If $\mathbf{S}$ is a solution to the stochastic dynamic system, by It\^o's formula it is easy to identify that $\mathcal{M}^f_G(t)$ is a continuous, local martingale.
 Conversely, suppose that $\mathcal{M}^f_G(t)$ is a continuous, local martingale for every $\{f_i(x)\in L^2(\mathbb{R}):i\in\mathcal{I}\}$ and $G(x)\in C^2(\mathbb{R}^4)$. By the standard stopping time argument, we have
 \beqnn
 \mathbf{S}(t)\ar=\ar \mathbf{S}(0)+\mathbf{A}(t)+\mathbf{M}(t),
 \eeqnn
 where $\mathbf{A}(t):=(A_{P_a},A_{P_b},A_{V_a},A_{V_b})$ is a  predictable, $\mathcal{S}$-valued process with locally bounded variations and $\mathbf{M}(t):=(M_{P_a},M_{P_b},M_{V_a},M_{V_b})$ is a continuous, $\mathcal{S}$-valued, local martingale. Moreover, for any $\{f_i(x)\in L^2(\mathbb{R}):i\in\mathcal{I}\}$, like the definition of $\mathbf{S}^f$ let
 $$\mathbf{A}^f:=(A_{P_a},A_{P_b},A^{f}_{V_a},A^{f}_{V_b})\quad \mbox{and}\quad \mathbf{M}^f:=(M_{P_a},M_{P_b},M^{f}_{V_a},M^{f}_{V_b}).$$
  By It\^o's formula, for any $G(x)\in C^2(\mathbb{R}^4)$
 \beqnn
 G(\mathbf{S}^f(t))\ar=\ar G(\mathbf{S}^f(0))+\int_0^t \frac{\partial G(\mathbf{S}^f(s))}{\partial S^f}d\mathbf{A}^f(s)+\frac{1}{2}\int_0^t \frac{\partial^2 G(\mathbf{S}^f(s))}{\partial |S^f|^2}d\langle \mathbf{M}^f,\mathbf{M}^f \rangle_s+\mbox{Local Mart.}\cr
  \ar=\ar G(\mathbf{S}^f(0))+\sum_{I\in\mathcal{I}}\int_0^t \frac{\partial G(\mathbf{S}^f(s))}{\partial P_I}dA_{P_I}(s) + \sum_{I\in\mathcal{I}}\int_0^t\frac{\partial G(\mathbf{S}^f(s))}{\partial V^{f_I}_I}dA^{f_I}_{V_I}(s)\cr
 \ar\ar+\sum_{I,I'\in\mathcal{I}}\frac{1}{2}\int_0^t   \frac{\partial^2 G(\mathbf{S}^f(s))}{\partial P_I \partial P_{I'}}d\langle M_{P_I},M_{P_{I'}} \rangle_s +\sum_{I,I'\in\mathcal{I}}\frac{1}{2}\int_0^t \frac{\partial^2 G(\mathbf{S}^f(s))}{\partial V_I^{f}\partial V_{I'}^{f}}d\langle M^{f}_{V_I},M_{V_{I'}}^{f} \rangle_s\cr
 \ar\ar +\sum_{I,I'\in\mathcal{I}}\frac{1}{2}\int_0^t \frac{\partial^2 G(\mathbf{X}^f(s))}{\partial P_I  \partial V_{I'}^{f}}d\langle M_{P_I},M_{V_{I'}}^{f} \rangle_s +\mbox{Local Mart.}
 \eeqnn
 By the uniqueness of canonical decompositions of special semi-martingales \cite[p.213]{DM}, for any $I,I'\in\mathcal{I}$ and $I\neq I'$
 \beqnn
 A_{P_a}(t)=\int_0^t \theta_a(s,\mathbf{S})ds,\quad
 A^{f}_{V_I}(t)= \int_0^t \eta_I^f(s,\mathbf{S}) ds
 \eeqnn
 and
 \beqnn
 \langle M_{P_I},M_{P_{I}} \rangle_t= 2\rho_I(\mathbf{S}(t))\sigma_I^2(t,\mathbf{S}),\quad
 \langle M_{P_a},M_{P_{b}} \rangle_t=  \langle M^{f}_{V_I},M_{V_{I'}}^{f}  \rangle_t = \langle M_{P_I},M_{V_{I'}}^{f} \rangle_t\equiv0.
 \eeqnn
 By the representation theorem for semi-martingales, the existence of weak solutions to
 (\ref{NLLOB01}) follows from, e.g. \cite[p.90]{IW}.
 \qed

We are now ready to prove the main result of this paper.

\medskip

 \textit{Proof for Theorem~\ref{T1}:} By the Skorokhod representation theorem, we may without loss of generality assume that
 $(\mathbf{S}^{(n)},\mathbf{D}^{(n)},\beta_a^{(n)},\beta_b^{(n)})\rightarrow(\mathbf{S},\mathbf{D},\beta_a,\beta_b)$ almost surely. We now proceed in five steps. (i) we prove that $\mu_{bM}(t)=\mu_{aL}(t)$ and that $\mu_{bL}(t)=\mu_{aM}(t)$; (ii) we show that $\mu_I(t)$ satisfies equations (\ref{Lintensity01}); (iii) we show that $\lambda_{IK}(t,x)$ can be given by (\ref{Lintensity02}); (iv) we show that $\beta_I(t)$ can be described by (\ref{Lintensity03}); (v) we identify that $\mathbf{S}(t)$ is a solution to the martingale problem associated with $\mathscr{A}^f_t$.

\medskip

 \noindent\textit{Step 1.}
  Since $$|\mu_{bM}(t)-\mu_{aL}(t)|=\lim_{n\to\infty}\Big||\delta_x^{(n)}|^2\mu_{bM}^{(n)}(t)-|\delta_x^{(n)}|^2\mu_{aL}^{(n)}(t)\Big|=\lim_{n\to\infty}|\delta_x^{(n)}\beta_a^{(n)}(t)|= 0$$
 and
 $$|\mu_{bL}(t)-\mu_{aM}(t)|=\lim_{n\to\infty}\Big||\delta_x^{(n)}|^2\mu_{bL}^{(n)}(t)-|\delta_x^{(n)}|^2\mu_{aM}^{(n)}(t)\Big|=\lim_{n\to\infty}|\delta_x^{(n)}\beta_b^{(n)}(t)|= 0,$$
  we have $\mu_{bM}(t)=\mu_{aL}(t):=\mu_a(t)$ and $\mu_{bL}(t)=\mu_{aM}(t):=\mu_b(t)$.

\medskip

  \noindent\textit{Step 2. Convergence of $\left\{ |\delta_x^{(n)}|^2 \mu^{(n)}_{IJ} \right \}_{n \geq 1}$.}
  For any $I\in\mathcal{I},J\in\mathcal{J}$, define
 \beqlb\label{eqn7.2}
 \mathcal{M}^{(n)}_{{IJ}}(t)
     \ar:=\ar \sum_{i\in\mathcal{I},j\in\mathcal{J}}\int_0^t \phi^{(n)}_{IJ,ij}(t-s)|\delta_x^{(n)}|^2\tilde{N}_{ij}^{(n)}(ds) \cr
     \ar\ar +\sum_{i\in\mathcal{I},k\in\mathcal{K}}\int_0^t\int_{\mathbb{R}}\int_{\mathbb{R}} \Phi^{(n)}_{IJ,ik}(y,t-s)\delta_v^{(n)}\tilde{M}_{ik}^{(n)}(ds,dy,dz)\cr
     \ar=\ar |\delta_x^{(n)}|^2\mu^{(n)}_{IJ}(t)- \hat{\mu}_{IJ}^{(n)}(t,\mathbf{S}^{(n)}(t-))
     -\sum_{i\in\mathcal{I},j\in\mathcal{J}}\int_0^t \phi^{(n)}_{IJ,ij}(t-s)|\delta_x^{(n)}|^2\mu^{(n)}_{ij}(s)ds \cr
     \ar\ar -\sum_{i\in\mathcal{I},k\in\mathcal{K}}\int_0^t\int_{\mathbb{R}} \Phi^{(n)}_{IJ,ik}(y,t-s)\delta_v^{(n)}\lambda^{(n)}_{ik}(s,y)dyds.
 \eeqlb
 We are going to show the almost sure convergence of  $\{\mathcal{M}^{(n)}_{{IJ}}(t)\}_{n\geq 1}$. To this end, for any $K>0$ we put $$\tau_K^{(n)}:=\inf\{t\in[0,T]:\|\mathbf{D}^{(n)}(t)\|_{\mathcal{D}_{1,2}^2}\geq K\}.$$
 It is easy to see that as $n\to\infty$
 $$\tau_K^{(n)}\to\tau_K:=\inf\{t\in[0,T]:\|\mathbf{D}(t)\|_{\mathcal{D}_{1,2}^2}\geq K\},\quad a.s.$$
  It is enough to prove
  the almost sure convergence of $\{\mathcal{M}^{(n)}_{{IJ}}(t\wedge\tau^{(n)}_K)\}_{n\geq 1}$. We analyse the different terms separately.

  \begin{itemize}
	\item  Since $\mathbf{D}^{(n)} \to \mathbf{D}$ a.s. and $\mathbf{D}$ is continuous,
\beqnn
\sup_{t\in[0,T]}\|\mathbf{D}^{(n)}(t)-\mathbf{D}(t)\|_{\mathcal{D}_{1,2}^2}\to 0
\eeqnn
(see \cite{Billingsley}, p.124) and
the first term on the right side of the last equality in (\ref{eqn7.2}) converges almost surely.  Convergence of the second term follows from (\ref{eqn2.8}): as $n\rightarrow \infty$,
 \beqnn
 \lefteqn{|\hat{\mu}_{IJ}^{(n)}(t\wedge\tau^{(n)}_K,\mathbf{S}^{(n)}((t\wedge\tau^{(n)}_K)-))-\hat{\mu}_{IJ}(t\wedge\tau_K,\mathbf{S}(t\wedge\tau_K))|}\ar\ar\cr
 \ar\ar\leq C_0(|\tau^{(n)}_K-\tau_K|+\|\mathbf{S}^{(n)}((t\wedge\tau^{(n)}_K)-)-\mathbf{S}(t\wedge\tau_K)\|_{\mathcal{S}^2})\to 0,\quad a.s.
 \eeqnn

 \item For the third term, we have  
 \beqnn
 \lefteqn{\Big|\int_0^{t\wedge\tau^{(n)}_K} \phi^{(n)}_{IJ,ij}(t-s)|\delta_x^{(n)}|^2\mu^{(n)}_{ij}(s)ds-\int_0^{t\wedge\tau_K} \phi_{IJ,ij}(t-s)\mu_{ij}(s)ds\Big|}\ar\ar\cr
 \ar\leq\ar \int_0^{t\wedge T} \Big|\phi^{(n)}_{IJ,ij}(t-s)|\delta_x^{(n)}|^2\mu^{(n)}_{ij}(s)-\phi_{IJ,ij}(t-s)\mu_{ij}(s)\Big|ds\cr
 \ar\ar + \Big|\int_{t\wedge\tau_K}^{t\wedge\tau^{(n)}_K} \phi^{(n)}_{IJ,ij}(t-s)|\delta_x^{(n)}|\mu^{(n)}_{ij}(s)ds\Big|\cr
 \ar\leq\ar \int_0^{t\wedge T} \Big|\phi^{(n)}_{IJ,ij}(t-s)|\delta_x^{(n)}|^2\mu^{(n)}_{ij}(s)-\phi_{IJ,ij}(t-s)\mu_{ij}(s)\Big|ds + C_0K|\tau_K-\tau^{(n)}_K|.
 \eeqnn
 The second term above tends to $0$ as $n\to \infty$. For the first term, by Condition~\ref{C3.4} and the dominated convergence theorem,
 \beqnn
 \lefteqn{\int_0^{t\wedge\tau_K}  \Big|\phi^{(n)}_{IJ,ij}(t-s)|\delta_x^{(n)}|^2\mu^{(n)}_{ij}(s)-\phi_{IJ,ij}(t-s)\mu_{ij}(s)\Big|ds}\ar\ar\cr
 \ar\leq\ar \int_0^{t\wedge\tau_K}  \phi^{(n)}_{IJ,ij}(t-s)\Big||\delta_x^{(n)}|^2\mu^{(n)}_{ij}(s)-\mu_{ij}(s)\Big|ds\cr
 \ar\ar +\int_0^{t\wedge\tau_K} \Big|\phi^{(n)}_{IJ,ij}(t-s)-\phi_{IJ,ij}(t-s)\Big|\mu_{ij}(s) ds\cr
 \ar\leq\ar C_0\int_0^{t\wedge\tau_K}  ||\delta_x^{(n)}|^2\mu^{(n)}_{ij}(s)-\mu_{ij}(s)|ds\cr
 \ar\ar +K\int_0^{t}|\phi^{(n)}_{IJ,ij}(t-s)-\phi_{IJ,ij}(t-s)|ds\to 0,\quad a.s.
 \eeqnn
 	\item For the fourth term, we also have
 \beqnn
 \lefteqn{\Big|\int_0^{t\wedge\tau^{(n)}_K}\int_{\mathbb{R}} \Phi^{(n)}_{IJ,ik}(y,t-s)\delta_v^{(n)}\lambda^{(n)}_{ik}(s,y)dyds-\int_0^{t\wedge\tau_K}\int_{\mathbb{R}} \Phi_{IJ,ik}(y,t-s)\lambda_{ik}(s,y)dyds\Big|}\ar\ar\cr
 \ar\leq\ar \Big|\int_{t\wedge\tau_K}^{t\wedge\tau^{(n)}_K}\int_{\mathbb{R}} \Phi^{(n)}_{IJ,ik}(y,t-s)\delta_v^{(n)}\lambda^{(n)}_{ik}(s,y)dyds\Big|\cr
 \ar\ar+ \Big|\int_0^{t\wedge T}\int_{\mathbb{R}} \Big[ \Phi^{(n)}_{IJ,ik}(y,t-s)\delta_v^{(n)}\lambda^{(n)}_{ik}(s,y)- \Phi_{IJ,ik}(y,t-s)\lambda_{ik}(s,y)\Big]dyds\Big|\cr
 \ar\leq\ar \int_0^{t\wedge T}\int_{\mathbb{R}} \Big| \Phi^{(n)}_{IJ,ik}(y,t-s)\delta_v^{(n)}\lambda^{(n)}_{ik}(s,y)- \Phi_{IJ,ik}(y,t-s)\lambda_{ik}(s,y)\Big|dyds +C_0K|\tau^{(n)}_K -\tau_K|.
 \eeqnn
The second term vanishes as $n\to \infty$. For the first term, by (\ref{CIntensity02}) and Lemma~\ref{BL01}
 \beqnn
 \lefteqn{\int_0^{t\wedge T}\int_{\mathbb{R}} \Big| \Phi^{(n)}_{IJ,ik}(y,t-s)\delta_v^{(n)}\lambda^{(n)}_{ik}(s,y)- \Phi_{IJ,ik}(y,t-s)\lambda_{ik}(s,y)\Big|dyds}\ar\ar\cr
 \ar\leq\ar  \sup_{y\in\mathbb{R},s\in[0,T]}\Big| \Phi^{(n)}_{IJ,ik}(y,s)- \Phi_{IJ,ik}(y,s)\Big|\int_0^{t\wedge T}\| \delta_v^{(n)}\lambda^{(n)}_{ik}(s,\cdot)\|_{L^1}ds\cr
 \ar\ar+ C_0\int_0^{t\wedge T} \|\delta_v^{(n)}\lambda^{(n)}_{ik}(s,\cdot)- \lambda_{ik}(s,\cdot)\|_{L^1}ds\cr
 \ar\leq\ar  K T\sup_{x\in\mathbb{R},s\in[0,T]}\Big| \Phi^{(n)}_{IJ,ik}(x,s)- \Phi_{IJ,ik}(x,s)\Big|+ C_0 \|\delta_v^{(n)}\lambda^{(n)}_{ik}(s,\cdot)- \lambda_{ik}(s,\cdot)\|_{L^1}\to 0,\quad a.s.
 \eeqnn
\end{itemize}
Since $K >0$ is arbitrary the limit $\mathcal{M}_{{IJ}}(t) := \lim_{n \to \infty} \mathcal{M}^{(n)}_{{IJ}}(t)$ exists and equals
 \beqnn
 \mathcal{M}_{{IJ}}(t)\ar =\ar \mu_{IJ}(t)-\hat\mu_{IJ}(t,\mathbf{S}(t))
     -\sum_{i\in\mathcal{I},j\in\mathcal{J}}\int_0^t \phi_{IJ,ij}(t-s)\mu_{ij}(s)ds \cr
     \ar\ar -\sum_{i\in\mathcal{I},k\in\mathcal{K}}\int_0^t\int_{\mathbb{R}} \Phi_{IJ,ik}(y,t-s)\lambda_{ik}(s,y)dyds, \quad \mbox{a.s.}
 \eeqnn
We are now going to prove that $ \mathcal{M}_{{IJ}}(t) = 0$ for all $t\geq 0$. By Cauchy-Schwarz inequality and  Condition~\ref{C3.4} ii), for any $t\in[0,T]$,
 \beqnn
 |\mathcal{M}^{(n)}_{{IJ}}(t)|^2\ar\leq\ar C_0+C_0\sup_{t\in[0,T]}||\delta_x^{(n)}|^2\mu^{(n)}_{IJ}(t) |^2 +C_0T\sum_{i\in\mathcal{I},j\in\mathcal{J}}\int_0^T||\delta_x^{(n)}|^2\mu^{(n)}_{ij}(s)|^2 ds \cr
 \ar\ar +C_0\sum_{i\in\mathcal{I},k\in\mathcal{K}}\int_0^T\Big|\int_{\mathbb{R}} \delta_v^{(n)}\lambda^{(n)}_{ik}(s,y)dy\Big|^2 ds\leq C_0\Big[1+\sup_{t\in[0,T]}\|\mathbf{D}^{(n)}(t)\|^2_{\mathcal{D}_1^2}\Big].
 \eeqnn
 Hence, it follows from Lemma~\ref{BL01} that $\sup_{n\geq 1}\mathbf{E}[|\mathcal{M}^{(n)}_{{IJ}}(t)|^2]<\infty$. In particular,
 \[
 	\{|\mathcal{M}^{(n)}_{{IJ}}(t)|^2\}_{n\geq 1} \quad \mbox{is uniformly integrable.}
\]
Thus, almost sure convergence implies convergence of $\{\mathcal{M}^{(n)}_{{IJ}}(t)\}_{n\geq 1}$ to $\mathcal{M}_{{IJ}}(t)$ in $L^1(\mathbf{P})$.
Hence, by Fatou's lemma, for any $K>0$ large enough,
 \beqlb\label{eqn7.3}
 \begin{split}
 \mathbf{E}&\left[|\mathcal{M}_{{IJ}}(t)|^2  \mathbf{1}_{\{|\mathcal{M}_{{IJ}}(t)|^2\leq K\}}\right]= \mathbf{E}\left[\lim_{n\to \infty} |\mathcal{M}^{(n)}_{{IJ}}(t)|^2\mathbf{1}_{\{|\mathcal{M}^{(n)}_{{IJ}}(t)|^2\leq K\}}\right] \leq \liminf_{n\to \infty}\mathbf{E}\left[ |\mathcal{M}^{(n)}_{{IJ}}(t)|^2\right]\cr
\leq& \liminf_{n\to \infty}  C_0\sum_{i\in\mathcal{I},j\in\mathcal{J}}\mathbf{E}\left[\Big|\int_0^t \phi^{(n)}_{IJ,ij}(t-s)|\delta_x^{(n)}|^2\tilde{N}_{ij}^{(n)}(ds)\Big|^2\right] \cr
 & +\liminf_{n\to \infty} C_0\sum_{i\in\mathcal{I},k\in\mathcal{K}}\mathbf{E}\left[\Big|\int_0^t\int_{\mathbb{R}}\int_{\mathbb{R}} \Phi^{(n)}_{IJ,ik}(y,t-s)\delta_v^{(n)}\tilde{M}_{ik}^{(n)}(ds,dy,dz)\Big|^2\right]\cr
=&\liminf_{n\to \infty}C_0 \sum_{i\in\mathcal{I},j\in\mathcal{J}}\int_0^t \mathbf{E}\left[ |\phi^{(n)}_{IJ,ij}(t-s)|^2|\delta_x^{(n)}|^4\mu_{ij}^{(n)}(s)\right]ds \cr
 & +\liminf_{n\to \infty} C_0\sum_{i\in\mathcal{I},k\in\mathcal{K}}\int_0^t\mathbf{E}\left[\int_{\mathbb{R}} \Big| \Phi^{(n)}_{IJ,ik}(y,t-s)\Big|^2|\delta_v^{(n)}|^2\lambda_{ik}^{(n)}(s,y)dyds\right]\cr
\leq& \lim_{n\to \infty} C_0|\delta_x^{(n)}|^2\sum_{i\in\mathcal{I},j\in\mathcal{J}}\int_0^t \mathbf{E}\left[ |\delta_x^{(n)}|^2\mu_{ij}^{(n)}(s)\right]ds \cr
 & +\lim_{n\to \infty} C_0\delta_v^{(n)}\sum_{i\in\mathcal{I},k\in\mathcal{K}}\int_0^t\mathbf{E}\left[ \|\delta_v^{(n)}\lambda_{ik}^{(n)}(s,\cdot)\|_{L^1}\right]ds=0.
 \end{split}
 \eeqlb
 The continuity of $\mathcal{M}_{{IJ}}(t)$ now implies
 $\mathbf{P}\{\mathcal{M}_{{IJ}}(t)=0;t\in[0,T]\}=1,$
 and hence
 \beqnn
 \mu_{IJ}(t)\ar=\ar\hat\mu_{IJ}(t,\mathbf{S}(t))
     +\sum_{i\in\mathcal{I},j\in\mathcal{J}}\int_0^t \phi_{IJ,ij}(t-s)\mu_{ij}(s)ds \cr
     \ar\ar +\sum_{i\in\mathcal{I},k\in\mathcal{K}}\int_0^t\int_{\mathbb{R}} \Phi_{IJ,ik}(y,t-s)\lambda_{ik}(s,y)dyds\qquad a.s.
 \eeqnn

 \noindent\textit{Step 3. Convergence of $\left\{ \delta_v^{(n)} \lambda^{(n)}_{IK} \right \}_{n \geq 1}$.} For any $f\in L^2(\mathbb{R})\cup L^\infty(\mathbb{R})$ and $I\in\mathcal{I},K\in\mathcal{K}$, let
 \beqnn
  \delta_v^{(n)}\lambda^{(n)f}_{IK}(t):=\int_\mathbb{R}\delta_v^{(n)}\lambda^{(n)}_{IK}(t,x)f(x)dx,\quad \lambda_{IK}^f(t):=\int_\mathbb{R}\lambda_{IK}(t,x)f(x)dx.
 \eeqnn
 It suffices to prove the almost convergence of $\left\{ \delta_v^{(n)} \lambda^{(n)f}_{IK} \right \}_{n \geq 1}$. Here we just consider the case with $f\in L^2(\mathbb{R})$. The case $f\in L^\infty(\mathbb{R})$ can be proved similarly.
 Let
 \beqnn
 \mathcal{M}^{(n)f}_{IK}(t)
  \ar:=\ar \sum_{i\in\mathcal{I},j\in\mathcal{J}}\int_0^t \psi_{IK,ij}(x,t-s)
     |\delta_x^{(n)}|^2\tilde{N}_{ij}^{(n)}(ds) \cr
     \ar\ar +\sum_{i\in\mathcal{I},k\in\mathcal{K}}\int_0^t\int_{\mathbb{R}}\int_{\mathbb{R}} \Psi_{IK,ik}(x,y,t-s)\delta_v^{(n)}\tilde{M}_{ik}^{(n)}(ds,dy,dz)\cr
 \ar=\ar \delta_v^{(n)}\lambda^{(n)f}_{IK}(t)- \int_\mathbb{R} \hat{\lambda}_{IK} (t,\mathbf{S}^{(n)}(t-),x)f(x)dx\cr
 \ar\ar
     -\sum_{i\in\mathcal{I},j\in\mathcal{J}}\int_0^t \int_{\mathbf{R}} \psi_{IK,ij}(x,t-s)f(x)dx|\delta_x^{(n)}|^2\mu_{ij}^{(n)}(s)dx \cr
     \ar\ar -\sum_{i\in\mathcal{I},k\in\mathcal{K}}\int_0^t\int_{\mathbb{R}}\int_{\mathbb{R}} \Psi_{IK,ik}(x,y,t-s)f(x)dx\delta_v^{(n)}\lambda_{ik}^{(n)}(s,y)dyds.
 \eeqnn
 Using the same arguments as in the previous step,
 \beqnn
  \mathcal{M}^{(n)f}_{IK}(t)\rightarrow  \mathcal{M}^{f}_{IK}(t)\ar=\ar \lambda^{f}_{IK}(t)- \int_\mathbb{R}\hat{\lambda}_{IK}(t,\mathbf{S}(t),x)f(x)dx\cr
 \ar\ar
     -\sum_{i\in\mathcal{I},j\in\mathcal{J}}\int_0^t \int_{\mathbf{R}}\psi_{IK,ij}(x,t-s)f(x)dx\mu_{ij}(s)ds\cr
     \ar\ar -\sum_{i\in\mathcal{I},k\in\mathcal{K}}\int_0^t\int_{\mathbb{R}}\int_{\mathbb{R}} \Psi_{IK,ik}(x,y,t-s)f(x)dx\lambda_{ik}(s,y)dyds \quad \mbox{a.s.}
 \eeqnn
 as well as
 \beqnn
  |\mathcal{M}^{(n)f}_{IK}(t)|^2\ar\leq\ar C_0|\delta_v^{(n)}\lambda^{(n)f}_{IK}(t)|^2+ C_0\Big|\int_\mathbb{R} \hat{\lambda}_{IK}(t,\mathbf{S}^{(n)}(t-),x)f(x)dx\Big|^2\cr
 \ar\ar
     +C_0\sum_{i\in\mathcal{I},j\in\mathcal{J}}\Big|\int_0^t \int_{\mathbf{R}} \psi_{IK,ij}(x,t-s)f(x)dx|\delta_x^{(n)}|^2\mu_{ij}^{(n)}(s)ds \Big|^2 \cr
     \ar\ar +C_0\sum_{i\in\mathcal{I},k\in\mathcal{K}}\Big|\int_0^t\int_{\mathbb{R}}\int_{\mathbb{R}} \Psi_{IK,ik}(x,y,t-s)f(x)dx\delta_v^{(n)}\lambda_{ik}^{(n)}(s,y)dyds\Big|^2\cr
     \ar\leq\ar C_0\|f\|^2_{L^2}\|\delta_v^{(n)}\lambda^{(n)}_{IK}(t,\cdot)\|^2_{L^2}+ C_0\|f\|^2_{L^2}\|\hat{\lambda}_{IK}(t,\mathbf{S}^{(n)}(t-),\cdot)\|^2_{L^2}\cr
     \ar\ar
     +C_0\|f\|^2_{L^2}\sum_{i\in\mathcal{I},j\in\mathcal{J}}\int_0^t \Big\|\psi_{IK,ij}(\cdot,t-s)\Big\|^2_{L^2}\Big||\delta_x^{(n)}|^2\mu_{ij}^{(n)}(s)\Big|^2 ds \cr
     \ar\ar +C_0T\|f\|^2_{L^2}\sum_{i\in\mathcal{I},k\in\mathcal{K}}\Big|\int_0^t\int_{\mathbb{R}}\| \Psi_{IK,ik}(\cdot, y,t-s)\|_{L^2}\delta_v^{(n)}\lambda_{ik}^{(n)}(s,y)dy ds\Big|^2\cr
     \ar\leq\ar C_0\|f\|^2_{L^2}\|\delta_v^{(n)}\lambda^{(n)}_{IK}(t,\cdot)\|^2_{L^2}+ C_0\|f\|^2_{L^2}
     +C_0\|f\|^2_{L^2}\sum_{i\in\mathcal{I},j\in\mathcal{J}}\int_0^t \Big||\delta_x^{(n)}|^2\mu_{ij}^{(n)}(s)\Big|^2 ds \cr
     \ar\ar +C_0\|f\|^2_{L^2}\sum_{i\in\mathcal{I},k\in\mathcal{K}}\int_0^t\| \delta_v^{(n)}\lambda_{ik}^{(n)}(s,\cdot)\|^2_{L^1} ds\leq C_0\Big[1+\sup_{t\in[0,T]}\|\mathbf{D}^{(n)}(t)\|_{\mathcal{D}_{1,2}^2}^2\Big].
 \eeqnn
 As in (\ref{eqn7.3}), it follows from Lemma~\ref{BL01}, \ref{BL04} and Fatou's lemma that
 $$
  \mathbf{P}\left\{\mathcal{M}^{f}_{IK}(t)=0,t\in[0,T]\right\}=1.$$

\noindent\textit{Step 4. Convergence of $\left\{ \beta^{(n)}_I\right \}_{n \geq 1}$.}
 As in Step~2, we can define
 \beqnn
 \mathcal{M}^{(n)}_a(t)\ar:=\ar \sum_{i\in\mathcal{I},j\in\mathcal{J}}\int_0^t \theta^{(n)}_{a,ij}(t-s)|\delta_x^{(n)}|^2\tilde{N}_{ij}^{(n)}(ds) \cr
     \ar\ar +\sum_{i\in\mathcal{I},k\in\mathcal{K}}\int_0^t\int_{\mathbb{R}}\int_{\mathbb{R}} \Theta^{(n)}_{a,ik}(y,t-s)\delta_v^{(n)}\tilde{M}_{ik}^{(n)}(ds,dy,dz)
     \cr
     \ar=\ar \beta^{(n)}_a(t)-\hat{\beta}_{a}^{(n)}(t,\mathbf{S}^{(n)}(t-))
     -\sum_{i\in\mathcal{I},j\in\mathcal{J}}\int_0^t \theta^{(n)}_{a,ij}(t-s)|\delta_x^{(n)}|^2\mu^{(n)}_{ij}(s)ds \cr
     \ar\ar -\sum_{i\in\mathcal{I},k\in\mathcal{K}}\int_0^t\int_{\mathbb{R}} \Theta^{(n)}_{a,ik}(y,t-s)\delta_v^{(n)}\lambda^{(n)}_{ik}(s,y)dyds
 \eeqnn
and prove the convergence of $\{\mathcal{M}^{(n)}_a(t)\}_{n\geq 1}$ to
 \beqnn
 \mathcal{M}_a(t)\ar=\ar \beta_a(t)-\hat{\beta}_{a}(t,\mathbf{S}(t))
     -\sum_{i\in\mathcal{I},j\in\mathcal{J}}\int_0^t \theta_{a,ij}(t-s)\mu_{ij}(s)ds \cr
     \ar\ar -\sum_{i\in\mathcal{I},k\in\mathcal{K}}\int_0^t\int_{\mathbb{R}} \Theta_{a,ik}(y,t-s)\lambda_{ik}(s,y)dyds
 \eeqnn
 and
 \beqnn
 \mathbf{E}[|\mathcal{M}_a(t)|^2\mathbf{1}_{\{|\mathcal{M}_a(t)|^2\leq K\}}]\leq \lim_{n\to\infty}  \mathbf{E}[|\mathcal{M}^{(n)}_a(t)|^2]=0,
 \eeqnn
for any $K>0$. From this, we can again deduce that $\mathbf{P}\{\mathcal{M}_a(t)=0,t\in[0,T]\}=1$ and hence that (\ref{Lintensity03}) holds.
 \medskip

 \noindent\textit{Step 5. $\mathbf{S}$ is a weak solution to (\ref{LLOB01}).}
In view of Theorem~\ref{ThmMart} it is enough to prove that $\mathbf{S}$ solves the martingale problem associated with $\mathscr{A}^f_t$. Since $C(\mathbb{R})$ is dense in $L^2(\mathbb{R})$, we may chose $C(\mathbb{R})$ as our set of test functions. For $f_a,f_b\in C(\mathbb{R})$, let $\mathbf{S}^{(n)f}:=(P^{(n)}_a,P^{(n)}_b,V_a^{(n)f},V_b^{(n)f})$, where
 \beqnn
 V_a^{(n)f}(t)\ar=\ar\int_\mathbb{R}V^{(n)}_a (0,x) f_a(x)dx +\int_0^t\int_{\mathbb{R}}\int_{\mathbb{R}_+}J^{(n)}_{aL}(s,y+P_a^{(n)}(s-),z)\delta_v^{(n)}M_{aL}^{(n)}(ds,dy,dz)\cr
   \ar\ar +\int_0^t\int_{\mathbb{R}}\int_{\mathbb{R}_+}J^{(n)}_{aC}(s,y+P_a^{(n)}(s-),z)\delta_v^{(n)} M_{aC}^{(n)}(ds,dy,dz),\cr
 V_b^{(n)f}(t)\ar=\ar\int_\mathbb{R}V^{(n)}_b (0,x) f_b(x)dx -\int_0^t\int_{\mathbb{R}}\int_{\mathbb{R}_+}J^{(n)}_{bL}(s,P_b^{(n)}(s-)-y,z)\delta_v^{(n)}M_{bL}^{(n)}(ds,dy,dz)\cr
   \ar\ar -\int_0^t\int_{\mathbb{R}}\int_{\mathbb{R}_+}J^{(n)}_{bC}(s,P_b^{(n)}(s-)-y,z)\delta_v^{(n)} M_{bC}^{(n)}(ds,dy,dz)
 \eeqnn
 and
 \beqnn
 J^{(n)}_{IL}(s,y,z)\ar=\ar (e^{z}-1)\int_{\Delta^{(n)}(y)}
 \frac{f_I(x)}{\delta_x^{(n)}}dx,\cr
 J^{(n)}_{IC}(s,y,z)\ar=\ar (e^{-z}-1)V_I^{(n)}(s-,y) \int_{\Delta^{(n)}(y)}\frac{f_I(x)}{\delta_x^{(n)}}dx.
 \eeqnn
 For any $x\in\mathbb{R}$, let $\mathbf{0}_1(x)=(x,0,0,0)$ and let $\mathbf{0}_i$ $(i=2,3,4)$ be defined similarly.
  By It\^o's formula, for any $G(x)\in C_b^2(\mathbb{R}^4)$,
 \beqnn
 \mathcal{M}^{(n)f}_G(t)\ar:=\ar
   G(\mathbf{S}^{(n)f}(t))-G(\mathbf{S}^{(n)f}(0)) - \int_0^t \frac{ \mathrm{D}_{\mathbf{0}_1(\delta^{(n)}_x)}G(\mathbf{S}^{(n)f}(s))}{\delta_x^{(n)}}\rho_{bM}^{(n)}(\mathbf{S}^{(n)}(s))\beta_a^{(n)}(s)ds\cr
 \ar\ar - \int_0^t \frac{ \mathrm{D}_{\mathbf{0}_1(\delta^{(n)}_x)}G(\mathbf{S}^{(n)f}(s))}{\delta_x^{(n)}}\frac{\rho_{bM}^{(n)}(\mathbf{S}^{(n)}(s))-\rho_{aL}^{(n)}(\mathbf{S}^{(n)}(s))}{\delta_x^{(n)}}|\delta_x^{(n)}|^2\mu^{(n)}_{aL}(s)ds\cr
 \ar\ar - \int_0^t\frac{ \mathrm{D}_{\mathbf{0}_1(\delta^{(n)}_x)}G(\mathbf{S}^{(n)f}(s))+ \mathrm{D}_{-\mathbf{0}_1(\delta^{(n)}_x)}G(\mathbf{S}^{(n)f}(s))}{|\delta_x^{(n)}|^2} |\delta_x^{(n)}|^2\rho_{aL}^{(n)}(\mathbf{S}^{(n)}(s))\mu^{(n)}_{aL}(s)ds\cr
 \ar\ar - \int_0^t \frac{ \mathrm{D}_{\mathbf{0}_2(\delta^{(n)}_x)}G(\mathbf{S}^{(n)f}(s))}{\delta_x^{(n)}}\rho_{bL}^{(n)}(\mathbf{S}^{(n)}(s))\beta_b^{(n)}(s)ds\cr
 \ar\ar - \int_0^t \frac{ \mathrm{D}_{\mathbf{0}_2(\delta^{(n)}_x)}G(\mathbf{S}^{(n)f}(s))}{\delta_x^{(n)}}\frac{\rho_{bL}^{(n)}(\mathbf{S}^{(n)}(s))-\rho_{aM}^{(n)}(\mathbf{S}^{(n)}(s))}{\delta_x^{(n)}}\delta_x^{(n)}\mu^{(n)}_{aM}(s)ds\cr
 \ar\ar- \int_0^t \frac{ \mathrm{D}_{\mathbf{0}_2(\delta^{(n)}_x)}G(\mathbf{S}^{(n)f}(s))+ \mathrm{D}_{-\mathbf{0}_2(\delta^{(n)}_x)}G(\mathbf{S}^{(n)f}(s))}{|\delta_x^{(n)}|^2} \rho_{aM}^{(n)}(\mathbf{S}^{(n)}(s))|\delta_x^{(n)}|^2\mu^{(n)}_{aM}(s)ds\cr
 \ar\ar - \int_0^t\int_{\mathbb{R}}\int_{\mathbb{R}_+} \frac{\mathrm{D}_{\mathbf{0}_3(J^{(n)}_{aL}(s,y,z)\delta_v^{(n)})}G(\mathbf{S}^{(n)f}(s))}{J^{(n)}_{aL}(s,y,z)\delta_v^{(n)}} J^{(n)}_{aL}(s,y,z)\delta_v^{(n)}\lambda^{(n)}_{aL}(s,y-P_a^{(n)}(s))\nu_{aL}(dz)dyds\cr
 \ar\ar -\int_0^t\int_{\mathbb{R}}\int_{\mathbb{R}_+}
 \frac{\mathrm{D}_{\mathbf{0}_3(J^{(n)}_{aC}(s,y,z)\delta_v^{(n)})}G(\mathbf{S}^{(n)f}(s))}{J^{(n)}_{aC}(s,y,z)\delta_v^{(n)}}J^{(n)}_{aC}(s,y,z)\delta_v^{(n)}\lambda^{(n)}_{aC}(s,y-P_a^{(n)}(s))\nu_{aC}(dz)dyds\cr
 \ar\ar - \int_0^t\int_{\mathbb{R}}\int_{\mathbb{R}_+} \frac{\mathrm{D}_{\mathbf{0}_4(-J^{(n)}_{bL}(s,y,z)\delta_v^{(n)})}G(\mathbf{S}^{(n)f}(s))}{J^{(n)}_{bL}(s,y,z)\delta_v^{(n)}}J^{(n)}_{bL}(s,y,z)\delta_v^{(n)}\lambda^{(n)}_{bL}(s,P_b^{(n)}(s)-y)\nu_{bL}(dz)dyds\cr
 \ar\ar -\int_0^t\int_{\mathbb{R}}\int_{\mathbb{R}_+}
 \frac{\mathrm{D}_{\mathbf{0}_4(-J^{(n)}_{bC}(s,y,z)\delta_v^{(n)})}G(\mathbf{S}^{(n)f}(s))}{J^{(n)}_{bC}(s,y,z)}J^{(n)}_{bC}(s,y,z)\delta_v^{(n)}\lambda^{(n)}_{bC}(s,P_b^{(n)}(s)-y)\nu_{bC}(dz)dyds
 \eeqnn
 is a martingale,  where $\mathrm{D}$ is a difference operator defined by $\mathrm{D}_xF(y):=F(y+x)-F(y)$.

 In proving that $\{\mathcal{M}^{f(n)}_G(t)\}_{n\geq 1}$ converges almost surely we may without loss of generality assume that $G$ along with its first and second derivative is bounded by 1.
 From Condition~\ref{C3.2} ii),
 \beqnn
 |\rho_{bM}^{(n)}(\mathbf{S}^{(n)}(t))-\rho_{bM}(\mathbf{S}(t))|\leq |\rho_{bM}^{(n)}(\mathbf{S}^{(n)}(t))-\rho_{bM}(\mathbf{S}^{(n)}(t))|+|\rho_{bM}(\mathbf{S}^{(n)}(t))-\rho_{bM}(\mathbf{S}(t))|\to 0.
 \eeqnn
 From the mean value theorem, there exists some $\xi^{(n)}_s\in[\mathbf{S}^{f(n)}(s),\mathbf{S}^{f(n)}(s)+\mathbf{0}_1(\delta^{(n)}_x)]$ such that
 \beqnn
 \Big|\frac{\mathrm{D}_{\mathbf{0}_1(\delta^{(n)}_x)}G(\mathbf{S}^{(n)f}(s))}{\delta_x^{(n)}}-G'(\mathbf{S}^{f}(s))\Big|
 =|\frac{\partial G(\xi^{(n)}_s)}{\partial P_a}-\frac{\partial G(\mathbf{S}^{f}(s))}{\partial P_a}|\leq \|\xi^{(n)}_s-\mathbf{S}^{f}(s)\|_{\mathcal{S}^2}\leq \delta_x^{(n)}.
 \eeqnn
Using similar arguments as above, from these two results and the dominated convergence theorem,
 \beqnn
 \int_0^t \frac{ \mathrm{D}_{\mathbf{0}_1(\delta^{(n)}_x)}G(\mathbf{S}^{(n)f}(s))}{\delta_x^{(n)}}\rho_{bM}^{(n)}(\mathbf{S}^{(n)}(s))\beta_a^{(n)}(s)ds \to \int_0^t \frac{\partial G(\mathbf{S}^{f}(s))}{\partial P_a}\rho_{bM}(\mathbf{S}(s))\beta_a(s)ds, \quad a.s.
 \eeqnn
 Applying the mean value theorem again, we also have
 \beqnn
 \frac{ \mathrm{D}_{\mathbf{0}_1(\delta^{(n)}_x)}G(\mathbf{S}^{(n)f}(s))+ \mathrm{D}_{-\mathbf{0}_1(\delta^{(n)}_x)}G(\mathbf{S}^{(n)f}(s))}{|\delta_x^{(n)}|^2}\to \frac{\partial^2 G(\mathbf{S}^{f}(s)))}{\partial P_a^2}
 \eeqnn
 and
 \beqnn
 \lefteqn{\int_0^t\frac{ \mathrm{D}_{\mathbf{0}_1(\delta^{(n)}_x)}G(\mathbf{S}^{(n)f}(s))+ \mathrm{D}_{-\mathbf{0}_1(\delta^{(n)}_x)}G(\mathbf{S}^{(n)f}(s))}{|\delta_x^{(n)}|^2} |\delta_x^{(n)}|^2\rho_{aL}^{(n)}(\mathbf{S}^{(n)}(s))\mu^{(n)}_{aL}(s)ds}\qquad\qquad\qquad\qquad\qquad\qquad\ar\ar\cr
 \ar\ar\to \int_0^t\frac{\partial^2 G(\mathbf{S}^{f}(s))}{\partial P_a^2} \rho_{aL}(\mathbf{S}(s))\mu_{aL}(s)ds,\quad a.s.
 \eeqnn
 In terms of the function $\tilde{\lambda}_{IK}$ defined in the proof of Theorem~\ref{T2}, we have that
 \beqnn
 \|\lambda_{aC}(s,\cdot-P_a^{(n)}(s))-\lambda_{aC}(s,\cdot-P_a(s))\|_{L^2} \ar\leq \ar \|\tilde{\lambda}_{aC}(s,\cdot-P_a^{(n)}(s))-\tilde{\lambda}_{aC}(s,\cdot-P_a(s))\|_{L^2}\cr
 \ar\ar + \|\hat\lambda_{aC}(s,\cdot-P_a^{(n)}(s))-\hat\lambda_{aC}(s,\cdot-P_a(s))\|_{L^2}.
 \eeqnn
 From (\ref{eqn2.8}) and (\ref{eqn5.15}),
 \beqnn
 \|\lambda_{aC}(s,\cdot-P_a^{(n)}(s))-\lambda_{aC}(s,\cdot-P_a(s))\|_{L^2}\leq C_0|P_a^{(n)}(s)-P_a(s)|\to 0,\quad a.s.
 \eeqnn
 Similar to the argument above, we have
 \beqnn
 \frac{\mathrm{D}_{\mathbf{0}_3(J^{(n)}_{aC}(s,y,z)\delta_v^{(n)})}G(\mathbf{S}^{(n)f}(s))}{J^{(n)}_{aC}(s,y,z)\delta_v^{(n)}}\to \frac{ \partial G(\mathbf{S}^{f}(s))}{\partial V_a^{f}},\quad a.s.
 \eeqnn
 Like the time stopping argument in Step~2, we can prove
 \beqnn
 \int_0^t\int_{\mathbb{R}}\Big|[V_I^{(n)}(s-,y)-V_I(s,y)] \int_{\Delta^{(n)}(y)}\frac{f_I(x)}{\delta_x^{(n)}}dx\Big|dyds\to 0,\quad a.s.
 \eeqnn
 Thus
  \beqnn
 \lefteqn{\int_0^t\int_{\mathbb{R}}\Big|\int_{\mathbb{R}_-}
 J^{(n)}_{aC}(s,y,z)\nu^{(n)}_{aC}(dz)-\alpha_{aC}V_I(s,y)f(y)\Big|dyds}\ar\ar\cr
 \ar\leq\ar \alpha_{aC}\int_0^t\int_{\mathbb{R}}\Big|[V_I^{(n)}(s-,y)-V_I(s,y)] \int_{\Delta^{(n)}(y)}\frac{f_I(x)}{\delta_x^{(n)}}dx\Big|dyds\cr
 \ar\ar +\alpha_{aC}\int_0^t\int_{\mathbb{R}}V_I(s,y) \Big| \int_{\Delta^{(n)}(y)}\frac{f_I(x)}{\delta_x^{(n)}}dx-f(y)\Big|dyds\cr
 \ar\leq\ar +\alpha_{aC}\int_0^t\int_{\mathbb{R}}\Big|[V_I^{(n)}(s-,y)-V_I(s,y)] \int_{\Delta^{(n)}(y)}\frac{f_I(x)}{\delta_x^{(n)}}dx\Big|dyds\cr
 \ar\ar +\alpha_{aC}\int_0^t\int_{\mathbb{R}}V_I(s,y) \Big| \int_{\Delta^{(n)}(y)}\frac{f_I(x)}{\delta_x^{(n)}}dx-f(y)\Big|dyds \to 0,\quad a.s.
 \eeqnn
 Putting the last three results together, we have
 \beqnn
 \lefteqn{\int_0^t\int_{\mathbb{R}}\int_{\mathbb{R}_-}
 \frac{\mathrm{D}_{\mathbf{0}_3(J^{(n)}_{aC}(s,y,z)\delta_v^{(n)})}G(\mathbf{S}^{f(n)}(s))}{J^{(n)}_{aC}(s,y,z)\delta_v^{(n)}}J^{(n)}_{aC}(s,y,z)\delta_v^{(n)}\lambda^{(n)}_{aC}(s,y-P_a^{(n)}(s))\nu_{aC}(dz)dyds}\qquad\qquad\qquad\qquad\qquad\ar\ar\cr
 \ar\to\ar  \alpha_{aL}\int_0^t\int_{\mathbb{R}}\frac{ \partial G(\mathbf{S}^{f}(s))}{\partial V_a^{f}} f_a(y)\lambda_{aL}(s,y-P_a(s))dyds,\quad a.s.
 \eeqnn
 The other terms can be analyzed similarly. Thus,
 \beqnn
 \mathcal{M}^{f(n)}_G(t)\overset{\rm a.s.}\to\mathcal{M}^{f}_G(t)
 \ar=\ar G(\mathbf{S}^{f}(t))-G(\mathbf{S}^{f}(0)) - \int_0^t \frac{ \partial G(\mathbf{S}^{f}(s))}{\partial P_a}\rho_{bM}(\mathbf{S}(s))\beta_a(s)ds\cr
 \ar\ar - \int_0^t \frac{ \partial G(\mathbf{S}^{f}(s))}{\partial P_a}\varrho_a(\mathbf{S}(s))\mu_{aL}(s)ds - \int_0^t\frac{\partial^2 G(\mathbf{S}^{f}(s))}{\partial P_a^2} \rho_{aL}(\mathbf{S}(s))\mu_{aL}(s)ds\cr
 \ar\ar - \int_0^t \frac{ \partial G(\mathbf{S}^{f}(s))}{\partial P_b}\rho_{bL}(\mathbf{S}(s))\beta_b(s)ds - \int_0^t \frac{ \partial G(\mathbf{S}^{f}(s))}{\partial P_b}\varrho_{b}(\mathbf{S}(s))\mu_{aM}(s)ds\cr
 \ar\ar- \int_0^t \frac{ \partial^2 G(\mathbf{S}^{f}(s))}{\partial P^2_b} \rho_{aM}(\mathbf{S}(s))\mu_{aM}(s)ds\cr
 \ar\ar - \int_0^t\int_{\mathbb{R}}\frac{ \partial G(\mathbf{S}^{f}(s))}{\partial V_a^{f}} \alpha_{aL}f_a(y)\lambda_{aL}(s,y-P_a(s))dyds\cr
 \ar\ar -\int_0^t\int_{\mathbb{R}}
 \frac{ \partial G(\mathbf{S}^{f}(s))}{\partial V_a^{f}}\alpha_{aC}f_a(y)V_a(s,y)\lambda_{aL}(s,y-P_a(s))dyds\cr
 \ar\ar - \int_0^t\int_{\mathbb{R}} \frac{ \partial G(\mathbf{S}^{f}(s))}{\partial V_b^{f}}\alpha_{bL}f_b(y)\lambda_{bL}(s,P_b(s)-y)dyds\cr
 \ar\ar - \int_0^t\int_{\mathbb{R}}
 \frac{ \partial G(\mathbf{S}^{f}(s))}{\partial V_b^{f}}\alpha_{bC}f_b(y)V_a(s,y)\lambda_{bC}(s,P_b(s)-y)dyds.
 \eeqnn
 Moreover, since $G\in C_b^2(\mathbb{R}^2)$, by H\"older inequality and Cauchy-Schwarz inequality, it is easy to see
  \beqnn
 \sup_{t\in[0,T]}|\mathcal{M}^{f(n)}_G(t)|^2
 \ar\leq\ar C_0 + C_0T\int_0^T \Big[|\beta_a^{(n)}(s)|^2+||\delta_x^{(n)}|^2\mu^{(n)}_{aL}(s)|^2+ |\beta_b^{(n)}(s)|^2+||\delta_x^{(n)}|^2\mu^{(n)}_{aM}(s)|^2\Big]ds\cr
 \ar\ar + C_0T\int_0^T\|\delta_v^{(n)}\lambda^{(n)}_{aL}(s,y)\|^2_{L^2}ds  + C_0T\int_0^T\|\delta_v^{(n)}\lambda^{(n)}_{bL}(s,y)\|^2_{L^2}ds  \cr
 \ar\ar +C_0T\int_0^T[\|V_{a}^{(n)}(s,\cdot) \|^4_{L^4}+
 \|V_{b}^{(n)}(s,\cdot)\|_{L^4}^4
 +\|\delta_v^{(n)}\lambda^{(n)}_{aC}(s,\cdot)\|_{L^4}^4+ \|\delta_v^{(n)}\lambda^{(n)}_{aC}(s,\cdot)\|_{L^4}^4]ds\cr
 \ar\leq\ar C_0\Big(1+\sup_{t\in[0,T]}\|\mathbf{D}^{(n)}\|_{\mathcal{D}_2^2}^2+\sup_{t\in[0,T],I\in\mathcal{I}}[|\beta_I^{(n)}(t)|^2+\|V_{I}^{(n)}(s,\cdot) \|^4_{L^4}+\|\delta_v^{(n)}\lambda^{(n)}_{IC}(s,\cdot)\|_{L^4}^4]\Big).
 \eeqnn
 This inequality along with Lemma~\ref{BL01}, \ref{BL04}, \ref{BL05}, \ref{BL06} and Proposition~\ref{BP03} proves that the sqeuence
 $\{\mathcal{M}^{f(n)}_G(t)\}_{n\geq 1}$ is uniformly integrable. Hence, since $\{\mathcal{M}^{f(n)}_G(t)\}_{n\geq 1}$ converges almost surely, it also converges to $\mathcal{M}^{f}_G(t)$ in $L^1(\mathbf{P})$.
 As a result, $\{\mathcal{M}^{f}_G(t):t\geq 0\}$ is a martingale.
 A standard stopping argument shows that $\{\mathcal{M}^{f}_G(t)\}$ is a local martingale for any $G\in C^2(\mathbb{R}^4)$. Thus $\mathbf{S}$ solves the martingale problem associated to $\mathscr{A}^f_t$.
 \qed

\bibliographystyle{abbrv}

\bibliography{References}

\end{document}